\documentclass[usenatbib]{mnras}

\usepackage{newtxtext,newtxmath}

\usepackage[T1]{fontenc}

\DeclareRobustCommand{\VAN}[3]{#2}
\let\VANthebibliography\thebibliography
\def\thebibliography{\DeclareRobustCommand{\VAN}[3]{##3}\VANthebibliography}

\usepackage{graphicx}	
\usepackage{color}
\usepackage{tabularx}
\usepackage{here}
\usepackage{float}
\usepackage{threeparttable}
\usepackage{tabu}
\usepackage{dcolumn}
\usepackage{enumitem}
\usepackage{booktabs}
\usepackage{rotating}
\usepackage{adjustbox}
\usepackage{blindtext}
\usepackage{pdflscape}
\usepackage{lipsum}

\newcommand{\oi}{[O\,{\sc i}]}
\newcommand{\oii}{[O\,{\sc ii}]}
\newcommand{\oiii}{[O\,{\sc iii}]}
\newcommand{\oiv}{[O\,{\sc iv}]}

\newcommand{\NI}{[N\,{\sc i}]}
\newcommand{\nii}{[N\,{\sc ii}]}
\newcommand{\niii}{N\,{\sc iii}]}

\newcommand{\sii}{[S\,{\sc ii}]}
\newcommand{\siii}{[S\,{\sc iii}]}

\newcommand{\hei}{He\,{\sc i}}
\newcommand{\heii}{He\,{\sc ii}}

\newcommand{\neiv}{[Ne\,{\sc iv}]}

\newcommand{\ariii}{[Ar\,{\sc iii}]}
\newcommand{\ariv}{[Ar\,{\sc iv}]}
\newcommand{\arv}{[Ar\,{\sc v}]}

\newcommand{\clii}{[Cl\,{\sc ii}]}
\newcommand{\cliii}{[Cl\,{\sc iii}]}
\newcommand{\cliv}{[Cl\,{\sc iv}]}
\newcommand{\kriv}{[Kr\,{\sc iv}]}
\newcommand{\fevi}{[Fe\,{\sc vi}]}
\newcommand{\fevii}{[Fe\,{\sc vi}]}

\newcommand{\mgv}{[Mg\,{\sc v}]}
\newcommand{\feiii}{[Fe\,{\sc iii}]}
\newcommand{\kiv}{[K\,{\sc iv}]}

\newcommand{\cii}{C\,{\sc ii}}
\newcommand{\ciii}{C\,{\sc iii}}
\newcommand{\civ}{C\,{\sc iv}}

\newcommand{\ha}{H$\alpha$}
\newcommand{\hb}{H$\beta$}
\newcommand{\hg}{H$\gamma$}
\newcommand{\hi}{H\,{\sc i}}

\newcommand{\kms}{km\,s$^{-1}$}

\newcommand{\te}{$T_{\rm e}$}
\newcommand{\Ne}{$n_{\rm e}$}

\title[Spatially resolved study of PN IC2165]
{Seimei KOOLS-IFU Mapping of the Gas and Dust Distributions in 
Galactic Planetary Nebulae: the Case of IC2165\thanks{This work is based on the program IDs 
19B-N-CN07, 19B-K-0014, 20B-N-CN06, and 20B-K-0015 (PI: M.~Otsuka).}}

\author[M.~Otsuka]{
Masaaki Otsuka$^{1}$\thanks{E-mail: otsuka@kusastro.kyoto-u.ac.jp} 
\\
$^{1}$Okayama Observatory, Kyoto University, Kamogata, Asakuchi, Okayama, 719-0232, Japan\\
}

\date{Accepted 2022 January 26. Received 2022 January 23; in original form 2021 November 19}

\pubyear{2022}

\begin{document}
\label{firstpage}
\pagerange{\pageref{firstpage}--\pageref{lastpage}}
\maketitle

\begin{abstract}
We investigated the physical and chemical properties of the gas and dust components in a carbon-rich planetary nebula (PN) IC2165 using two-dimensional (2-D) emission-line maps with superior resolution. The extinction map is generated in a self-consistent and assumption-free manner. The circumstellar gas-to-dust mass ratio (GDR) map ranges radially from 1210 in the central nebula filled with hot gas plasma to 120 near the ionisation front. The determined GDR is comparable to $\sim$400, which is commonly adopted for carbon-rich asymptotic giant branch (AGB) stars, and $\sim$100 for ISM. Except for the inner regions, the GDR in IC2165 is nearly the same as in such AGB stars, indicating that most dust grains withstand the harsh radiation field without being destroyed. The gas and dust mass distributions concentrated in the equatorial plane may be related to the nonisotropic mass loss during the AGB phase and nebula shaping. 
The spatial distributions of electron densities/temperatures and ionic/elemental abundances were investigated herein. We determined 13 elemental abundances using PSF-matched spatially integrated multiwavelength spectra extracted from the same aperture. 
Their values are consistent with values predicted by a theoretical model for stars of initially 1.75\, M$_{\sun}$ and $Z=0.003$. 
Finally, we constructed the photoionisation model using our distance measurement to be consistent with all derived quantities, including the GDR and gas and dust masses and post-AGB evolution. 
Thus, we demonstrate the capability of Seimei/KOOLS-IFU and how the spatial variation of the gas and dust components in PNe derived from IFU observations can help understand the evolution of the circumstellar/interstellar medium.
\end{abstract}


\section{Introduction}
\label{S-intro}

\begin{figure*}
 \begin{tabular}{@{}c@{\hspace{-4pt}}c@{\hspace{-4pt}}c@{}}
\includegraphics[width=0.34\textwidth]{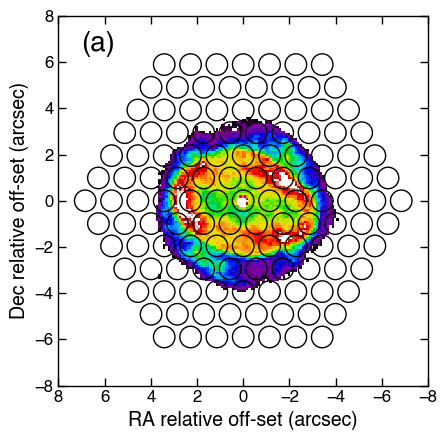}&
\includegraphics[width=0.34\textwidth]{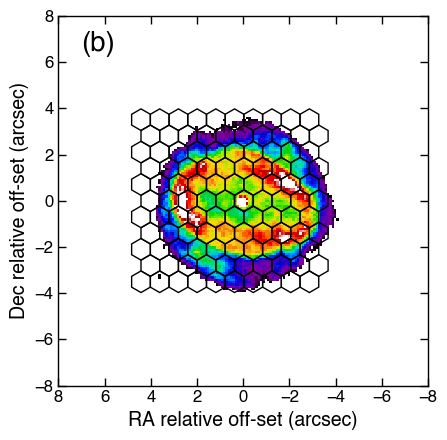}&
\includegraphics[width=0.34\textwidth]{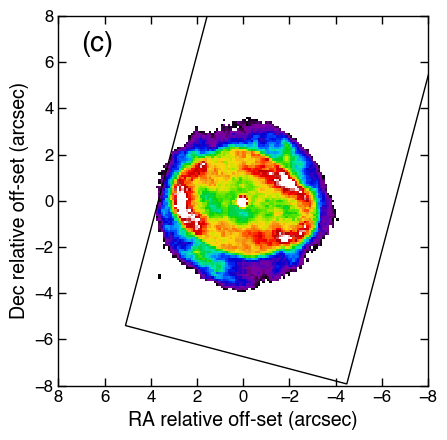}
 \end{tabular}
\caption{({\it left}) A dither position in KOOLS-IFU VPH blue/red observations of IC2165 overlaid on the \emph{HST}/WFPC2 F547M image 
(in the range from 15 and 100\,$\%$ of the maximum intensity).
Each circle indicates the position and dimension of 127 fibres. 
({\it centre}) A dither position in KOOLS-IFU VPH 683 observation overlaid on the same image.
Each hexagon indicates the position and dimension of 110 fibres. 
Both KOOLS-IFU and VIMOS (27{\arcsec}$\times$27{\arcsec} square field centred on the PN, not presented here) cover the entire nebula. 
({\it right}) A portion of the \emph{IUE} slit-entrance (9.9{\arcsec}$\times$22{\arcsec} in the entire size) 
overlaid on the same image. 
}
\label{F:specpos}
\end{figure*}

Planetary nebula (PN) is the final evolutionary stage of stars that had initial values of $1-8$\,M$_{\sun}$. 
PN comprises a central star and an extended circumstellar envelope (CSE) formed by a substantial mass loss during the asymptotic giant branch (AGB) phase. 
The CSE is partially ionised by UV photons emitted from a hot central star evolving into a white dwarf.
The physical and chemical characteristics of the CSE are determined through internal and external processes that occurred before and after the PN phase. The metal abundances of the CSE reflect the surrounding chemical environment (where the progenitor formed), AGB nucleosynthesis, and mass loss. The stellar-synthesised metals play an important role in the shaping of CSE because the dust grains formed from these metals control stellar mass loss. The metal-enriched ejecta from the PN progenitors alter the chemical properties of the surrounding interstellar medium (ISM), eventually becoming a part of the ISM of the host galaxies. Thus, the chemical evolution of the Universe is the direct consequence of material cycling between stellar mass loss and ISM. 
Therefore, the physical and chemical properties of the CSEs 
of PNe is an area of active research.

Three-dimensional (3-D) spectroscopy has validated spatial variation of the physical and chemical properties of the CSE of the PN. 
For instance, \citet{2013AaA...560A.102M} investigated the shell and rim structures of NGC3242 and determined electron temperature ({\te}) and density ({\Ne}), 
and elemental abundance distributions. \citet{Walsh:2016aa} revealed the dust-to-gas ratio distribution of NGC7009.
\citet{Garca-Rojas:2021aa} performed detailed plasma and abundance analyses of three PNe, demonstrating a high abundance discrepancy.
\citet{Danehkar:2021aa} demonstrated the morphological and kinematical structures of PNe.

However, several observations of PNe have been conducted by placing a narrow-entrance slit on a few positions or only a single position of the extended nebulae depending on the observers’ research motives and the capabilities of the instruments. It is not guaranteed that physical and chemical parameters derived from such observations correspond to representative values in target objects. 
Moreover, the physical and chemical properties in the PN’s CSE were believed to be spatially homogeneous based on one-dimensional (1-D) spectroscopy, 
but this is not in accordance with the results obtained by the recent 3-D spectroscopy. 
PNe and even stellar evolution may have been incorrectly understood using the spatially integrated observational data.

3-D spectroscopy is extremely useful for investigating the spatial distribution of gas and dust properties. 
It assists in providing a comprehensive understanding regarding 
the amount of mass of different types of gas and dust returned by the stars to the ISM 
and the time and the directions in which these gas and dust were returned. 
Owing to these reasons, we have performed a 3-D spectroscopic survey of Galactic PNe. 
This study focuses on the PN IC2165 using data from the Kyoto Okayama Optical Low-dispersion Spectrograph with optical-fiber integral field unit \citep[KOOLS-IFU;][]{Matsubayashi:2019aa} attached to one of the two Nasmyth foci of the Kyoto University Seimei 3.8-m telescope \citep{Kurita:2020aa} and archived data (Table\,\ref{T-log}).

IC2165 has been actively studied since the pioneering work of \cite{1942ApJ....95..356W} because of its high surface brightness. 
IC2165 is a carbon-rich PN \citep[i.e., number density ratio $n$(C)/$n$(O) $>1$, e.g., ][]{Hyung:1994ab,Pottasch:2004aa,Bohigas:2013aa,Miller:2019aa} 
comprising carbon-rich dust, exhibiting a featureless continuum in the infrared wavelengths and the 11\,{\micron} emission owing to carbonaceous (graphite or amorphous carbon) 
and silicon carbide grains \citep{1983MNRAS.203P...9R,Delgado-Inglada:2014aa}. 
Although C richness and a simple morphology \citep[][see Fig.\,\ref{F:specpos}]{Hua:1985aa} 
suggest a low-mass progenitor star of initially $\sim$1.5--3\,M$_{\sun}$, 
its origin remains unclear since the pre-Gaia distance is poorly constrained, ranging between 2.03 and 3.90\,kpc \citep{Cahn:1992aa,Frew:2013aa,Phillips:2004aa,Stanghellini:2010aa,Tajitsu:1998aa}, and its average value is $3.06 \pm 0.80$\,kpc. 
However, the distance is not improved even after using Gaia distance \citep[2.83$^{+0.97}_{-0.92}$\,kpc;][]{Chornay:2021aa}. In most cases, the initial mass of PNe has been evaluated by detecting the corresponding post-AGB evolution track of the observed effective temperature and luminosity. The distance is required for determining luminosity and any mass estimate. Thus, we need to recalculate the distance that maintains consistency between the observation results and the AGB theory.
\citet{Miller:2019aa} concluded that IC2165 is a chemically uniform PN by analysing the Hubble Space Telescope (\emph{HST}) spectra of the northern part of the elliptical nebula’s rim obtained with slits of a width of 0.2{\arcsec} and 0.5{\arcsec}. 
The spatial distribution of the physical and chemical parameters in the entire nebula remains unknown.

Therefore, this study of IC2165 aims at 
(1) investigating spatial distributions of the gas and dust, 
(2) obtaining the representative values of electron densities/temperatures and elemental abundances using the 
spatially integrated multiwavelength spectra extracted from the same aperture, 
(3) clarifying the origin and evolution of this PN, and 
(4) constructing the photoionisation model using our distance measurement results for consistency with all the observed/derived quantities in our 
2-D and 1-D analyses and the AGB/post-AGB evolution. Through these investigations into Galactic PNe, 
we deeply understand the PN evolution and also material recycling between stars and ISM in galaxies. 
The sections of this paper are organised as follows. 
In \S\,\ref{S:obs}, we explain the observations and the archival data used as well as their data reduction.
In \S\,\ref{S-Res}, we perform the 2-D and 1-D spectral analyses. 
We present discussions of the study in \S\,\ref{S-discuss} and summarise the current work in \S\,\ref{S-sum}.

\section{Dataset and reduction} \label{S:obs}

\subsection{Seimei/KOOLS-IFU optical spectra}
\label{S:kools}

\begin{table}
\centering
\caption{Summary of spectroscopic observations of IC2165. 
\label{T-log}}
\begin{tabularx}{\columnwidth}{@{}@{\extracolsep{\fill}}l@{\hspace{2pt}}l@{\hspace{-2pt}}c@{\hspace{-8pt}}D{p}{-}{-1}@{\hspace{-2pt}}c@{\hspace{-2pt}}c@{}}
    \midrule
Date        & Tel./Inst. &Grism &\multicolumn{1}{c}{Range}      &\multicolumn{1}{c}{$R$}              &Seeing \\
   &            &      &\multicolumn{1}{c}{({\AA})}    & &({\arcsec})\\ 
    \midrule
2019/12/19 & Seimei/KOOLS-IFU  &VPH-blue          &4200 p 8530        &$\sim$500     &1.2-1.3\\
2019/12/19 & Seimei/KOOLS-IFU  &VPH-red           &5600 p 10100       &$\sim$800     &1.2-1.3\\
2020/12/09 & Seimei/KOOLS-IFU  &VPH-683           &6500 p 7400        &$\sim$2200    &1.1-1.4\\
2007/09/03 & VLT/VIMOS         &HR-blue           &4150 p 6200        &$\sim$2550    &1.3-2.0\\
1981/03/26 & \emph{IUE}/SWP    &                  &1150 p 1978        &$\sim$300     &\\
1981/03/26 & \emph{IUE}/LWR    &                  &1852 p 3348        &$\sim$400     &\\
    \midrule	     
\end{tabularx}
\end{table}

\subsubsection{VPH-blue and -red low-resolution spectroscopic mapping}
\label{Kools-obs1}

KOOLS-IFU had sparse hexagonally-packed 127 fibres; the diameter of each fibre is 0.93{\arcsec} in the observation.
The field of view (FoV) has approximately 15{\arcsec} diameter on the sky (Fig.\,\ref{F:specpos}a). 

The sky conditions were very stable and the sky was very clear throughout the night 
(outside temperature of $4.5-5.2$\,$^{\circ}$C, 
relative humidity of $76-78$,$\%$, and 
pressure of 985.5\,hpa. These parameters are necessary to calculate the offset value at every wavelength owing to atmospheric dispersion). 
During the observations, the seeing evaluated by the Shack Hartmann (SH) wavefront sensor was $\sim$$1.2-1.3${\arcsec} at the full width at half maximum (FWHM). 
We used the VPH-blue and VPH-red grisms (with the order-sorting filter O56) whose usable spectral range was $4200-8530$\,{\AA} and $5600-10100$\,{\AA}, respectively. 
The airmass was $1.48-1.52$ (1.49 on average) in the VPH-blue and $1.48-1.56$ (1.53 on average) in the VPH-red settings. 
The detector used in both the VPH-blue/red and VPH-683 observations was a single Hamamatsu CCD with an effective size of 2048$\times$1640 pixels. 

To compensate for the gaps between fibres and obtain spatially resolved spectral images, we employed 5 positions $\times$ 5 points dithering with a step size of 0.45{\arcsec} for each position.
We applied a 60\,sec 
exposure at each 25 on- and 3 off-source positions. We observed a bright reference star HD44996 ($m_{V} = 6.12$, B2.5Ve) in close proximity to IC2165 prior to science exposure at each position to regulate the fibre position on IC2165 and correct the flux variations caused by atmospheric extinction and slight sky changes during observations by comparing the flux of this reference star.
For flux calibration and telluric absorption removal, we observed HR3454 ($m_{V} = 4.30$, B3V) at an airmass of 1.38 in both the VPH-blue and -red using a five points cross-pattern with a step size of 0.45{\arcsec}.

\subsubsection{VPH-683 moderate-resolution spectroscopic mapping}
\label{S-683}

The observation aims at effectively resolving the blends of {\ha} and {\nii}\,6548/83\,{\AA} lines.
KOOLS-fibre IFU's array was modified in October 2020; currently, it contains dense, square-packed 110 microlenses and fibres with the shape of a regular hexagon and a 0.81{\arcsec} across corner length.
 Consequently, the FoV is approximately 8{\arcsec}$\times$8{\arcsec} square field (Fig.\,\ref{F:specpos}b).

The sky conditions were very stable and the sky was clear throughout the night ($7.0-7.5$\,$^{\circ}$C, $80-85$\,$\%$, and 981.3\,hpa). 
During the observation, the seeing measured using the SH wavefront sensor was $\sim$$1.1-1.4${\arcsec} at FWHM.
We used the VPH-683 grism with the O56 filter, which exhibits a usable spectral range of $6500-7400$\,{\AA}.
The airmass was in the range of $1.48-1.63$ (1.51 on average). We used 9 positions $\times$ 6 points dithering with a step size of 0.48{\arcsec} along RA and 0.61{\arcsec} along DEC for each position.
We applied a 60\,sec exposure at each of the 54 on- and 3 off-source positions.
We observed HD44996 before each science exposure for the same reason as described in \S\,\ref{Kools-obs1}.

\subsubsection{Reduction of the KOOLS-IFU data}
\label{S-reduc}

We reduced the KOOLS-IFU data using NOAO/{\sc IRAF\footnote{IRAF is distributed by the National Optical Astronomy Observatories, which are operated by the Association of Universities for Research in Astronomy, Inc., under a cooperative agreement with the National Science Foundation.}} reduction packages and our programs, which performed overscan subtraction, residual bias pattern subtraction, bad pixel masking, cross-talk correction, scattered light subtraction between fibres, illumination and sensitivity correction, flux calibration, telluric removal, and conversion to heliocentric wavelength. Wavelength calibration was performed using $>40$ Hg/Ne/Xe lines that entirely covered the effective wavelengths. 
We adopted a constant plate scale of 2.5\,{\AA} in the VPH-blue/red spectra and 1.0\,{\AA} in the VPH-683 spectrum. 
For the VPH-blue, -red, and -683, the RMS errors in wavelength determination were 0.18, 0.08, and 0.09\,{\AA}, respectively.
The average spectral resolution ($R$), defined as $\lambda$/(Gaussian FWHM), measured from the Hg/Ne/Xe comparison lines was $\sim$500 for the VPH-blue spectrum, $\sim$800 for the VPH-red spectrum, and $\sim$2200 for the VPH-683 spectrum. 

We obtained an RA-DEC-wavelength datacube using our codes and image reconstruction based on the drizzle technique, and we corrected the spatial displacements due to atmospheric dispersion at each wavelength by following \citet{1967ApOpt...6...51O}. 
We adopted a constant plate scale of 0.412{\arcsec} per pixel. 
Using the fully processed standard stars' datacube, we evaluated the off-center value in RA and DEC relative to a reference position: the offset did not exceed 0.01{\arcsec} ($\sim$0.03\,pixel) in both RA and DEC for all the wavelengths in the KOOLS and ESO VLT/VIMOS data (\S\,\ref{S-ESO}) $0.000\pm0.002${\arcsec} (one-$\sigma$). 
A detailed explanation for the KOOLS-IFU data flux calibration is presented in Appendix \S\,\ref{A:flux}.

\subsection{ESO VLT/VIMOS optical spectrum}
\label{S-ESO}

We use the archived spectra data obtained using ESO VLT 8.2-m/Visible Multi-Object Spectrograph \citep[VIMOS;][]{2003SPIE.4841.1670L}. 
The VIMOS data are used to detect important diagnostic lines that are either too weak to detect or are beyond the wavelength ranges covered by KOOLS-IFU.
The observation was performed on September 3, 2007 (PI: H.~Monteiro, ID:079.D-0117(B)) under the seeing of $\sim$$1.32-2.03${\arcsec} at 5000\,{\AA} measured at 
an airmass of $1.42-1.61$ (1.51 on average) using a differential image motion monitor.  

VIMOS has square-packed 40$\times$40 fibres, each side of which is 0.67{\arcsec}. The detectors are four E2V 44-82 CCDs, and the effective size of each detector is 2048$\times$4096 pixels. The FoV is 27{\arcsec}$\times$27{\arcsec} square field centred on the PN. 
The used grism was HR-blue, covering a wavelength range of $4150-6200$\,{\AA}. 
Exposure times were $4 \times 480$\,s. The resultant spectral resolution was $\sim$2550. 
We adopted a constant plate scale of 0.54\,{\AA}. 
The flux standard star LTT2415 ($m_{V} = 12.38$, sdG) was observed at an airmass of $1.46-1.52$ (1.49 on average). 

We obtained the RA-DEC-wavelength datacube using the KOOLS-IFU data reduction method, and the spatial displacements due to atmospheric dispersion at each wavelength were corrected. In terms of spatial resolution, we used a constant plate scale of 0.412{\arcsec}.

\subsection{HST UV-optical images}
\label{S:HST}

\begin{table}
\caption{\emph{HST}/WFPC2 band flux density of IC2165.}
\label{T:HST}
\begin{tabularx}{\columnwidth}{@{\extracolsep{\fill}}clcD{p}{\pm}{-1}@{}}
\midrule
Band & Part & $\lambda_{\rm c}$ ({\AA}) & \multicolumn{1}{c}{$F_{\lambda}$ (erg\,s$^{-1}$\,cm$^{-2}$\,{\AA}$^{-1}$)}\\ 
\midrule
F185W & CSPN & 1977.1 & 2.21{\times}10^{-15} ~p~ 1.50{\times}10^{-17} \\ 
F218W & CSPN & 2204.4 & 1.35{\times}10^{-15} ~p~ 1.39{\times}10^{-17} \\ 
F255W & CSPN & 2600.4 & 1.45{\times}10^{-15} ~p~ 1.30{\times}10^{-17} \\ 
F336W & CSPN & 3359.5 & 9.83{\times}10^{-16} ~p~ 7.79{\times}10^{-18} \\ 
F439W & CSPN & 4312.1 & 5.18{\times}10^{-16} ~p~ 4.56{\times}10^{-18} \\ 
F547M & CSPN & 5483.9 & 2.82{\times}10^{-16} ~p~ 2.51{\times}10^{-18} \\ 
F185W & CSPN+PN & 1977.1 & 7.56{\times}10^{-14} ~p~ 2.73{\times}10^{-16} \\ 
F218W & CSPN+PN & 2204.4 & 2.99{\times}10^{-14} ~p~ 2.16{\times}10^{-16} \\ 
F255W & CSPN+PN & 2600.4 & 3.65{\times}10^{-14} ~p~ 1.32{\times}10^{-16} \\ 
F336W & CSPN+PN & 3359.5 & 7.36{\times}10^{-14} ~p~ 1.44{\times}10^{-17} \\ 
F439W & CSPN+PN & 4312.1 & 4.04{\times}10^{-14} ~p~ 2.33{\times}10^{-17} \\ 
F547M & CSPN+PN & 5483.9 & 2.25{\times}10^{-14} ~p~ 6.22{\times}10^{-18} \\
\midrule
\end{tabularx}
\end{table}

From Mikulski Archive for Space Telescopes (MAST), we downloaded the Hubble Space Telescope/Wide Field Planetary Camera2 (\emph{HST}/WFPC2) dataset of the six photometry bands (PI: J.~Westphal, ID:6281) taken on 
1996 November 26. We performed aperture photometry of the central star of the PN (CSPN) and 
the central star plus nebula (CSPN+PN). The respective F255W 
(\mbox{$\lambda_{\rm c} = 2600.4$\,{\AA}}) and 
F547M 
(\mbox{$\lambda_{\rm c} = 5483.9$\,{\AA}}) band flux densities of CSPN+PN 
are used to determine a scaling factor for KOOLS-IFU, VIMOS, and \emph{IUE} spectra 
(\S\,\ref{S:iue}). The results are summarised in Table\,\ref{T:HST}.

\subsection{International Ultraviolet Explorer (\emph{IUE}) UV spectra}
\label{S:iue}

We analysed the \emph{IUE} SWP13585 ($1150-1978$\,{\AA}, $R$$\sim$300, and PI: P.~J~Harrington) and LWR10218 spectra ($1852-3348$\,{\AA}, $R$$\sim$400, and same PI) to mainly measure the [{\cii}], [{\ciii}]+{\ciii}], [N\,{\sc iii, iv}], and [O\,{\sc iv}] lines and their abundances. We downloaded these spectra from MAST. The dimension and location of the slit entrance (9.9{\arcsec}~$\times$~22{\arcsec}) are shown in Fig.\,\ref{F:specpos}(c). We normalised the flux density of the overlapping SWP and LWR spectra to obtain a single $1150-3348$\,{\AA} spectrum.

\subsection{PSF of KOOLS-IFU, VIMOS, and \emph{IUE}}
\label{S:PSF}

For correctly combining the use of KOOLS-IFU, VIMOS, and \emph{IUE} data, we performed 
point-spread function (PSF) matching at every wavelength because the spatial resolution 
is different depending on the instruments. 
This is an essential step for both 2-D and 1-D spectral analyses. 
KOOLS-IFU fibre sampling size is undersampling for the astronomical targets and 
produces effective PSF FWHM of $\gtrsim 2${\arcsec} in the resultant images. 
This is larger than typical atmospheric seeing size ($1.5-2${\arcsec}) at the 
Okayama observatory. So, we perform PSF deconvolution in order to obtain the spatially-resolved 
emission images. To do so, we need to establish the FWHM-wavelength relationship. 
Due to the spatial resolution of KOOLS-IFU and observatory site seeing, the central star is hardly resolved in compact objects 
such IC2165. In general, the use of the central star as PSF is not preferred: 
because the central star is embedded in the nebula (so we cannot measure the profile in low intensity level) 
and low signal-to-noise ratio (SNR) due to its faintness. 
Therefore, for KOOLS-IFU, we establish the FWHM-wavelength relationships via Moffat function 
fitting FWHM measurements of the HR3454 and HD44996 images at each wavelength.

\begin{eqnarray}
I(r) &=& \left[1 + \left(r/{\alpha}\right)^{2}\right]^{-\beta} \\
{\alpha} &=& {\rm FWHM}\left[2^{\left(1/{\beta}\right)} - 1 \right]^{-0.5}, 
\label{E:moffat}
\end{eqnarray}

\noindent where $I(r)$ is the radial profile of these PSF standard stars and $\beta$ is the shape parameter. 
$\beta$ is $5.15 \pm 0.08$ in the VPH-blue/red spectra, and its value is $5.17 \pm 0.05$ in the VPH-683 spectrum. 
The effect of FWHM degradation due to airmass must then be considered.
Measurements based on the KOOLS-IFU spectral images taken during two nights 
in 2020 Jan and Feb reveal that its quality, defined by the FWHM of point sources, 
is degraded by airmass with a power of $0.42 \pm 0.01$. 
During both nights, the sky conditions were similar to those observed during the observation runs for IC2165. 
Thus, during the exposure to IC2165, the expected PSF's FWHM in arcsec can be expressed 
as a function of $\lambda$ in {\AA} using the following equation. 
The covered PSF's FWHM in IC2165 is in the range from 2.34{\arcsec} at 9900\,{\AA} to 2.51{\arcsec} at 4200\,{\AA}.

\begin{eqnarray}
{\rm FWHM} = \left\{ \begin{array}{ll}
(2.59\pm0.01)\,(\lambda/5500)^{-0.094 \pm 0.01} & \mbox{in blue}\\
(2.51\pm0.02)\,(\lambda/5500)^{-0.119 \pm 0.02} & \mbox{in red}\\
(2.58\pm0.01)\,(\lambda/5500)^{-0.070 \pm 0.02} & \mbox{in 683}
\end{array} \right.
\label{E:psf}
\end{eqnarray}

For VIMOS, we determined the FWHM-wavelength relationship by fitting the Moffat function to 
the standard star LTT2415 images at each wavelength using Eq.\,(\ref{E:moffat}).
Because the airmass of IC2165 is not significantly different from that of LTT2415,
 we do not consider image quality degradation due to airmass. $\beta$ is $4.56 \pm 0.02$. 
FWHM can be expressed in the following manner.

\begin{eqnarray}
{\rm FWHM} = (1.45 \pm 0.01)\,(\lambda/5500)^{-0.188 \pm 0.006} & \mbox{in VIMOS}
\label{E:psf2}
\end{eqnarray}

For \emph{IUE}, we adopt the FWHM-wavelength relations presented in Fig.\,3(b) of \citet{Cassatella:1985aa}. 
They can be approximately expressed as 

\begin{eqnarray}
{\rm FWHM} = \left\{ \begin{array}{ll}
2.54\,(\lambda/5500) + 1.18 & \mbox{in SWP}\\
1.66\,(\lambda/10^{3})^{2} - 9.49\,(\lambda/10^{3}) + 17.54 &\mbox{in LWR}\\
\end{array} \right.
\label{E:psf2}
\end{eqnarray}

\subsection{Absolute flux calibrated spectra}
\label{S:ABS}

\begin{figure}
\centering
\includegraphics[width=\columnwidth]{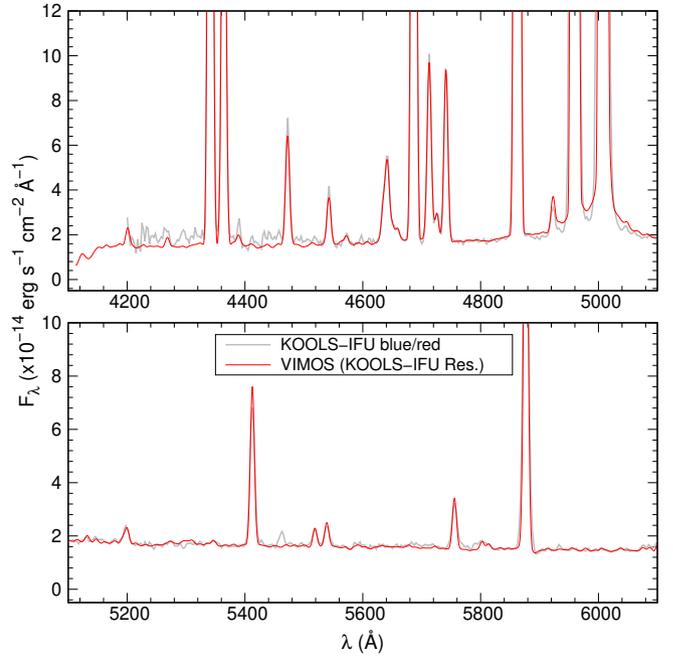}
\vspace{-10pt}
\caption{
Comparison between the KOOLS-IFU VPH blue/red (grey line) and the VIMOS spectra (red line) 
extracted in the \emph{IUE} aperture. 
The spectral resolution of the VIMOS spectrum is matched with that of the KOOLS-IFU spectrum with a Gaussian kernel. 
\label{F:koolsspec3}
}
\end{figure}

The absolute flux calibration for KOOLS-IFU was performed by comparing the PSF-matched \emph{HST}/WFPC2 F547M image with the corresponding pseudo image generated from the KOOLS-IFU data and the WFPC2/F547M filter transmission curve.
Thus, we adopted a constant scaling factor of 1.015 and 0.992 in the reduced VPH-blue/red data, respectively. 
We scaled the VPH-683 data to match with the absolute flux calibrated VPH-blue/red data: the adopted scaling factor is 1.046. Similarly, by comparing the flux calibrated VIMOS spectra to the absolute flux calibrated KOOLS-IFU spectra, we determined a constant scaling factor of 1.051.
Finally, for the \emph{IUE} spectrum, we determined a scaling factor of 1.152 
to be consistent with the band flux density in the \emph{HST} WFPC2/F255W image 
convoluted by the Gaussian kernel with FWHM at 2600\,{\AA} = 4.08{\arcsec} using Eq.\,(\ref{E:psf2}).

Fig.\,\ref{F:koolsspec3} displays the VIMOS spectrum overlaid on the KOOLS-IFU VPH-blue/red spectra. 
These are spatially integrated 1-D spectra which are extracted from the \emph{IUE} aperture after their PSFs were matched. 
The VIMOS spectrum exhibits the same spectral resolution as the KOOLS-IFU spectrum with a Gaussian kernel. 
Both spectra are nearly identical, which is apparent. 
Thus, we can confidently combine the KOOLS-IFU and VIMOS data for further analysis.

\begin{figure*}
  \begin{adjustbox}{addcode={\begin{minipage}{\width}}
{\caption{
Selected 2-D emission-line images of IC2165. In each panel, spatial resolution evaluated by the Gaussian PSF FWHM measured from the PSF deconvolved 
standard stars is indicated by the filled grey circle with the radius of a half of Gaussian PSF FWHM. 
\label{F-maps}
      }\end{minipage}},rotate=90,center}
\includegraphics[width=\textheight]{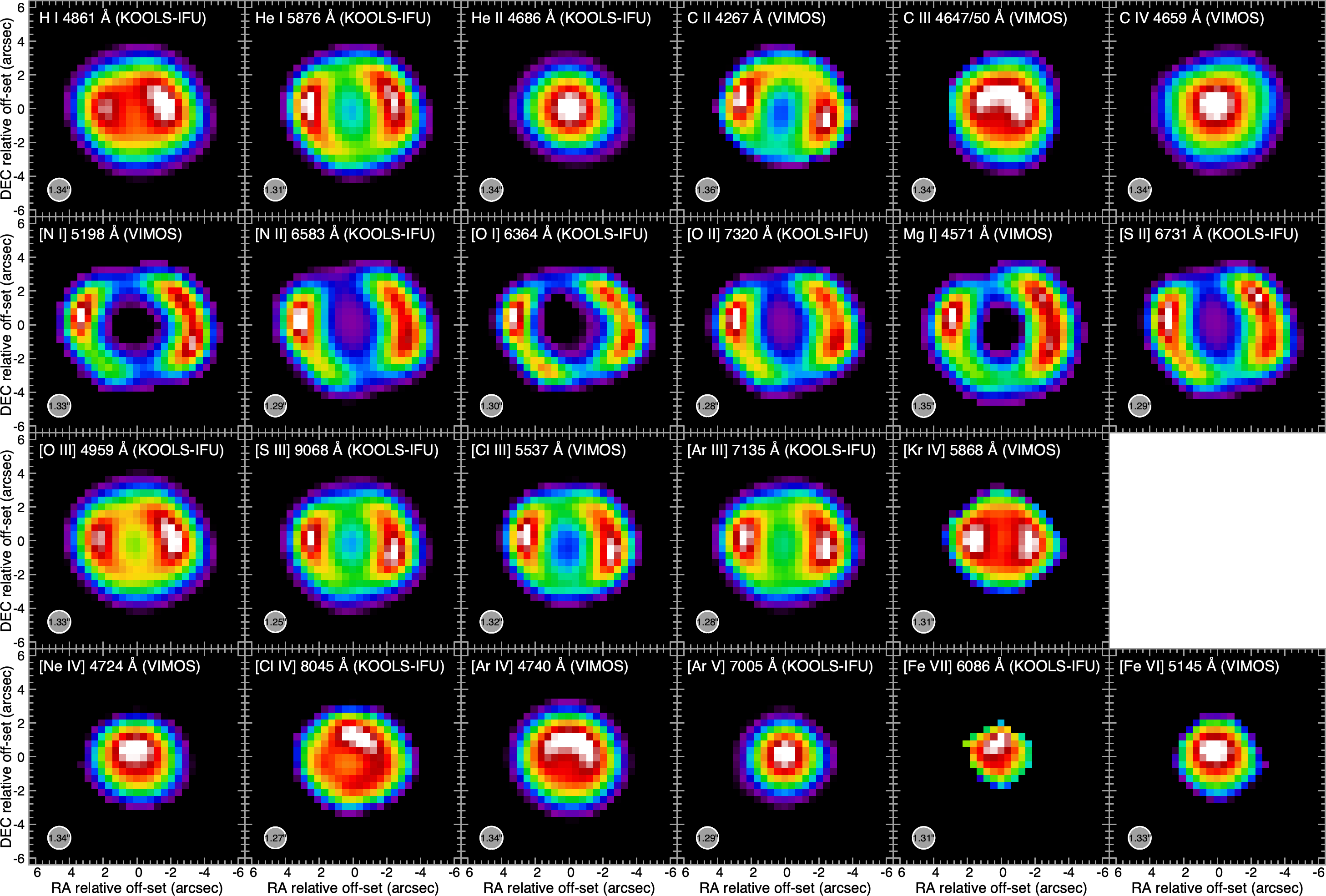}
  \end{adjustbox}
\end{figure*}

\begin{figure*}
\centering
\includegraphics[width=\textwidth]{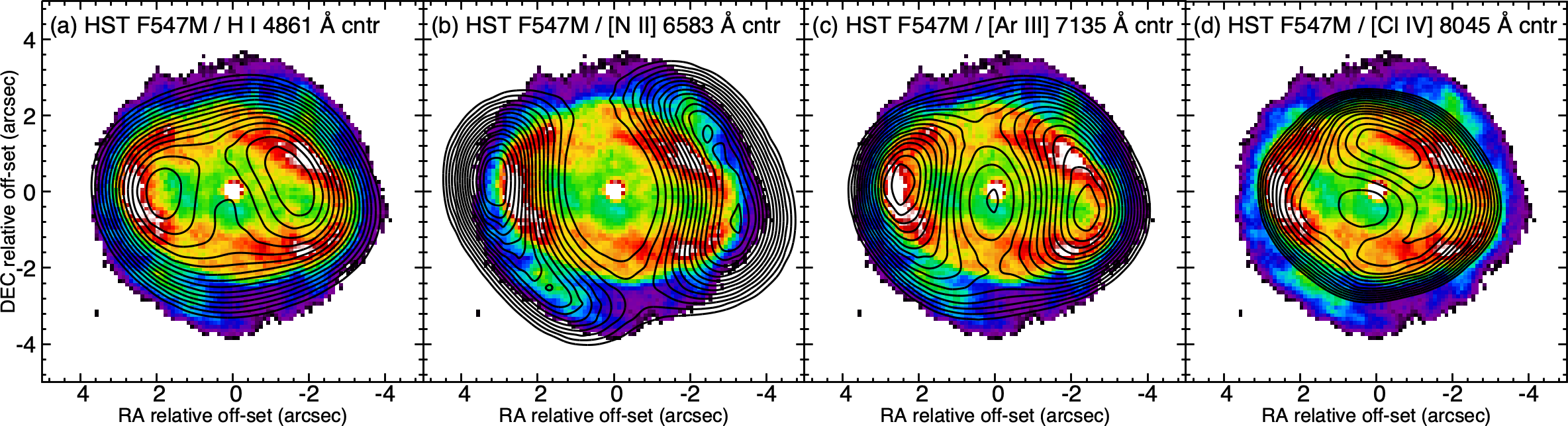}
\vspace{-10pt}
\caption{The intensity contour of the {\hi}\,4861\,{\AA}, {\nii}\,6583\,{\AA}, {\ariii}\,7135\,{\AA}, and {\arv}\,7005\,{\AA} line maps 
(grey lines) overlaid on the \emph{HST}/F547M image. The 15 contours of the corresponding lines 
with a constant interval (between 25 and 100\,$\%$ of the maximum intensity) are drawn on each panel. 
\label{F:koolsAhst}
}
\end{figure*}

\section{Results}
\label{S-Res}

\subsection{Analysis of the spatially resolved 2-D emission-line maps}
\label{S:2Dana}

\subsubsection{Emission-line distributions}

In Fig.\,\ref{F-maps}, we present the selected emission-line maps created through 
the Richardson--Lucy PSF deconvolution algorithm \citep{Richardson:1972aa,Lucy:1974aa} for each wavelength 
and subsequently performed line fitting in each spaxel (i.e., pixel in space) using the automated line fitting algorithm code \citep[{\sc ALFA},][]{Wesson:2016aa}. 
During the PSF deconvolution and line fitting processes, we only considered the signals that are 
greater than four-$\sigma$ above the background and a line flux of 10$^{-18}$\,erg\,s$^{-1}$\,cm$^{-2}$.
Our target was to obtain the seeing-limited emission images with their Gaussian PSF FWHM of $\sim$$1.2-1.3${\arcsec} (Table\,\ref{T-log}), corresponding to the Moffat FWHM of $\sim$1.2{\arcsec}. 
We determined the required number of iterations by plotting the iteration number versus the Gaussian FWHM of the standard stars in the range of $4200-10000$\,{\AA}. 
We confirmed that the total flux is preserved both before and after the deconvolution.
In the resultant maps, each spaxel has a SNR of $\ge 5$.
The filled grey circle with the radius of the half Gaussian PSF FWHM in each panel represents the spatial resolution measured using the Gaussian PSF FWHM from the PSF-deconvolved standard stars.
The center of each panel corresponds to the position of the central star listed in the Guide Star Catalog (GSC), 
Version 2.3.2 \citep[][RA(2000.0) = 06:21:42.77520, DEC(2000.0) = --12:59:13.9524]{Lasker:2008aa}.

In Fig.\,\ref{F:koolsAhst}, we show the intensity contour maps of the {\hi}\,4861\,{\AA}, {\nii}\,6583\,{\AA}, {\ariii}\,7135\,{\AA}, and {\cliv}\,8045\,{\AA} lines overlaid on the \emph{HST}/F547M image. 
Our emission-line maps successfully trace and reproduce the characteristic nebular structures and their dimension confirmed in the \emph{HST} image. 
Our {\nii}\,6583\,{\AA} line contour is almost identical to the result obtained by \citet{Hua:1985aa} using the {\nii}\,6583\,{\AA} image with a narrowband filter (FWHM = 5.5\,{\AA}). 
Hence, we verify our observation and image reconstruction.

According to Fig.\,\ref{F:koolsAhst}, IC2165 appears to comprise three distinct nebular structures: 
(1) an elliptical nebula surrounding the central star with its major axis along the position angle (PA) of $\sim$80$^{\circ}$ (red colour), 
(2) earlobe structures that are symmetrically distributed externally to this elliptical nebula (green), and 
(3) a slightly elongated halo along with the PA of $\sim$100$^{\circ}$ that envelops these two structures (blue and purple).

In Fig.\,\ref{F-maps}, we present the emission-line distributions according to the types of emissions and ionisation degree. 
The first row of Fig.\,\ref{F-maps} shows the maps of recombination lines (RLs), 
the second row shows the maps of neutral and singly ionised collisionally excited lines (CELs), 
the third row displays the maps of doubly ionised CELs, and 
the fourth row presents the maps of triply and more ionised CELs, excluding the {\kriv}\,5868\,{\AA} 
(ionisation potential (IP) = 36.9\,eV), which is grouped into a doubly ionised emission owing to its similar distribution.

Comparison of emission-line maps with the \emph{HST}/F547M image suggests the following. 
Doubly excited and {\hi} and {\hei} lines are mainly emitted from the inner elliptical nebula. 
The regions near the central star could be filled with highly excited gases like {\heii} and C\,{\sc iii, iv}. 
The spatial distribution of neutral and singly ionised CELs is higher than that of the {\hi} lines.
The ionisation boundaries and photodissociation regions (PDRs) could be formed around the regions indicated in purple. 
One finds that the {\nii} line is more widely extended than {\ha}, despite that both IPs are similar.
This relates with recombination rate at lower temperature. 
The ionisation radius is inversely proportional to the recombination rate.
Accordingly, the {\nii} distribution tends to be higher than that of {\hi}. 
The charge exchange could be considerable.

The ionisation structure is explained by the radiation strength radially decreasing as the distance from the central star. 
The distribution of doubly or less ionised emission lines is concentrated along the elliptical nebula's major axis, whereas 
the distribution of highly ionised emissions is enhanced at the ends of its minor axis.

\subsubsection{Circumstellar/interstellar extinction distribution}
\label{S-chb}

\begin{figure*}
\centering
\includegraphics[width=0.88\textwidth]{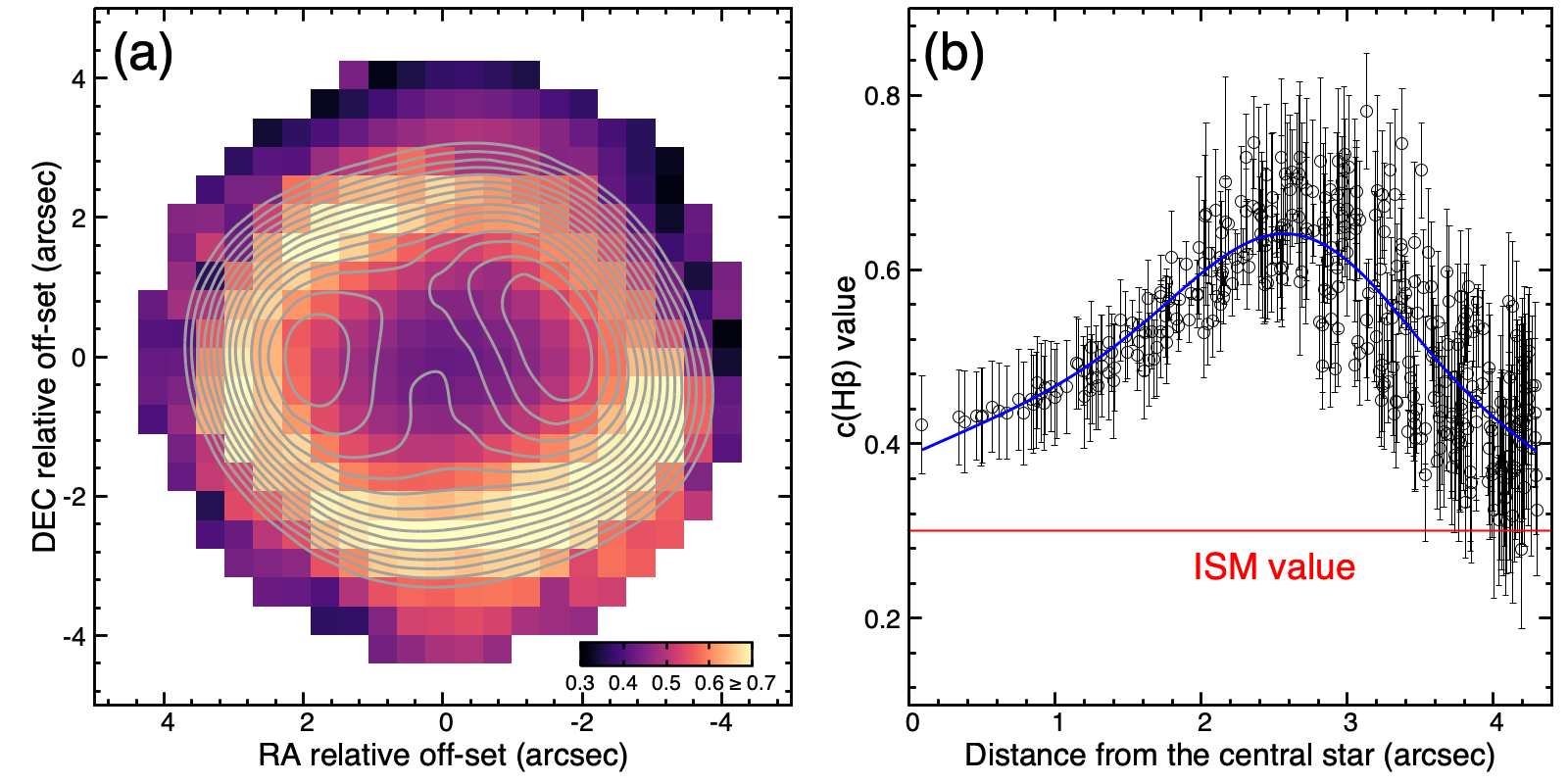}
\caption{
({\it left}) The $c$({\hb}) map, which is masked with a circular area of the radius of 4.3{\arcsec}. 
The grey lines are the 15 contours of the observed {\hi}\,4861\,{\AA} line as presented in Fig.\,\ref{F:koolsAhst}(a). 
({\it right}) The $c$({\hb}) radial plot. 
The error bar corresponds to the one-$\sigma$. 
The horizontal red line is the representative line-of-sight ISM $c$({\hb}) value 
towards IC2165 (0.30). The blue line is a smooth function fitting for this plot.
\label{F:chb}
}
\end{figure*}

The observed line flux $F$($\lambda$) is reddened via interstellar/circumstellar dust. 
We obtain the extinction-free line flux $I$($\lambda$) using Eq.\,(\ref{eq-1}):

\begin{eqnarray}
  I(\lambda) &=& F(\lambda)~\cdot~10^{c({\rm H\beta})(1 + f(\lambda))},
   \label{eq-1}
\end{eqnarray}

\noindent where $f$($\lambda$) is the extinction function at $\lambda$ computed using the reddening law of \citet{Cardelli:1989aa}. 
Herein, we adopt the selective extinction value $R_{V}$ of 3.1. 
$c$({\hb}) is the value of $\log_{10}$$F$({\hb})/$I$({\hb}).

Previous estimates of the distance to IC2165, $D$, range between 2.03 and 3.90\,kpc (\S\,\ref{S-intro}). 
According to the 3-D dust map of \citet{2018MNRAS.478..651G}, the ISM $E(g-r)$ towards IC2165 is $0.23^{+0.03}_{-0.02}$ in $D$ of $2-5$\,kpc. 
Using this value and the reddening law of \citet{Cardelli:1989aa}, the line of sight ISM $c$({\hb}) towards IC2165 is estimated to be $0.27-0.33$. 
We adopt the representative ISM $c$({\hb}) of 0.30.

The spectral resolution of KOOLS-IFU and VIMOS is insufficient to isolate the {\hi} line and deblend the individual lines in the {\hi} and {\heii} line complexes. 
Without deblending the {\hi} and {\heii} lines, $c$({\hb}) and dereddened {\hi} line fluxes will be under/overestimated, resulting in derivations of physical and chemical parameters based on incorrect dereddened line-flux normalisation. 
Eventually, we might misinterpret IC2165. 
It is also true for all PNe studies that depict the {\heii} lines based on low--moderate dispersion spectra. 
Therefore, we separate the {\hi} line fluxes from the complexes of the {\hi}/{\heii} lines and derive $c$({\hb}) through following processes. 
We iterate steps 1--9 until the difference between the $c$({\hb}) values derived from steps 1 and 9 is $<10^{-3}$. 
Regardless of the first guess value, this stopping criterion is attained within four-time iterations. 
In all the steps, we refer to the emissivity of the {\hi} and {\heii} lines that are theoretically calculated by \citet{Storey:1995aa} under the assumption of Case B.

\begin{enumerate}[label = {[{\bf Step-\arabic*}]},align=left]
\item adopts a uniform $c$({\hb}) value map (0.40) as the first guess. 
\item obtains the extinction-free line maps using Eq.(\ref{eq-1}) with the $c$({\hb}) map assumed in {\bf Step-1}. 
\item obtains the {\te}({\oiii}) and {\Ne}({\cliii}) maps.
\item calculates the theoretical intensity ratio of the {\heii} at 4859, 6560, 8746, 8859, and 9225\,{\AA} to the {\heii}\,5411\,{\AA} line under {\te}({\oiii}) and {\Ne}({\cliii}).
\item synthesises the extinction-free maps of these six {\heii} maps using the extinction- free {\heii}\,5411\,{\AA} map.
\item performs reddening of these extinction-free {\heii} maps using Eq.(\ref{eq-1}) with the $c$({\hb}) map assumed in {\bf Step-1}.
\item subtracts each reddened {\heii} map from each reddened 
{\hi} at 4861, 6563, 8751, 8862, and 9229\,{\AA} and {\heii} line complex to obtain these reddened {\hi} maps.
\item obtains the theoretical Balmer and Paschen {\hi} ratio maps with respect to {\hb} and {\ha} using the {\te}({\oiii}) and {\Ne}({\cliii}) maps.
\item obtains the average $c$({\hb}) amongst the four $c$({\hb}) maps obtained by comparing the reddened Paschen {\hi}-to-{\ha} and {\ha}-to-{\hb} ratio maps with the maps obtained in {\bf Step-8}. 
\end{enumerate}

We adopt the {\te}({\oiii}) and {\Ne}({\cliii}) values to calculate the theoretical {\hi} and {\heii} line ratios because the {\oiii} and {\cliii} line maps are very similar to the {\hi} maps (see Fig.\,\ref{F-maps}). 
We generate the {\te}({\oiii}) and {\Ne}({\cliii}) maps in every spaxel by plotting {\te}({\oiii}) and {\Ne}({\cliii}) diagnostic curves (these curves are the functions of {\te} and {\Ne}) and determining their intersection. 
Each {\te}({\oiii}) and {\Ne}({\cliii}) diagnostic curve is obtained from $I$(4959\,{\AA} + 5007\,{\AA})/$I$(4363\,{\AA}) and $I$(5517\,{\AA})/$I$(5537\,{\AA}), respectively. 
Here, we do not subtract the recombination contributions of O$^{3+}$ to the {\oiii}\,4363\,{\AA} line using Eq.\,(3) proposed by \citet{Liu:2000aa} because in 1-D spectrum analysis (see \S\,\ref{S-plasma}), we find that the recombination contribution from O$^{3+}$ to {\oiii}\,4361\,{\AA} is very small, i.e, $\sim$1\,$\%$ of the observed {\oiii}\,4361\,{\AA} flux. 
Besides, the O$^{3+}$ CELs are not within the optical wavelengths. 
The resultant {\te}({\oiii}) and {\Ne}({\cliii} maps are discussed in \S\,\ref{S-tene}. 
The adopted effective collision strengths $\Omega$({\te}) and transition probabilities $A_{ji}$ ($i,j$: energy level, $E_{j} > E_{i}$) are presented in Appendix Table\,\ref{T-atomf}.

In Fig.\,\ref{F:chb}(a), we present the $c$({\hb}) map and overlay the intensity contours of the {\hi}\,4861\,{\AA} line on it. 
The one-$\sigma$ uncertainty of the $c$({\hb}) map is less than 13\,$\%$, including the uncertainties originated from {\te}({\oiii}), {\Ne}({\cliii}), and {\hi} as well as the {\heii}\,5411\,{\AA} line fluxes. 
$c$({\hb}) exhibits large spatial variations; $c$({\hb}) is $0.55 \pm 0.06$ on average, and it varies between 0.33 and 0.78 within 4{\arcsec} from the central star. Higher $c$({\hb}) values are distributed for tracing the inner elliptical nebula. 
The average $c$({\hb}) value within the radius of 1.0{\arcsec} centred on the central star is $0.44 \pm 0.02$, which is considerably lower than the $c$({\hb}) value measured in the nebula, indicating that the regions nearby the central star are harsh environment 
for the survival of dust grains.

Using the $c$({\hb}) map and its radial plot (Fig.\ref{F:chb}(b)), we can find the spatial distribution of each circumstellar and ISM dust grain. 
Fig.\ref{F:chb}(b) indicates that the $c$({\hb}) value decreases after its peak-out at $\sim$2.4{\arcsec} (minor radius) and $\sim$3.2{\arcsec} (major radius), and eventually to be the ISM value. 
Within $\sim$4{\arcsec} of the central star, the $c$({\hb}) value is larger than the value of the ISM representative of 0.3, indicating that circumstellar dust dominates over ISM dust in that region.

\subsubsection{Electron temperature and density distributions}
\label{S-tene}

\begin{figure*}
\centering
\includegraphics[width=\textwidth]{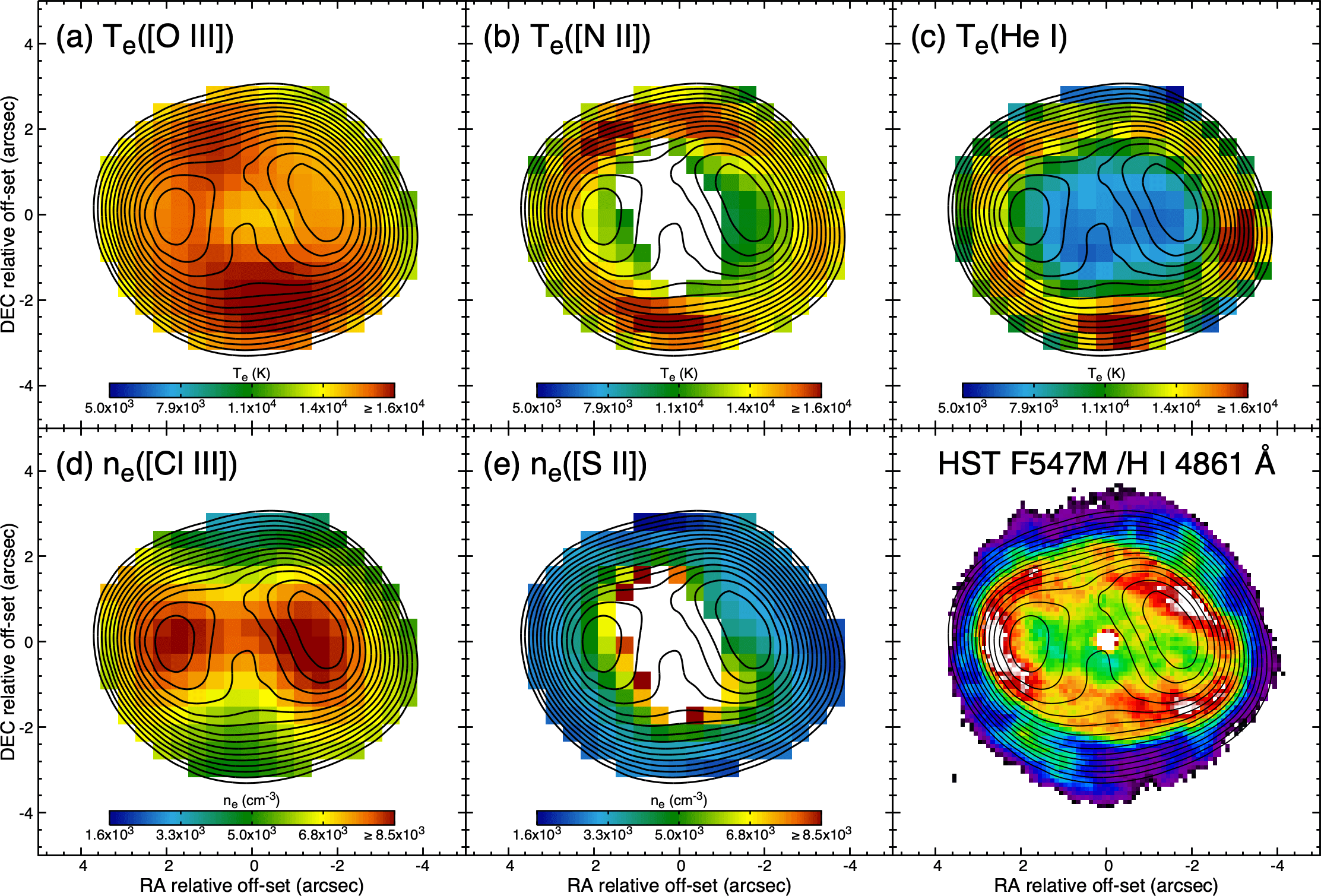}
\vspace{-12pt}
\caption{The electron temperature ({\te}) and density ({\Ne}) maps of IC2165. 
In the last panel, the {\hb} intensity contours as presented in Fig.\,\ref{F:koolsAhst}(a) are superimposed 
on the \emph{HST}/F547M image to help us figure out which regions each contour 
covers.
\label{F:2D-nete}
}
\end{figure*}

If the diagnostic lines have different degrees of ionisation, the resulting {\te} and {\Ne} will be different even if they are observed 
at the same location on the projected sky. 
So, it is not easy to imagine that there will be a convergence to a single {\te} and a single {\Ne}. 
This is because along the line of sight, we determine the temperature and density structure of the 3-D extended objects from the hot ionised gas regions to the warm neutral/ionised gas regions and subsequently to PDRs. 
Therefore, diagnostic lines must be used that are appropriate for each temperature and density region.
IFU observations effectively provide spatial variations of {\te} and {\Ne} across the entire PNe.
In this section, we investigate the 2-D {\te} and {\Ne} distributions.

In the CEL analysis, we generate {\te} and {\Ne} maps in every spaxel by plotting {\te} and {\Ne} diagnostic curves and determining their intersection. 
In the neutral/low-ionisation zone (whose IP is $\lesssim10$\,eV), {\te}({\nii}), {\Ne}({\sii}) maps are determined from their curves. 
{\te}({\nii}) and {\Ne}({\sii}) diagnostic curves are obtained from $I$(6548\,{\AA} + 6583\,{\AA})/$I$(5755\,{\AA}) and $I$(6716\,{\AA})/$I$(6731\,{\AA}), respectively. 
In the moderate-ionisation zone (IP $\sim$$20-55$\,eV), {\te}({\oiii}) and {\Ne}({\cliii}), maps are derived from their curves. 
We do not derive the {\te} maps in the high-ionisation zone (IP $\gtrsim40$\,eV) because of the insufficient SNR of the auroral transition lines {\cliv}\,5323\,{\AA}, {\ariii}\,5191\,{\AA}, and {\ariv}\,7263\,{\AA} in each spaxel.

Regarding the calculation of the {\te}({\oiii}) map, as explained in \S\,\ref{S-chb}, the recombination contributions are not eliminated from O$^{3+}$ to the {\oiii}\,4363\,{\AA} line. 
The difference of {\te}({\oiii}) between the {\oiii}\,4363\,{\AA} correct and uncorrected cases is $\lesssim 100$\,K. 
The removal of the O$^{3+}$ component does not affect the {\Ne}({\cliii) map derivation. 
However, our 1-D spectra analysis (\S\,\ref{S-plasma}) demonstrates that the contribution from N$^{2+}$ to the {\nii}\,5755\,{\AA} line is enormous, $\sim$18\,$\%$ of $I$({\nii}\,5755\,{\AA}) using Eq.\,(1) of \citet{Liu:2000aa}. However, no N$^{2+}$ CELs are observed in the optical wavelengths. 
Therefore, we scale the {\nii}\,5755\,{\AA} line flux using a constant factor of 0.817, corresponding to the ratio of {\nii} $I_{\rm P}$(5755\,{\AA})/$I_{\rm T}$(5755\,{\AA}) listed in Table\,\ref{T:line}. 
Here, $I_{\rm P}$ and $I_{\rm T}$ mean the pure collisionally excited and the recombination line contaminated {\nii}\,5755\,{\AA} line fluxes, respectively.

The RL {\te}({\hei}) map is derived from the {\hei} $I$(7281\,{\AA})/$I$(6678\,{\AA}) ratio map. 
Under Case B, we use the derived {\Ne}({\cliii}) map and the {\te}--{\Ne} logarithm interpolate function of the effective recombination coefficients based on a previous report \citet{Benjamin:1999aa}. 
This interpolate function is valid in {\te} = $5000-20000$\,K and {\Ne} = 10$^{2-6}$\,cm$^{-3}$. 
The {\Ne}({\cliii}) map value is within this {\Ne} range. 
We synthesise the {\heii} 6680\,{\AA} map based on the theoretical ratio of {\heii} 6680\,{\AA}/5411\,{\AA} using the {\te}({\oiii) and {\Ne}({\cliii}) maps. 
Subsequently, we subtract it from the complex of the {\hei}\,6678\,{\AA} plus {\heii}\,6680\,{\AA} lines.

The resulting {\te} and {\Ne} maps are shown in Fig.\,\ref{F:2D-nete}.
The {\hb} contours are superimposed on the \emph{HST}/F547M image in the last panel to help us determine which regions each contour covers. The uncertainties of {\te}({\oiii}), {\te}({\nii}), {\te}({\hei}), {\Ne}({\cliii}), and {\Ne}({\sii}) in spaxel are typically 8, 10, 24, 37, and 41\,$\%$, respectively. 
These values include the uncertainty propagated by the uncertainties of both line-flux measurement and $c$({\hb}). 
{\te} and {\Ne} derived from the diagnostic lines with a higher ionisation degree are generally more compactly distributed in space, and their mean value is higher than those of {\te} and {\Ne} derived from the diagnostic lines with a lower ionisation degree.
The most reasonable explanation for this trend is related to the consumption of high-energy photons in the inner region of the nebula.

{\te}({\oiii}) is in the range from 11740 to 16780\,K, and {\Ne}({\cliii}) ranges from 3350 to 8800\,cm$^{-3}$. 
{\te}({\nii}) ranges from 9750 to 17100\,K, and {\Ne}({\sii}) ranges from 1670 to 8970\,cm$^{-3}$. 
{\te}({\hei}) ranges from 5140 to 19240\,K.
The average values of {\te}({\oiii}), {\te}({\nii}), {\te}({\hei}), {\Ne}({\cliii}), and {\Ne}({\sii}) are 14910\,K (standard deviation (std) = 1000\,K), 13490\,K (1630\,K), 11020\,K (3070\,K), 6700\,cm$^{-3}$ (1210\,cm$^{-3}$), and 3770\,cm$^{-3}$ (1500\,cm$^{-3}$), respectively. 
As previously stated, the uncertainty of {\te}({\nii} would be larger in practice. 
Even on this basis, all the {\te} and {\Ne} maps reflect the nebula ionisation/cooling proceeding.

The {\te}({\oiii}) and {\Ne}({\cliii}) distributions are maximised at approximately 2{\arcsec} away from the central star. 
The {\te}({\oiii}) distribution attains its peak at the edges of the minor axis of the inner elliptical nebula. 
{\Ne}({\cliii}) attains its peaks at the ends of the major axis. 
The {\te}({\oiii}) and {\Ne}({\cliii}) distributions imply that the nebula has a constant gas pressure condition and the cooling proceeds mainly in a forward direction from the major axis. 
A high {\te}({\oiii}) is an evidence for the peak of the intensity of {\ciii}, {\civ} and CELs ionised at a triple or a greater degree at the ends of the minor axis.

The {\te}({\nii}) and {\Ne}({\sii}) distributions could not be measured around the central star because the N$^{+}$ and S$^{+}$ gases are concentrated around the ionisation front. 
Compared with {\te}({\oiii}) and {\Ne}({\cliii}), the {\te}({\nii}) and {\Ne}({\sii}) distributions are considerably extended along the major axis. 
This could be related to the ionisation/cooling process. 
As observed in {\te}({\oiii}), {\te}({\nii}) also tends to be high along the minor axis. 
{\Ne}({\sii}) demonstrates a relatively small variation outside the inner elliptical nebula.

{\te}({\hei}) is comparable to an extent with {\te}({\nii}) because they have the same ionisation degree: the difference with {\te}({\nii}) is its value of $\sim$$1000-2000$\,K on average. 
Notably, cold temperature regions ranging from $\sim$$5000-8000$\, K exist in close proximity to the central star.
This is supported by the recent spatially resolved VLT/MUSE studies of Galactic PNe: \citet{Walsh:2018aa} for NGC7009, \citet{2020AaA...634A..47M} for NGC3132, and \citet{Garca-Rojas:2021aa} for Hf2-2, M1-42, and NGC6778. 
The {\te}({\hei}) distribution is different from a commonly accepted view that the temperature gradient develops from the central star along the direction in which ionisation/cooling proceeds. 
It is evident that {\te}({\hei}) is lower than {\te}({\nii}), despite the fact that both are singly ionised species.

\begin{figure*}
\centering
\includegraphics[scale=0.18]{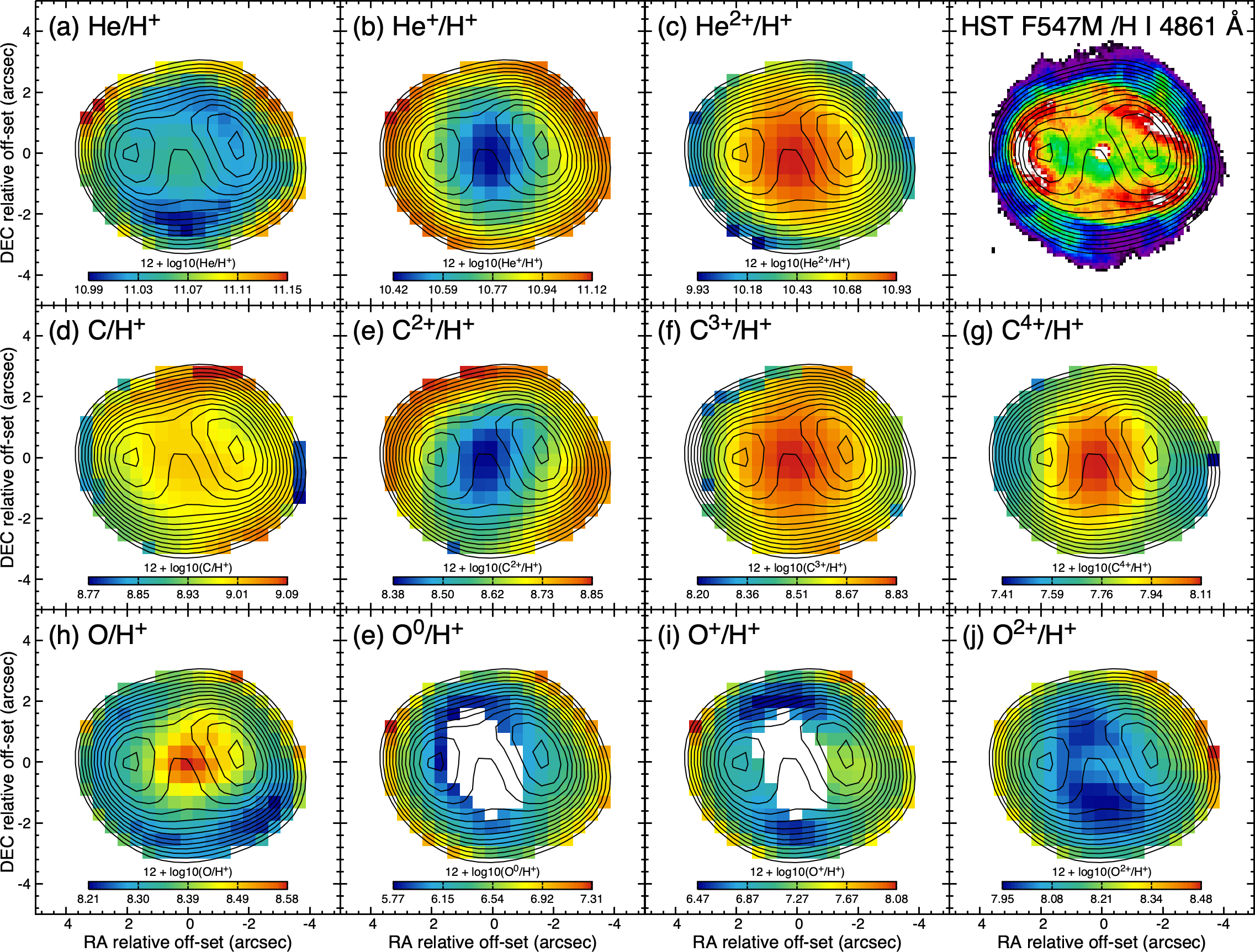}
\caption{The ionic and elemental abundance maps of He, C, and O. 
The black lines on each panel are the intensity contours of the observed {\hi}\,4861\,{\AA} line as presented in Fig.\,\ref{F:koolsAhst}(a).
\label{F:ion-map1}}
\vspace{6pt}
\includegraphics[scale=0.18]{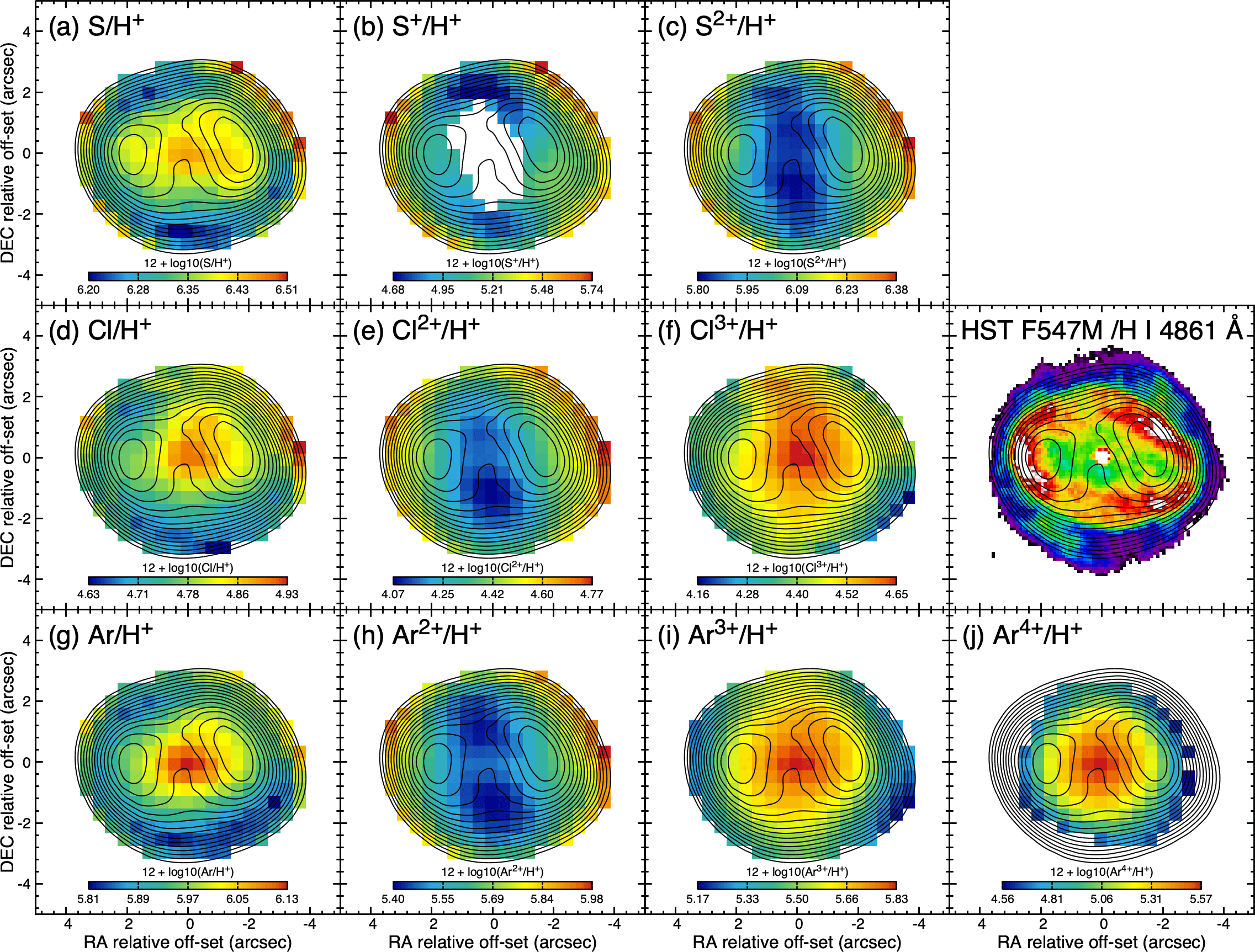}
\caption{The ionic and elemental abundance maps of S, Cl, and Ar. 
The black lines on each panel are the intensity contours of the observed {\hi}\,4861\,{\AA} line as presented in Fig.\,\ref{F:koolsAhst}(a).
\label{F:ion-map2}
}
\end{figure*}

\begin{figure}
\centering
\includegraphics[scale=0.18]{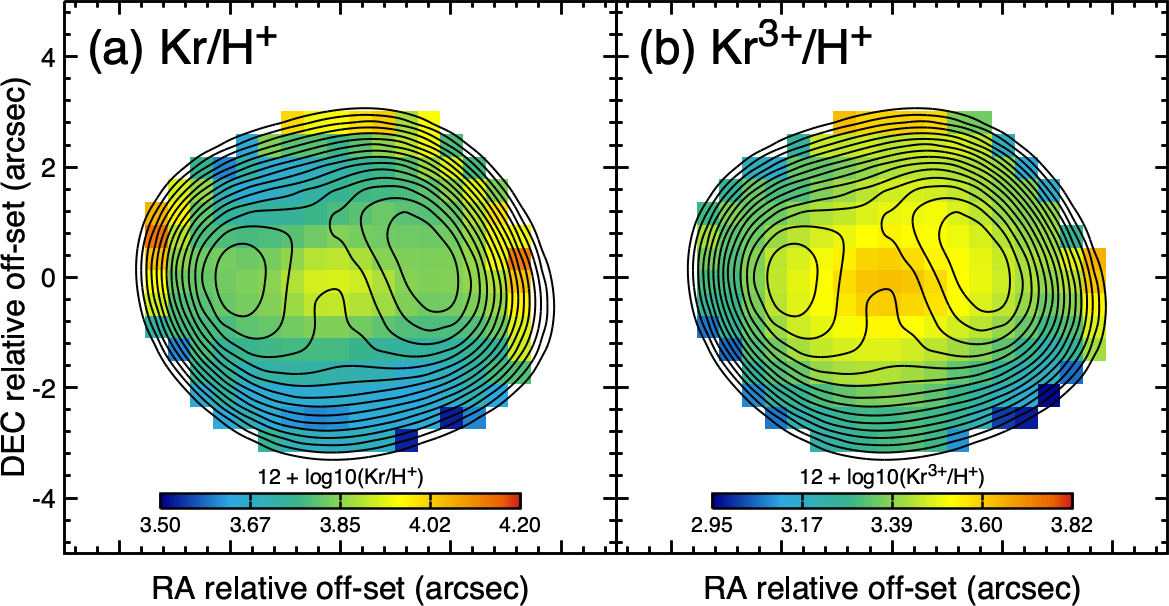}
\caption{The ionic and elemental abundance maps of Kr. 
The black lines on each panel are the intensity contours of the observed {\hi}\,4861\,{\AA} line as presented in Fig.\,\ref{F:koolsAhst}(a).
\label{F:ion-map3}
}
\end{figure}

\subsubsection{Ionic abundance distributions}

We derive the ionic abundance maps from the 2-D emission-line, 
$c$({\hb}), and 2-D {\te}/{\Ne} maps to investigate the spatial distribution of elements.
Here, we define the number density ratio of X to H as the abundance of element X. 

The RL He$^{+}$ map is obtained using the {\te}({\hei}) and {\Ne}({\cliii}) maps. 
We select the electron temperature for the RL He$^{2+}$ and C$^{2+,3+,4+}$ maps by referring to the result regarding the elemental He map in PN NGC7009 reported by \citet{Walsh:2018aa}. 
These four RL maps are obtained using the {\te}({\oiii}) and {\Ne}({\cliii}) maps. 
We use the effective recombination coefficients of \citet{Davey:2000aa} and \citet{Pequignot:1991aa} for the C$^{2+}$ and C$^{3+,4+}$ calculations, respectively.
The CEL ionic abundance maps are generated by solving atomic multiple energy-level population models using the {\te} and {\Ne} maps: 
{\te}({\nii}) and {\Ne}({\sii}) maps are used for the O$^{0,+}$ and S$^{+}$ and {\te}({\oiii}) and {\Ne}({\cliii}) maps for the O$^{2+}$, S$^{2+}$, Cl$^{2+,3+}$, 
Ar$^{2+,3+,4+}$, and Kr$^{3+}$ maps, respectively. 
When we detect more than one line for a target ion, we use the intensity-weighted average of the derived abundances.
For the O$^{+}$ map, we subtracted out the contribution from the O$^{2+}$ recombination using Eq.(2) of \citet{Liu:2000aa}. 
This work would be the first report on the spatial distribution of the slow neutron capture element krypton (Kr) amongst the PNe. 
To obtain the Kr$^{3+}$ map, we subtract the contribution from the {\heii}\,5869\,{\AA} line using the theoretical ratio of the {\heii}\,5869\,{\AA} to 5411\,{\AA} lines. 
Nonetheless, deriving elemental abundance maps requires the maps of singly and more highly ionised species. 
In this sense, the N$^{0,+}$ maps alone are inadequate. Therefore, we did not calculate the N$^{0,+}$ maps.

The resultant ionic abundance maps are presented in Figs.\,\ref{F:ion-map1} to \ref{F:ion-map3}. 
The intensity scale is in the units of 12 + $\log_{10}$\,(X$^{\rm m+}$/H$^{+}$), where $\log_{10}$\,H$^{+}$ is 12. 
The one-$\sigma$ uncertainty in each spaxel is typically 15\,$\%$ (0.07\,dex) in He$^{2+}$, 20\,$\%$ (0.08\,dex) in He$^{+}$, C$^{2+,3+,4+}$, S$^{2+}$, Cl$^{3+}$, and Ar$^{2+,4+}$, 25\,$\%$ (0.11\,dex) in O$^{0,2+}$, S$^{+}$, Cl$^{2+}$, and Ar$^{3+}$, 
30\,$\%$ (0.13\,dex) in Kr$^{3+}$, and 40\,$\%$ (0.18\,dex) in O$^{+}$. 
As discussed in \S\,\ref{S-PI}, the photoionisation model of IC2165 predicts that the ionisation fraction of the neutral hydrogen is 0.95\,$\%$ of the total H within the volume of interest. Therefore, the derived X$^{\rm m+}$/H$^{+}$ maps are almost the same as those of the X$^{\rm m+}$/(H$^{0}$ + H$^{+}$). 
The statistics of the ionic abundance maps are summarised in Table\,\ref{T-2Dabund}.

The ionic abundance maps depict the stratification of ionisation caused by local radiation strength. 
Therefore, it would be meaningless to combine the ionic abundances of an element derived from different locations for obtaining the elemental abundance. 
The degree of ionisation determines the spatial extent. 
The ionic abundances derived from the species in the stage of triply or greater ionised CELs and {\heii}, {\ciii}, and {\civ} lines are considerably enhanced in the inner elliptical nebula. 
Amongst them, highly ionised species such as C$^{3+,4+}$ and Ar$^{3+,4+}$ are concentrated near to the central star. 
Conversely, the ionic abundances from the lower ionised species are enhanced in their surroundings. 
The ionic abundances derived from the neutral to doubly ionised CELs and {\hei} and {\cii} lines are especially low along the minor axis of the inner elliptical nebula, and they are increased as the distance from the central star along the major axis increases. 
Such ionic abundance distributions could be explained if the metal distribution is weighted along the major axis and if nebula cooling mainly proceeds along this axis. 
CEL ionic abundances are sensitive to {\te} and are generally lower as {\te} increases. 
Certainly, {\te} along the minor axis is higher than those in other regions, and {\te} along the major axis is vice versa.
The RL ionic abundances exhibit small variations in spatial distributions compared with the CEL ones in the same ionisation stage, e.g., the C$^{2+}$ versus O$^{2+}$. 
This would be related to the {\te}-insensitive emissivity of the RLs.

\begin{table}
\renewcommand{\arraystretch}{0.80}
\caption{The statistics of the 2-D ionic (X$^{\rm m+}$/H$^{+}$) and elemental (X/H$^{+}$) abundance distributions. \label{T-2Dabund}}
\begin{tabularx}{\columnwidth}{@{}@{\extracolsep{\fill}}
l@{\hspace{-12pt}}
D{.}{.}{-1}
@{\hspace{-12pt}}
D{.}{.}{-1}
@{\hspace{-12pt}}
D{.}{.}{-1}
@{\hspace{-12pt}}
D{.}{.}{-1}@{}}
\midrule
X$^{\rm m+}$/H$^{+}$&
\multicolumn{1}{r}{Ave.}& 
\multicolumn{1}{r}{Std. Dev.}& 
\multicolumn{1}{r}{Min.} &
\multicolumn{1}{r}{Max.}\\
\midrule
He$^{+}$/H$^{+}$ & 7.14{\times}10^{-2} & 2.68{\times}10^{-2} & 2.63{\times}10^{-2} & 1.31{\times}10^{-1} \\ 
He$^{2+}$/H$^{+}$ & 4.13{\times}10^{-2} & 2.01{\times}10^{-2} & 8.56{\times}10^{-3} & 8.55{\times}10^{-2} \\ 
C$^{2+}$/H$^{+}$ & 4.49{\times}10^{-4} & 1.25{\times}10^{-4} & 2.39{\times}10^{-4} & 7.14{\times}10^{-4} \\ 
C$^{3+}$/H$^{+}$ & 4.52{\times}10^{-4} & 1.31{\times}10^{-4} & 2.64{\times}10^{-5} & 6.69{\times}10^{-4} \\ 
C$^{4+}$/H$^{+}$ & 7.02{\times}10^{-5} & 2.35{\times}10^{-5} & 2.57{\times}10^{-5} & 1.30{\times}10^{-4} \\ 
O$^{0}$/H$^{+}$ & 3.31{\times}10^{-6} & 2.82{\times}10^{-6} & 5.85{\times}10^{-7} & 1.51{\times}10^{-5} \\ 
O$^{+}$/H$^{+}$ & 1.53{\times}10^{-5} & 1.12{\times}10^{-5} & 2.92{\times}10^{-6} & 5.65{\times}10^{-5} \\ 
O$^{2+}$/H$^{+}$ & 1.40{\times}10^{-4} & 3.83{\times}10^{-5} & 8.94{\times}10^{-5} & 2.70{\times}10^{-4} \\ 
S$^{+}$/H$^{+}$ & 1.44{\times}10^{-7} & 7.79{\times}10^{-8} & 4.83{\times}10^{-8} & 4.04{\times}10^{-7} \\ 
S$^{2+}$/H$^{+}$ & 1.10{\times}10^{-6} & 3.74{\times}10^{-7} & 6.36{\times}10^{-7} & 2.39{\times}10^{-6} \\ 
Cl$^{2+}$/H$^{+}$ & 2.49{\times}10^{-8} & 9.87{\times}10^{-9} & 1.18{\times}10^{-8} & 5.92{\times}10^{-8} \\ 
Cl$^{3+}$/H$^{+}$ & 2.96{\times}10^{-8} & 6.31{\times}10^{-9} & 1.44{\times}10^{-8} & 4.44{\times}10^{-8} \\ 
Ar$^{2+}$/H$^{+}$ & 4.37{\times}10^{-7} & 1.52{\times}10^{-7} & 2.54{\times}10^{-7} & 9.63{\times}10^{-7} \\ 
Ar$^{3+}$/H$^{+}$ & 3.59{\times}10^{-7} & 1.23{\times}10^{-7} & 1.48{\times}10^{-7} & 6.75{\times}10^{-7} \\ 
Ar$^{4+}$/H$^{+}$ & 1.39{\times}10^{-7} & 8.47{\times}10^{-8} & 3.61{\times}10^{-8} & 3.68{\times}10^{-7} \\ 
Kr$^{3+}$/H$^{+}$ & 2.56{\times}10^{-9} & 7.46{\times}10^{-10} & 3.30{\times}10^{-10} & 4.44{\times}10^{-9} \\ 
\midrule
X/H$^{+}$&
\multicolumn{1}{r}{Ave.}& 
\multicolumn{1}{r}{Std. Dev.}& 
\multicolumn{1}{r}{Min.} &
\multicolumn{1}{r}{Max.}\\
\midrule
He/H$^{+}$ & 1.12{\times}10^{-1} & 9.54{\times}10^{-3} & 9.72{\times}10^{-2} & 1.42{\times}10^{-1} \\ 
C/H$^{+}$ & 9.60{\times}10^{-4} & 1.03{\times}10^{-4} & 5.90{\times}10^{-4} & 1.24{\times}10^{-3} \\ 
O/H$^{+}$ & 2.37{\times}10^{-4} & 4.48{\times}10^{-5} & 1.62{\times}10^{-4} & 3.79{\times}10^{-4} \\ 
S/H$^{+}$ & 2.24{\times}10^{-6} & 3.22{\times}10^{-7} & 1.59{\times}10^{-6} & 3.21{\times}10^{-6} \\ 
Cl/H$^{+}$ & 6.05{\times}10^{-8} & 8.23{\times}10^{-9} & 4.29{\times}10^{-8} & 8.57{\times}10^{-8} \\ 
Ar/H$^{+}$ & 8.93{\times}10^{-7} & 1.42{\times}10^{-7} & 6.49{\times}10^{-7} & 1.34{\times}10^{-6} \\ 
Kr/H$^{+}$ & 6.48{\times}10^{-9} & 1.96{\times}10^{-9} & 1.36{\times}10^{-9} & 1.74{\times}10^{-8} \\ 
\midrule
\end{tabularx}

\end{table}

\begin{table}
\caption{The 2-D ionisation correction factor ICF(X) for element X. \label{T-icf-2D}}
\centering
\renewcommand{\arraystretch}{0.80}
\begin{tabularx}{\columnwidth}{@{}@{\extracolsep{\fill}}l@{\hspace{3pt}}ll@{}}
 \midrule 
X      &Expression                      &ICF(X)               \\
 \midrule 
He     &ICF(X)$\times$(He$^{+}$ + He$^{2+}$)&1                                \\
C      &ICF(X)$\times$(C$^{2+}$ + C$^{3+}$ + C$^{4+}$) &1                       \\
O      &ICF(X)$\times$(O$^{+}$ + O$^{2+}$) &0.822\,S/(S$^{+}$ + S$^{2+}$)               \\
S      &ICF(X)$\times$S$^{2+}$+ S$^{+}$&1 + 0.896\,(Ar$^{3+}$ + Ar$^{4+}$)/Ar$^{2+}$                         \\
Cl     &ICF(X)$\times$(Cl$^{2+}$ + Cl$^{3+}$)&0.993\,Ar/(Ar$^{2+}$ + Ar$^{3+}$)      \\
Ar     &ICF(X)$\times$Ar$^{2+}$ + Ar$^{3+}$ + Ar$^{4+}$&$1 + 0.117\,({\rm S^{+}}/{\rm S^{2+}})$ \\ 
Kr     &ICF(X)$\times$Kr$^{3+}$ &Ar/Ar$^{3+}$ \\ 
 \midrule 
\end{tabularx}
\end{table}

\subsubsection{Elemental abundance distributions}
\label{S-2D-abund-el}

The elemental abundance maps are derived by implementing the ionisation correction factor ICF(X) to element X.
ICFs recover the ionic abundances in unobserved ionisation stages covered by the obtained spectra.
We determine ICF(X) based on the ionisation fraction derived using our photoionisation model (\S\,\ref{S-PI}).
The spatial distributions of the ionic abundance maps are almost consistent at the same stage of ionisation and type of emission. 
This implies that it does not prefer to use ICF maps based on CELs for generating the RL elemental abundance maps. Thus, ICF(He) and ICF(C) are 1.
The derived elemental abundance maps are displayed in the first column of Figs.\,\ref{F:ion-map1} to \ref{F:ion-map3}. 
The one-$\sigma$ uncertainty in each spaxel is typically 13\,$\%$ (0.06\,dex) in He, C, and Ar, 20\,$\%$ (0.09\,dex) in S, 30\,$\%$ (0.13\,dex) in Cl, 35\,$\%$ (0.15\,dex) in O, 40\,$\%$ (0.18\,dex) in Kr. We summarise the statistics of the elemental abundance maps and the adopted equation of ICF(X) for each element X in Tables\,\ref{T-2Dabund} and \ref{T-icf-2D}, respectively.

All elemental abundances spatially vary with distance from the central star. 
Notably, elemental abundances are highest near the central star and exhibit a large gradient along the major axis of the elliptical nebula. 
The nebula's rim is more metal-rich than the surrounding regions, whereas it appears to be slightly metal-poor along its minor axis. 
These findings imply that the AGB-synthesised elements had been distributed mainly along the major axis. 
We discuss the spatial variation of elemental abundances in \S\,\ref{S-discuss1}.

\subsubsection{Gas-to-dust mass ratio (GDR) distribution}
\label{S-GDR}

\begin{figure*}
\centering
\includegraphics[width=\textwidth]{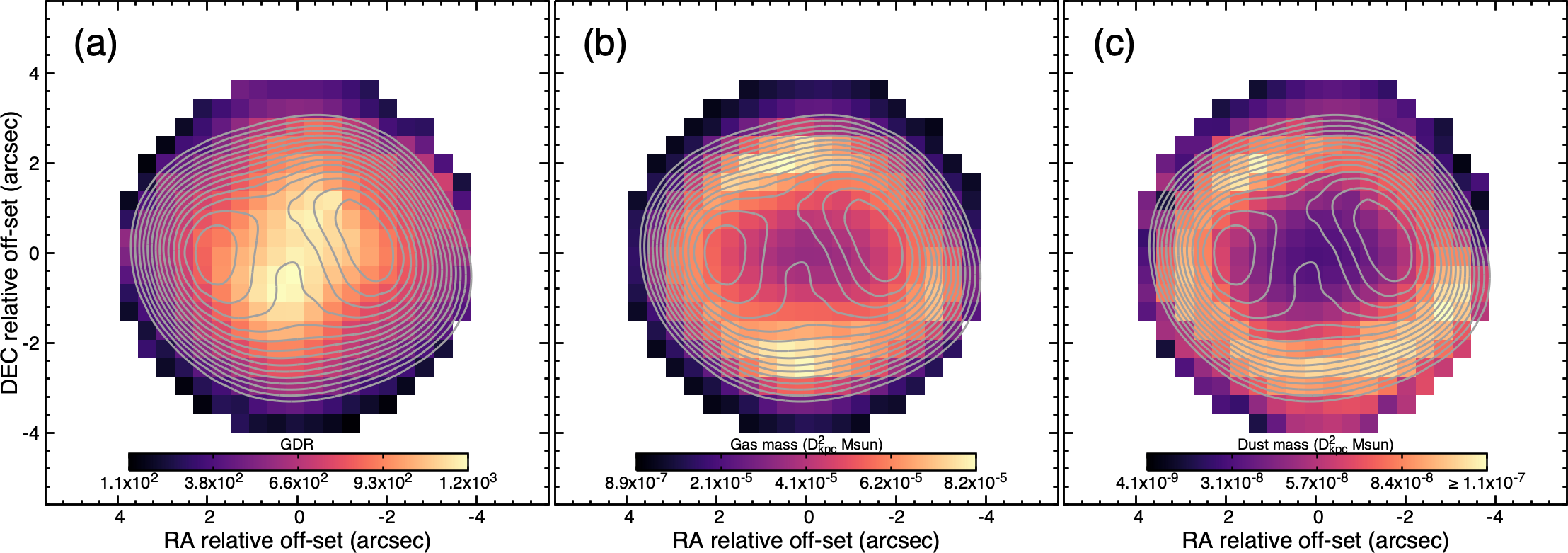}
\vspace{-10pt}
\caption{
({\it from left to right}) The gas-to-dust mass ratio (GDR), gas, and dust mass maps, respectively. 
The grey lines on each panel 
are the intensity contours of the observed {\hi}\,4861\,{\AA} line with a constant interval. 
The maps are masked with a circular area of the radius of 4.0{\arcsec}.
\label{F:GDR}
}
\end{figure*}

The $c$({\hb}) value is equal to the sum of line-of-sight ISM and circumstellar extinction. The subtraction of a constant of 0.3 from the residual value of this $c$({\hb}) due to ISM (\S\,\ref{S-chb}) provides the extinction due to circumstellar media. Considering this, we derive the map of the ``circumstellar'' GDR, $\psi$ based on the assumption that extinction is primarily caused by dust grains. We examine the gas and dust distributions using our images only. 
\citet{1983MNRAS.203P...9R} reported the SiC emission on featureless dust continuum which may be attributed to graphite grains. 
From our 1-D analysis regarding abundance, it is revealed that calcium (Ca) is strongly depleted, suggesting that calcite (CaCO$_{3}$) 
might be formed; CaCO$_{3}$ exhibits a broad emission around 100{\micron} \citep[e.g.,][]{Kemper:2002aa}. 
To facilitate a more general discussion, we use a single size amorphous carbon grain with a radius of 0.1\,{\micron}.
We establish the following equations for determining GDR by referring to \citet{2007pcim.book.....K}. 
When gas and dust grains coexist in the same volume and the respective total masses are $m_{\rm g}$ and $m_{\rm d}$, $\psi$ can be expressed by 
\begin{eqnarray}
\psi = \frac{m_{\rm g}}{m_{\rm d}} = \frac{\sum \left( n_{\rm g} \mu_{\rm g} \right)}{\frac{4}{3} \pi a^{3} n_{\rm d} \rho_{\rm d}}
, \label{Eq-1}
\end{eqnarray}
\noindent where $n_{\rm g}$ and $\mu_{\rm g}$ are the number density and atomic weight of the target gas, respectively. 
$a$, $n_{\rm d}$, and $\rho_{\rm d}$ are the radius, number density, and density of the grain, respectively. 
Using Eq.\,(\ref{Eq-1}), we obtain
\begin{eqnarray}
n_{\rm d} = \frac{\sum \left(\mu_{g} n_{\rm g}\right)}{\frac{4}{3} \pi a^{3} \rho_{\rm d} \psi}.
\end{eqnarray}
The optical depth at {\hb} 4861.33\,{\AA} $\tau$({\hb}) due to dust absorption along line of sight can be represented as 
\begin{eqnarray}
\tau({\rm H}\beta) &=& \int \pi a^2 Q_{\rm ext}({\rm H}\beta,a) n_{\rm d}~ds \nonumber\\
                   &=& \frac{ \pi a^2 Q_{\rm ext}({\rm H}\beta,a) }{\frac{4}{3} \pi a^{3} \rho_{\rm d} \psi}\sum \mu_{g} \int n_{\rm g}~ds, 
\end{eqnarray}
\noindent where $Q_{\rm ext}$({\hb},$a$) is the extinction cross section of the grain with the radius of $a$ at {\hb}. 
As $\tau$({\hb}) is equal to $c$({\hb})/$\log\,e$, $\psi$ can be expressed by 
\begin{eqnarray}
\psi = \frac{3(\log\,e) Q_{\rm ext}({\rm H}\beta,a)}
{4 a \rho_{\rm d} (c({\rm H}\beta)_{\rm tot.} - c({\rm H}\beta)_{\rm ISM})} \sum \mu_{g} \int n_{\rm g}~ds,
\label{Eq-2}
\end{eqnarray}
\noindent where $c$({\hb})$_{\rm tot.}$ is the sum of the line-of-sight ISM and circumstellar extinction (i.e., $c$({\hb}) map, Fig.\,\ref{F:chb}), 
and $c$({\hb})$_{\rm ISM}$ is the extinction due to ISM (i.e., a constant of 0.3). 
Eq.\,(\ref{Eq-2}) is valid for the regions where $c({\rm H}\beta)_{\rm tot.}$ is larger than $c({\rm H}\beta)_{\rm ISM}$.
The observed luminosity of the gas component emitted at a distance of $D$ can be expressed as 
\begin{eqnarray}
\int n_{\rm e} n_{\rm g} J_{\rm g}(T_{\rm e},n_{\rm e})~dV &=& 4 \pi D^{2} I_{\rm g},
\end{eqnarray}
\noindent where $J_{\rm g}$({\te},{\Ne}) is the volume emissivity of the gas component per $n_{\rm g}n_{\rm e}$.
The volume element $dV$ is equal to $A$$ds$, where $A$ in cm$^{2}$ is the actual area per spaxel, which is equal to $(0.412 \times 4.848{\times}10^{-6}\,D)^{2}$ ($D$ is in cm here). Therefore, we obtain
\begin{eqnarray}
\int n_{\rm g} ~ds           &=& \frac{4 \pi D^{2} I_{\rm g}}{A n_{\rm e} J_{\rm g}(T_{\rm e},n_{\rm e})},
\end{eqnarray}
\noindent where $I_{\rm g}$ is the observed extinction-free line flux of the gas component that is corrected using Eq.\,(\ref{eq-1}) with the $c$({\hb})$_{\rm tot.}$ value. 
Hence, we can calculate $\psi$ using Eq.\,(\ref{Eq-3}), where $C$ is a dimensionless constant $2.507\times10^{11}$, which is equal to $(0.412 \times 4.848{\times}10^{-6})^{-2}$.
\begin{eqnarray}
\psi 
&=& \frac{ 3 (\log\,e) Q_{\rm ext}({\rm H}\beta,a) C}
{ 4 a \rho_{\rm d} (c({\rm H}\beta)_{\rm tot.} - c({\rm H}\beta)_{\rm ISM})}
\sum 
\frac{4 \pi I_{g} \mu_{g} }{n_{\rm e} J_{\rm g}(T_{\rm e},n_{\rm e}) }.
\label{Eq-3}
\end{eqnarray}

We derive the total gas mass map from the {\hi}, {\hei}, {\heii}, C\,{\sc ii}, C\,{\sc iii}, C\,{\sc iv}, {\oi}, {\oii}, and {\oiii} maps. 
These elements are the three most abundant elements in IC2165 (Tables\,\ref{T-2Dabund} and \ref{T-elem}). 
{\te}({\oiii}) and {\Ne}({\cliii}) are adopted for calculating the H$^{+}$, C$^{2+,3+,4+}$, and O$^{2+}$ gas masses. 
{\te}({\hei}) and {\Ne}({\cliii}) are adopted for determining the He$^{+}$ gas mass. 
{\te}({\nii}) and {\Ne}({\sii}) are adopted for determining the O$^{0,+}$ gas masses. 
Both neutral hydrogen and molecular gas masses are not included in our DGR. 
The neutral atomic {\hi} line at 21\,cm and molecular lines such as H$_{2}$ and CO have not been detected thus far \citep{1995MNRAS.273..801G,1996ApJ...462..777K,1996A&A...315..284H}. 
$Q_{\rm ext}$({\hb},0.1\,{\micron}) is calculated as 4.46 using the Mie theory \citep{1983asls.book.....B} with the optical data of \citet{Martin:1991aa} for the AC-type amorphous carbon (soot produced by striking an arc between two amorphous carbon electrodes under a controlled Ar atmosphere). We adopt $\rho_{\rm d}$ of 1.85\, g\,cm$^{-3}$.

The obtained circumstellar GDR map (Fig.\,\ref{F:GDR}a) exhibits large spatial fluctuations. 
As expected, GDR is enhanced in the vicinity of the central nebula because few dust grains can survive the harsh radiation received from the central star. 
Because majority of the dust grains may be destroyed in that region, the GDR may be relatively high. 
The GDR within a radius of 2{\arcsec} is in the range from 852 to 1205, and its average and standard deviations (std) are 1040 and 97, respectively. 
The GDR map indicates that the bulk of the dust grains resides outside this region. 
Indeed, the average GDR value outside the radius of 2{\arcsec} is 565 (std = 220). 
GDR appears to decrease in direct proportion to nebular excitation or the strength of local radiation.
GDR varies radially from 1205 in the central nebula to 116 near the ionisation front and 678 (std = 284) on average within the radius of 4{\arcsec}.
The obtained GDR is consistent with $\sim$400, as generally applied for C-rich AGB stars \citep[$386 \pm 90$ on average amongst 18 C-rich stars;][]{1985ApJ...293..273K}, and 100 for ISM \citep{1974A&A....35..361K}. 
Except for the inner hot gas regions, the GDR in IC2165 is nearly identical to that in C-rich AGB stars, implying that a majority of the dust grains survive without being destroyed.

\begin{table}
\centering
\renewcommand{\arraystretch}{0.8}
\caption{The mass of each gas and dust component derived from the 2-D emission maps. 
}
\label{T-gmass}
\begin{tabularx}{\columnwidth}{@{}@{\extracolsep{\fill}}ll@{}}
    \midrule

Gas comp. parameter & Value ($D_{\rm kpc}^2$ M$_{\sun}$)\\
 \midrule	  
$m_{\rm g}$(H$^{+}$)  & 8.79$\times10^{-3}$ $\pm$ 3.66$\times10^{-3}$\\  
$m_{\rm g}$(He$^{+}$) & 1.93$\times10^{-3}$ $\pm$ 8.14$\times10^{-4}$\\ 
$m_{\rm g}$(He$^{2+}$) & 1.40$\times10^{-3}$ $\pm$ 5.68$\times10^{-4}$\\ 
$m_{\rm g}$(C$^{2+}$) & 5.14$\times10^{-5}$ $\pm$ 2.47$\times10^{-5}$\\ 
$m_{\rm g}$(C$^{3+}$) & 4.83$\times10^{-5}$ $\pm$ 2.14$\times10^{-5}$\\ 
$m_{\rm g}$(C$^{4+}$) & 7.46$\times10^{-6}$ $\pm$ 3.28$\times10^{-6}$\\ 
$m_{\rm g}$(O$^{0}$) & 8.63$\times10^{-7}$ $\pm$ 4.36$\times10^{-7}$\\ 
$m_{\rm g}$(O$^{+}$) & 8.13$\times10^{-6}$ $\pm$ 6.51$\times10^{-6}$\\ 
$m_{\rm g}$(O$^{2+}$) & 2.11$\times10^{-5}$ $\pm$ 9.65$\times10^{-6}$\\ 
$\sum m_{\rm g}({\rm H^{+},He^{+,2+},C^{2+,3+,4+},O^{0,+,2+})}$ & 1.23$\times10^{-2}$ $\pm$ 3.83$\times10^{-3}$\\ 
    \midrule
Dust comp. parameter & Value ($D_{\rm kpc}^2$ M$_{\sun}$)\\
 \midrule	  	  
$m_{\rm d}$(AC) & 1.79$\times10^{-5}$ $\pm$ 8.64$\times10^{-6}$\\
  \midrule	  
Carbon comp. parameter & Value ($D_{\rm kpc}^2$ M$_{\sun}$)\\
 \midrule	 
$\sum m_{\rm g}({\rm C^{2+,3+,4+})}$ + $m_{\rm d}$(AC) & 1.25$\times10^{-4}$ $\pm$ 3.57$\times10^{-5}$\\ 
    \midrule	  
\end{tabularx}
\end{table}

\begin{figure*}
\centering
\includegraphics[width=0.96\textwidth]{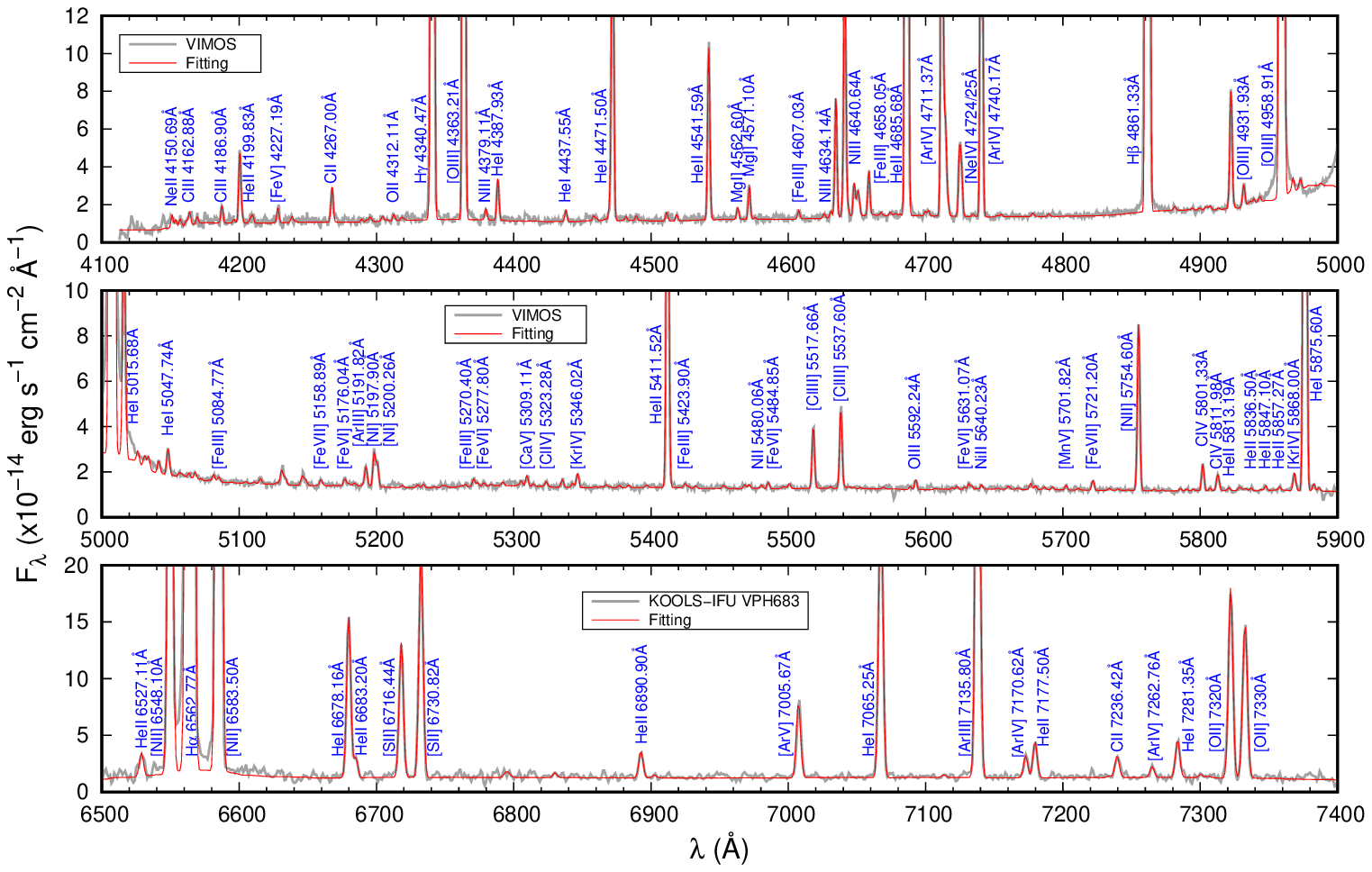}
\caption{
 The VIMOS ({\it upper and middle panels}) and KOOLS-IFU VPH-683 ({\it bottom panel}) 1-D spectra extracted in the \emph{IUE} aperture (grey line) 
and the fitting result using {\sc ALFA} code (red line). Prominent lines amongst the identified emission-lines are indicated by blue ink. 
Note that {\hb} and {\oiii}\,4959/5007\,{\AA} lines in the VIMOS spectrum are saturated.
\label{F:koolsspec2}
}

\vspace{6pt}

\centering
\includegraphics[width=0.96\textwidth]{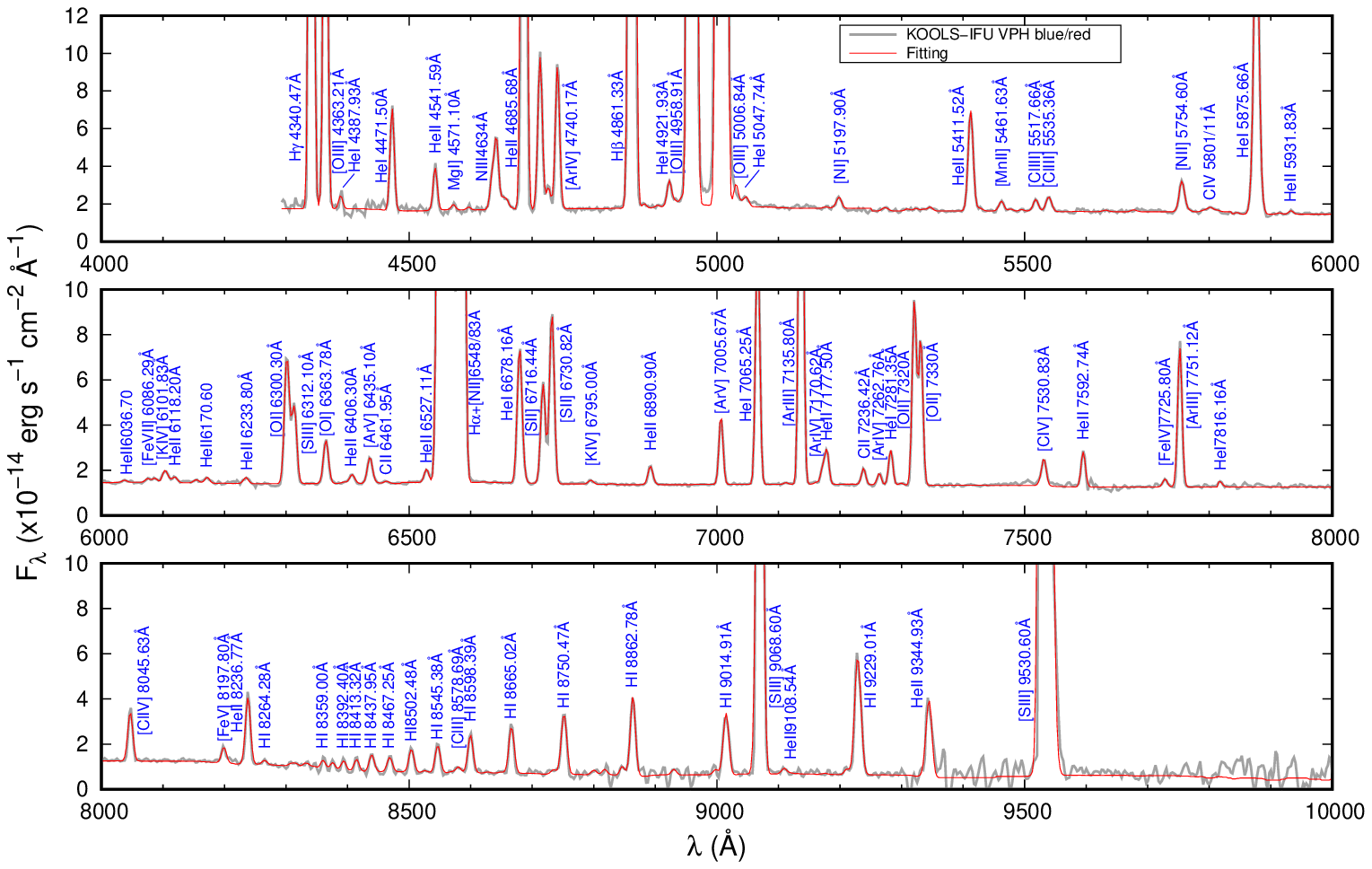}
\caption{
The KOOLS-IFU VPH-blue/red 1-D spectrum extracted in the \emph{IUE} aperture (grey line) and the fitting result using {\sc ALFA} code (red line). 
Prominent lines amongst the identified emission-lines are indicated by blue ink. 
\label{F:koolsspec1}
}
\end{figure*}

\begin{table*}
\caption{The measured {\hi} and {\heii} line fluxes of the 1-D spectrum extracted in the \emph{IUE} aperture by our deblending procedure. 
}
\renewcommand{\arraystretch}{0.7}
\centering
\begin{tabularx}{\textwidth}{@{}@{\extracolsep{\fill}}c
D{p}{\pm}{-1}
c
D{.}{.}{-1}
D{p}{\pm}{-1}
D{p}{\pm}{-1}
D{.}{.}{-1}
D{.}{.}{-1}
D{.}{.}{-1}
@{}}
\midrule
$\lambda_{\rm lab.}$ &
\multicolumn{1}{c}{$F$($\lambda$)} &
Ion &
\multicolumn{1}{c}{$f$($\lambda$)} &
\multicolumn{1}{c}{$F$($\lambda$)} &
\multicolumn{1}{c}{$I$($\lambda$)} &
\multicolumn{1}{c}{Theo. $I$($\lambda$)} &
\multicolumn{1}{c}{Theo. $I$($\lambda$)} \\
({\AA})&
\multicolumn{1}{c}{(erg\,s$^{-1}$\,cm$^{-2}$)}&&&
\multicolumn{1}{c}{[$F$({\hb}) = 100]}&
\multicolumn{1}{c}{[$I$({\hb}) = 100]}&
\multicolumn{1}{c}{[$I$({\hb}) = 100]}&
\multicolumn{1}{c}{[$I$({\heii}\,5411\,{\AA}) = 3.633]}\\
\midrule
2385.40 & 1.58{\times}10^{-13} ~p~ 2.18{\times}10^{-14} & {\heii} & 1.223 & 1.254 ~p~ 0.173 & 6.117 ~p~ 0.711 &        &4.065\\ 
2511.21 & 2.59{\times}10^{-13} ~p~ 1.69{\times}10^{-14} & {\heii} & 0.971 & 2.054 ~p~ 0.136 & 7.227 ~p~ 0.711 &        &6.245\\ 
2733.30 & 5.30{\times}10^{-13} ~p~ 1.21{\times}10^{-14} & {\heii} & 0.722 & 4.208 ~p~ 0.108 & 10.721 ~p~ 0.641 &        &10.434\\ 
3203.10 & 3.06{\times}10^{-13} ~p~ 2.12{\times}10^{-14} & {\heii} & 0.463 & 11.330 ~p~ 0.271 & 20.637 ~p~ 0.868 &        &19.811\\ 
4338.67 & 1.16{\times}10^{-13} ~p~ 5.56{\times}10^{-15} & {\heii} & 0.157 & 0.921 ~p~ 0.045 & 1.129 ~p~ 0.057 &        &1.129\\ 
4340.47 & 4.80{\times}10^{-12} ~p~ 1.44{\times}10^{-13} & {\hg} & 0.157 & 38.109 ~p~ 1.230 & 46.680 ~p~ 1.602 & 47.254 \\ 
4859.32 & 2.92{\times}10^{-13} ~p~ 1.28{\times}10^{-14} & {\heii} & 0.001 & 2.314 ~p~ 0.105 & 2.315 ~p~ 0.106 &  &2.318 \\ 
4861.33 & 1.26{\times}10^{-11} ~p~ 1.48{\times}10^{-13} & {\hb} & 0.000 & 100.000 ~p~ 1.174 & 100.000 ~p~ 1.174 & 100.000& \\ 
5411.52 & 5.39{\times}10^{-13} ~p~ 4.68{\times}10^{-15} & {\heii}  & -0.126 & 4.276 ~p~ 0.062 & 3.633 ~p~ 0.063 &   &3.633 \\ 
6560.10 & 1.15{\times}10^{-12} ~p~ 4.32{\times}10^{-14} & {\heii} & -0.297 & 9.131 ~p~ 0.359 & 6.213 ~p~ 0.281 &  &6.250 \\ 
6562.77 & 5.06{\times}10^{-11} ~p~ 3.69{\times}10^{-13} & {\ha} & -0.298 & 401.663 ~p~ 5.551 & 273.150 ~p~ 7.156 & 279.077\\ 
8594.87 & 5.32{\times}10^{-15} ~p~ 1.76{\times}10^{-16} & {\heii} & -0.554 & 0.042 ~p~ 0.001 & 0.021 ~p~ 0.001 &  &0.021  \\ 
8598.39 & 1.75{\times}10^{-13} ~p~ 9.13{\times}10^{-15} & P14 & -0.554 & 1.385 ~p~ 0.074 & 0.675 ~p~ 0.046 & 0.644 &  \\ 
8661.47 & 6.69{\times}10^{-15} ~p~ 2.20{\times}10^{-16} & {\heii} & -0.560 & 0.053 ~p~ 0.002 & 0.026 ~p~ 0.001 &  &0.026  \\ 
8665.02 & 2.15{\times}10^{-13} ~p~ 1.39{\times}10^{-14} & P13 & -0.560 & 1.705 ~p~ 0.112 & 0.825 ~p~ 0.064 & 0.805 &  \\ 
8746.00 & 8.66{\times}10^{-15} ~p~ 2.84{\times}10^{-16} & {\heii} & -0.568 & 0.069 ~p~ 0.002 & 0.033 ~p~ 0.002 &  &0.033  \\ 
8750.47 & 2.70{\times}10^{-13} ~p~ 1.34{\times}10^{-14} & P12  & -0.568 & 2.139 ~p~ 0.109 & 1.024 ~p~ 0.068 & 1.025 &  \\ 
8859.15 & 1.14{\times}10^{-14} ~p~ 3.71{\times}10^{-16} & {\heii} & -0.578 & 0.090 ~p~ 0.003 & 0.043 ~p~ 0.002 &  &0.043  \\ 
8862.78 & 3.52{\times}10^{-13} ~p~ 2.63{\times}10^{-14} & P11& -0.578 & 2.791 ~p~ 0.211 & 1.320 ~p~ 0.115 & 1.333 &  \\ 
9225.23 & 2.17{\times}10^{-14} ~p~ 7.01{\times}10^{-16} & {\heii} & -0.605 & 0.172 ~p~ 0.006 & 0.078 ~p~ 0.004 &  &0.079  \\ 
9229.01 & 6.34{\times}10^{-13} ~p~ 3.44{\times}10^{-14} & P9& -0.605 & 5.026 ~p~ 0.279 & 2.294 ~p~ 0.164 & 2.455 &  \\ 
\midrule
\end{tabularx}
\label{T-chb}
\end{table*}

\begin{table*}
\small
\centering
\caption{
The identified emission-lines in the 1-D \emph{IUE}, KOOLS-IFU, and VIMOS 
spectra extracted from the \emph{IUE} aperture. 
These line fluxes are normalised to $I$({\hb}) = $4.61{\times}10^{-11} \pm 3.49{\times}10^{-12}$\,erg\,s$^{-1}$\,
cm$^{-2}$, where $I$({\hb}) is 100. S means the source of the data: I, V, K is \emph{IUE}, ESO VIMOS, and KOOLS-IFU, respectively.
\label{T:line}
}
\renewcommand{\arraystretch}{0.90}
\begin{tabular}{@{\extracolsep{\fill}}
r@{\hspace{7pt}}
l@{\hspace{-1pt}}
D{.}{.}{-1}@{\hspace{5pt}}
D{.}{.}{-1}@{\hspace{2pt}}
D{.}{.}{-1}@{\hspace{5pt}}
c@{\hspace{7pt}}
r@{\hspace{7pt}}
l@{\hspace{-1pt}}
D{.}{.}{-1}@{\hspace{5pt}}
D{.}{.}{-1}@{\hspace{2pt}}
D{.}{.}{-1}@{\hspace{5pt}}
c@{\hspace{7pt}}
r@{\hspace{7pt}}
l@{\hspace{-1pt}}
D{.}{.}{-1}@{\hspace{5pt}}
D{.}{.}{-1}@{\hspace{2pt}}
D{.}{.}{-1}@{\hspace{5pt}}
c@{\hspace{7pt}}
@{}}
\midrule
\multicolumn{1}{c}{$\lambda_{\rm lab.}$({\AA})} &Ion&\multicolumn{1}{c}{$f$($\lambda$)}&
\multicolumn{1}{c}{$I$($\lambda$)}&\multicolumn{1}{c}{${\delta}$\,$I$($\lambda$)}&S&
\multicolumn{1}{c}{$\lambda_{\rm lab.}$({\AA})} &Ion&\multicolumn{1}{c}{$f$($\lambda$)}&
\multicolumn{1}{c}{$I$($\lambda$)}&\multicolumn{1}{c}{${\delta}$\,$I$($\lambda$)}&S&
\multicolumn{1}{c}{$\lambda_{\rm lab.}$({\AA})} &Ion&\multicolumn{1}{c}{$f$($\lambda$)}&
\multicolumn{1}{c}{$I$($\lambda$)}&\multicolumn{1}{c}{${\delta}$\,$I$($\lambda$)}&S\\
\midrule
1240.81 &  N\,{\sc v}  & 1.905 & 58.710 & 8.810 &  I   & 4958.91 &   [O\,{\sc iii}]   & -0.026 & 420.501 & 5.728 &  K   & 6548.10 &   [N\,{\sc ii}]   & -0.296 & 14.567 & 0.612 &  K   \\ 
1404.28 &  [O\,{\sc iv}]  & 1.430 & 53.868 & 5.974 &  I   & 5006.84 &   [O\,{\sc iii}]   & -0.038 & 1260.607 & 16.643 &  K   & 6560.10 &  {\heii}  & -0.297 & 6.213 & 0.281 &  K   \\ 
1483.32 &  [N\,{\sc iv}]  & 1.308 & 45.711 & 4.608 &  I   & 5015.68 &   {\hei}   & -0.040 & 2.131 & 0.435 & V   & 6562.77 &  {\hi}  & -0.298 & 273.150 & 7.156 &  K   \\ 
1602.37 &  [Ne\,{\sc iv}]  & 1.198 & 5.113 & 0.677 &  I   & 5030.63 &   Fe\,{\sc ii}   & -0.044 & 0.100 & 0.029 & V   & 6583.50 &   [N\,{\sc ii}]   & -0.300 & 43.340 & 1.789 &  K   \\ 
1661.92 &  O\,{\sc iii}]  & 1.168 & 42.094 & 3.833 &  I   & 5047.74 &   {\hei}   & -0.048 & 0.210 & 0.019 & V   & 6678.16 &  {\hei}  & -0.313 & 2.815 & 0.081 &  K   \\ 
1751.83 &  [N\,{\sc iii}]  & 1.154 & 29.244 & 2.690 &  I   & 5067.52 &   Ni\,{\sc iii}   & -0.052 & 0.059 & 0.018 & V   & 6683.20 &  {\heii}  & -0.313 & 0.343 & 0.036 &  K   \\ 
1906.68 &  [C\,{\sc iii}] & 1.257 & 879.855 & 83.434 &  I   & 5080.96 &   Ni\,{\sc ii}   & -0.056 & 0.057 & 0.014 & V   & 6716.44 &   [S\,{\sc ii}]   & -0.318 & 2.343 & 0.075 &  K   \\ 
1908.73 & +  C\,{\sc iii}]  & * & * & * & I & 5084.77 &   [Fe\,{\sc iii}]   & -0.057 & 0.044 & 0.013 & V   & 6730.82 &   [S\,{\sc ii}]   & -0.320 & 3.928 & 0.116 &  K   \\ 
2321/30  &  O\,{\sc iii}]  & 1.382 & 5.174 & 0.233 &  I   & 5130.46 &   Ne\,{\sc ii}   & -0.067 & 0.119 & 0.017 & V   & 6795.00 &   [K\,{\sc iv}]   & -0.328 & 0.076 & 0.010 &  K   \\ 
2322-28  &  [C\,{\sc ii}]  & 1.385 & 58.312 & 6.834 &  I   & 5132.94 &   C\,{\sc ii}   & -0.068 & 0.060 & 0.017 & V   & 6890.90 &   {\heii}   & -0.341 & 0.392 & 0.017 &  K   \\ 
2385.40 &  {\heii}  & 1.223 & 6.117 & 1.014 &  I   & 5145.75 &   [Fe\,{\sc vi}]   & -0.071 & 0.076 & 0.019 & V   & 7005.67 &   [Ar\,{\sc v}]   & -0.356 & 1.248 & 0.041 &  K   \\ 
2423.12 &  [Ne\,{\sc iv}]  & 1.135 & 68.473 & 5.927 &  I   & 5158.89 &   [Fe\,{\sc vii}]   & -0.074 & 0.047 & 0.013 & V   & 7065.25 &   {\hei}   & -0.364 & 4.817 & 0.148 &  K   \\ 
2511.21 &  {\heii}  & 0.971 & 7.227 & 0.711 &  I   & 5176.04 &   [Fe\,{\sc vi}]   & -0.077 & 0.055 & 0.009 & V   & 7135.80 &   [Ar\,{\sc iii}]   & -0.374 & 9.978 & 0.312 &  K   \\ 
2733.30 &  {\heii}  & 0.722 & 10.721 & 0.641 &  I   & 5191.82 &   [Ar\,{\sc iii}]   & -0.081 & 0.151 & 0.015 & V   & 7170.62 &   [Ar\,{\sc iv}]   & -0.378 & 0.262 & 0.017 &  K   \\ 
2783.15 &  [Mg\,{\sc v}]  & 0.685 & 5.883 & 0.512 &  I   & 5197.90 &   [N\,{\sc i}]   & -0.082 & 0.385 & 0.048 &  K   & 7177.50 &   {\heii}   & -0.379 & 0.599 & 0.023 &  K   \\ 
2800.00 &  Mg\,{\sc ii}  & 0.673 & 2.288 & 0.285 &  I   & 5200.26 &   [N\,{\sc i}]   & * &  *  &  *  &  K   & 7236.42 &   C\,{\sc ii}   & -0.387 & 0.288 & 0.018 &  K   \\ 
2835.46 &  O\,{\sc iii}  & 0.650 & 10.198 & 0.697 &  I   & 5270.40 &   [Fe\,{\sc iii}]   & -0.098 & 0.056 & 0.011 & V   & 7237.17 &   C\,{\sc ii}   & * &  *  &  *  &  K   \\ 
3132.50 &  O\,{\sc iii}  & 0.498 & 54.410 & 2.169 &  I   & 5277.80 &   [Fe\,{\sc vi}]   & -0.099 & 0.023 & 0.007 & V   & 7237.26 &   [Ar\,{\sc iv}]   & * &  *  &  *  &  K   \\ 
3203.10 &  {\heii}  & 0.463 & 20.637 & 0.868 &  I   & 5309.11 &   [Ca\,{\sc v}]   & -0.106 & 0.076 & 0.010 & V   & 7262.76 &   [Ar\,{\sc iv}]   & -0.391 & 0.198 & 0.013 &  K   \\ 
4150.69 &   Ne\,{\sc ii}   & 0.215 & 0.145 & 0.046 & V   & 5323.28 &   [Cl\,{\sc iv}]   & -0.108 & 0.051 & 0.006 & V   & 7281.35 &   {\hei}   & -0.393 & 0.630 & 0.027 &  K   \\ 
4162.88 &   C\,{\sc iii}   & 0.211 & 0.144 & 0.034 & V   & 5335.18 &   [Fe\,{\sc vi}]   & -0.111 & 0.057 & 0.011 & V   & 7318.92 &   [O\,{\sc ii}]   & -0.398 & 3.121 & 0.128 &  K   \\ 
4163.25 &   C\,{\sc iii}   & * &   *   &   *   & V   & 5346.02 &   [Kr\,{\sc iv}]   & -0.113 & 0.103 & 0.013 & V   & 7319.99 &   [O\,{\sc ii}]   & * &   *   &   *   &  K   \\ 
4163.32 &   [K\,{\sc v}]   & * &   *   &   *   & V   & 5366.80 &   [Mn\,{\sc vi}]   & -0.117 & 0.022 & 0.006 & V   & 7329.67 &   [O\,{\sc ii}]   & -0.400 & 2.734 & 0.112 &  K   \\ 
4186.90 &   C\,{\sc iii}   & 0.204 & 0.185 & 0.035 & V   & 5411.52 &   {\heii}   & -0.126 & 3.633 & 0.063 &  K   & 7330.73 &   [O\,{\sc ii}]   & * &   *   &   *   &  K   \\ 
4199.83 &   {\heii}   & 0.200 & 0.769 & 0.029 & V   & 5423.90 &   [Fe\,{\sc iii}]   & -0.128 & 0.028 & 0.008 & V   & 7331.40 &   [Ar\,{\sc iv}]  &  *  & 0.034 & 0.002 &  K   \\ 
4199.98 &   N\,{\sc ii}   & * &   *   &   *   & V   & 5424.22 &   [Fe\,{\sc vi}]   & * &   *   &   *   & V   & 7320/30  &   [O\,{\sc ii}]$_{\rm R}$  &  *  & 0.169 & 0.002 &  K   \\ 
4200.10 &   N\,{\sc iii}   & * &   *   &   *   & V   & 5480.06 &   N\,{\sc ii}   & -0.138 & 0.024 & 0.008 & V   & 7320/30  &   [O\,{\sc ii}]$_{\rm P}$  &  *  & 5.652 & 0.170 &  K   \\ 
4227.19 &   [Fe\,{\sc v}]   & 0.192 & 0.162 & 0.040 & V   & 5484.85 &   [Fe\,{\sc vi}]   & -0.139 & 0.042 & 0.008 & V   & 7525.96 &  {\hei}  & -0.426 & 0.095 & 0.015 &  K   \\ 
4227.69 &   N\,{\sc ii}   & * &   *   &   *   & V   & 5500.17 &   Fe\,{\sc ii}   & -0.142 & 0.033 & 0.009 & V   & 7530.83 &  [C\,{\sc iv}]  & -0.426 & 0.376 & 0.053 &  K   \\ 
4267.00 &   C\,{\sc ii}   & 0.180 & 0.380 & 0.033 & V   & 5517.66 &   [Cl\,{\sc iii}]   & -0.145 & 0.402 & 0.013 & V   & 7592.74 &   {\heii}   & -0.435 & 0.627 & 0.058 &  K   \\ 
4312.11 &   O\,{\sc ii}   & 0.165 & 0.069 & 0.020 & V   & 5537.60 &   [Cl\,{\sc iii}]   & -0.149 & 0.519 & 0.027 & V   & 7725.80 &   [Fe\,{\sc iv}]   & -0.452 & 0.145 & 0.036 &  K   \\ 
4338.67 &  {\heii}  & 0.157 & 1.129 & 0.057 &  K   & 5592.24 &   O\,{\sc iii}   & -0.158 & 0.064 & 0.008 & V   & 7751.12 &   [Ar\,{\sc iii}]   & -0.455 & 2.463 & 0.116 &  K   \\ 
4340.47 &  {\hi}  & 0.157 & 46.680 & 1.602 &  K   & 5631.07 &   [Fe\,{\sc vi}]   & -0.165 & 0.035 & 0.006 & V   & 7816.16 &   {\hei}   & -0.464 & 0.103 & 0.029 &  K   \\ 
4363.21 &   [O\,{\sc iii}]$_{\rm T}$  & 0.149 & 21.820 & 0.984 &  K   & 5640.23 &   Ni\,{\sc ii}   & -0.166 & 0.027 & 0.006 & V   & 8045.63 &   [Cl\,{\sc iv}]   & -0.492 & 0.907 & 0.064 &  K   \\ 
*  &   [O\,{\sc iii}]$_{\rm R}$  &  *  & 0.224 & 0.086 &  K   & 5701.82 &   [Mn\,{\sc v}]   & -0.176 & 0.027 & 0.005 & V   & 8197.80 &   [Fe\,{\sc v}]   & -0.511 & 0.298 & 0.044 &  K   \\ 
*  &   [O\,{\sc iii}]$_{\rm P}$  &  *  & 21.596 & 0.987 &  K   & 5721.20 &   [Fe\,{\sc vii}]   & -0.179 & 0.062 & 0.007 & V   & 8236.77 &   {\heii}   & -0.515 & 1.238 & 0.080 &  K   \\ 
4379.11 &   N\,{\sc iii}   & 0.144 & 0.116 & 0.032 & V   & 5721.46 &   [Fe\,{\sc vii}]   & * &   *   &   *   & V   & 8237.15 &   {\hei}   & * &  *  &  *  &  K   \\ 
4387.93 &  {\hei}  & 0.142 & 0.598 & 0.075 &  K   & 5754.60 &   [N\,{\sc ii}]$_{\rm T}$  & -0.185 & 1.155 & 0.038 &  K   & 8264.28 &   {\hi}   & -0.518 & 0.076 & 0.025 &  K   \\ 
4437.55 &   {\hei}   & 0.126 & 0.115 & 0.018 & V   & *  &  [N\,{\sc ii}]$_{\rm R}$  &  *  & 0.211 & 0.081 &  K   & 8265.71 &   {\hei}   & * &  *  &  *  &  K   \\ 
4471.50 &   {\hei}   & 0.115 & 4.164 & 0.174 &  K   & *  &  [N\,{\sc ii}]$_{\rm P}$   &  *  & 0.944 & 0.089 &  K   & 8267.94 &   {\hi}   & * &  *  &  *  &  K   \\ 
4541.59 &   {\heii}   & 0.093 & 1.716 & 0.153 &  K   & 5801.33 &   C\,{\sc iv}   & -0.192 & 0.164 & 0.006 & V   & 8333.78 &   {\hi}   & -0.526 & 0.063 & 0.020 &  K   \\ 
4562.60 &   Mg\,{\sc i}]   & 0.087 & 0.108 & 0.015 & V   & 5801.51 &   C\,{\sc iv}   & * &   *   &   *   & V   & 8345.47 &   {\hi}   & -0.527 & 0.103 & 0.033 &  K   \\ 
4571.10 &   Mg\,{\sc i}]   & 0.084 & 0.299 & 0.028 & V   & 5806.57 &   {\heii}   & -0.193 & 0.017 & 0.005 & V   & 8359.00 &   {\hi}   & -0.529 & 0.171 & 0.055 &  K   \\ 
4607.03 &   [Fe\,{\sc iii}]   & 0.073 & 0.075 & 0.022 & V   & 5811.98 &   C\,{\sc iv}   & -0.193 & 0.083 & 0.004 & V   & 8374.48 &   {\hi}   & -0.531 & 0.157 & 0.051 &  K   \\ 
4634.14 &   N\,{\sc iii}   & 0.065 & 1.067 & 0.043 & V   & 5812.14 &   C\,{\sc iv}   & * &   *   &   *   & V   & 8392.40 &   {\hi}   & -0.533 & 0.177 & 0.049 &  K   \\ 
4640.64 &   N\,{\sc iii}   & 0.063 & 2.293 & 0.051 & V   & 5813.19 &   {\heii}   & -0.194 & 0.028 & 0.003 & V   & 8413.32 &   {\hi}   & -0.535 & 0.221 & 0.048 &  K   \\ 
4647.42 &   C\,{\sc iii}   & 0.061 & 0.281 & 0.022 & V   & 5836.50 &   {\heii}   & -0.197 & 0.017 & 0.004 & V   & 8437.95 &   {\hi}   & -0.537 & 0.311 & 0.042 &  K   \\ 
4649.13 &   O\,{\sc ii}   & 0.061 & 0.067 & 0.019 & V   & 5847.10 &   {\heii}   & -0.199 & 0.029 & 0.007 & V   & 8467.25 &   {\hi}   & -0.541 & 0.262 & 0.043 &  K   \\ 
4650.25 &   C\,{\sc iii}   & 0.060 & 0.181 & 0.017 & V   & 5857.27 &   {\heii}   & -0.200 & 0.020 & 0.005 & V   & 8499.00 &   {\heii}   & -0.544 & 0.419 & 0.048 &  K   \\ 
4651.47 &   C\,{\sc iii}   & 0.060 & 0.060 & 0.017 & V   & 5867.74 &   [Kr\,{\sc iv}]   & -0.202 & 0.091 & 0.011 & V   & 8502.48 &   {\hi}   & * &  *  &  *  &  K   \\ 
4658.05 &   [Fe\,{\sc iii}]   & 0.058 & 0.130 & 0.025 & V   & 5869.02 &   {\heii}   & -0.202 & 0.023 & 0.006 & V   & 8545.38 &   {\hi}   & -0.549 & 0.496 & 0.039 &  K   \\ 
4658.30 &   C\,{\sc iv}   & 0.058 & 0.262 & 0.031 & V   & 5875.66 &   {\hei}   & -0.203 & 10.095 & 0.210 &  K   & 8578.69 &   [Cl\,{\sc ii}]   & -0.552 & 0.113 & 0.030 &  K   \\ 
4658.64 &   C\,{\sc iv}   & * &  *  &  *  & V   & 5882.14 &   {\heii}   & -0.204 & 0.706 & 0.081 &  K   & 8581.87 &   {\hei}   & * &  *  &  *  &  K   \\ 
4685.68 &   {\heii}   & 0.050 & 49.681 & 1.847 &  K   & 5931.83 &   {\heii}   & -0.211 & 0.105 & 0.020 &  K   & 8594.87 &  {\heii}  & -0.554 & 0.021 & 0.001 &  K   \\ 
4701.53 &   [Fe\,{\sc iii}]   & 0.045 & 0.043 & 0.014 & V   & 6036.70 &   {\heii}   & -0.226 & 0.054 & 0.017 &  K   & 8598.39 &  {\hi}  & -0.554 & 0.675 & 0.046 &  K   \\ 
4711.37 &   [Ar\,{\sc iv}]   & 0.042 & 4.836 & 0.074 & V   & 6074.10 &   {\heii}   & -0.232 & 0.099 & 0.015 &  K   & 8661.47 &  {\heii}  & -0.560 & 0.026 & 0.001 &  K   \\ 
4713.17 &   {\hei}   & 0.042 & 0.416 & 0.047 & V   & 6086.29 &   [Fe\,{\sc vii}]   & -0.233 & 0.090 & 0.014 &  K   & 8665.02 &  {\hi}  & -0.560 & 0.825 & 0.064 &  K   \\ 
4714.17 &   [Ne\,{\sc iv}]   & 0.041 & 0.566 & 0.048 & V   & 6101.83 &   [K\,{\sc iv}]   & -0.235 & 0.285 & 0.013 &  K   & 8746.00 &  {\heii}  & -0.568 & 0.033 & 0.002 &  K   \\ 
4724.15 &   [Ne\,{\sc iv}]   & 0.038 & 0.555 & 0.023 & V   & 6118.20 &   {\heii}   & -0.238 & 0.142 & 0.024 &  K   & 8750.47 &   {\hi}   & -0.568 & 1.024 & 0.068 &  K   \\ 
4725.67 &   [Ne\,{\sc iv}]   & 0.038 & 0.382 & 0.019 & V   & 6170.60 &   {\heii}   & -0.245 & 0.152 & 0.020 &  K   & 8859.15 &  {\heii}  & -0.578 & 0.043 & 0.002 &  K   \\ 
4740.17 &   [Ar\,{\sc iv}]   & 0.034 & 5.502 & 0.093 & V   & 6233.80 &   {\heii}   & -0.254 & 0.152 & 0.032 &  K   & 8862.78 &  {\hi}  & -0.578 & 1.320 & 0.115 &  K   \\ 
4859.32 &  {\heii}  & 0.001 & 2.315 & 0.106 &  K   & 6300.30 &   [O\,{\sc i}]   & -0.263 & 3.293 & 0.110 &  K   & 9014.91 &   {\hi}   & -0.590 & 1.101 & 0.079 &  K   \\ 
4861.33 &  {\hi}  & 0.000 & 100.000 & 1.174 &  K   & 6310.80 &   {\heii}  & -0.264 & 0.167 & 0.003 &  K   & 9068.60 &   [S\,{\sc iii}]   & -0.593 & 16.035 & 0.850 &  K   \\ 
4904.45 &   O\,{\sc ii}   & -0.012 & 0.026 & 0.008 & V   & 6312.10 &   [S\,{\sc iii}]   & -0.264 & 1.712 & 0.176 &  K   & 9225.23 &  {\heii}  & -0.605 & 0.078 & 0.004 &  K   \\ 
4906.83 &   O\,{\sc ii}   & -0.012 & 0.036 & 0.007 & V   & 6363.78 &   [O\,{\sc i}]   & -0.271 & 1.095 & 0.037 &  K   & 9229.01 &  {\hi}  & -0.605 & 2.294 & 0.164 &  K   \\ 
4906.83 &   [Fe\,{\sc iv}]   & * &   *   &   *   & V   & 6406.30 &   {\heii}   & -0.277 & 0.223 & 0.016 &  K   & 9344.93 &   {\heii}   & -0.613 & 1.317 & 0.183 &  K   \\ 
4907.20 &   N\,{\sc iii}   & * &   *   &   *   & V   & 6435.10 &   [Ar\,{\sc v}]   & -0.281 & 0.663 & 0.022 &  K   & 9530.60 &   [S\,{\sc iii}]   & -0.625 & 39.778 & 2.894 &  K   \\ 
4921.93 &   {\hei}   & -0.016 & 0.951 & 0.054 &  K   & 6461.95 &   C\,{\sc ii}   & -0.284 & 0.056 & 0.011 &  K   & 9542.06 &   {\heii}   & -0.625 & 8.408 & 2.198 &  K   \\ 
4931.23 &   [O\,{\sc iii}]   & -0.019 & 0.204 & 0.015 & V   & 6527.11 &  {\heii}  & -0.293 & 0.439 & 0.079 &  K   & 10049.37 &   {\hi}   & -0.656 & 5.337 & 0.360 &  K   \\ 
\midrule
\end{tabular}
\end{table*}

\subsubsection{Gas and dust mass distributions}
\label{S-gasmass}

We directly derive the total gas mass map by simply adding the H$^{+}$, He$^{+,2+}$, C$^{2+,3+,4+}$, and O$^{0,1+,2+}$ gas masses using Eq.\,(\ref{Eq-4}). 
We obtain the total dust mass map by dividing the resultant gas mass map by the GDR map. 
\begin{eqnarray}
m_{\rm g} &=& \sum \frac{4 \pi D^{2} I_{g} \mu_{g} }{n_{\rm e} J_{\rm g}(T_{\rm e},n_{\rm e}) }.
\label{Eq-4}
\end{eqnarray}
Fig.\,\ref{F:GDR}(b) and (c) display the gas and dust mass distributions, respectively. 
These mass maps are similar regarding their increase with increasing distance from the central star.
The gas and dust masses seem to concentrate in the elliptical nebula's minor axis, which would be regarded as the equatorial plane. 
Such mass distributions may be related to the nonisotropic mass loss during the AGB phase and nebula shaping. 
Each gas and dust mass is listed in Table\,\ref{T-gmass}, where $D_{\rm kpc}$ is the distance in the unit of kpc and the uncertainty corresponds to one-$\sigma$.
Later, we will discuss the gas and dust masses with our distance measurement.

\subsection{Analysis of the 1-D spectrum extracted from the \emph{IUE} aperture}
\label{S-1d}

\subsubsection{Extinction correction}

The {\heii} contribution to {\hi} is very small in terms of line flux. 
However, as we learn from the 2-D analysis, both {\hi} and {\heii} line distributions are very different from each other. 
Therefore, in either 1-D or 2-D analysis, without subtracting {\heii} line contamination, one cannot obtain the accurate $c$({\hb}) and $I$({\hb}) values required for extinction correction, plasma analyses, and abundance derivations (especially, the ions with similar IP as {\heii}; e.g., C\,{\sc iv}, {\ariv}, and {\arv}). 
For example, if we ignore the {\heii} line contribution, we obtain the similar $c$({\hb}) = $0.57 \pm 0.03$ but a larger $I$({\hb}) = $(4.79\pm 0.35){\times}10^{-12}$\,erg\,s$^{-1}$\,cm$^{-2}$. 
Owing to the overestimated $I$({\hb}), the derived ionic abundances become $\sim$$5-10$\,$\%$ smaller compared with the case of the {\heii} contamination being considered. 
The underestimation of each ionic abundance may be slight, but it results in a large difference in the calculation of the ICFs and ultimately elemental abundances.

Following our method explained in \S\,\ref{S-chb}, we obtain $c$({\hb}) = $0.56 \pm 0.03$, where {\Ne}({\cliii}) = 6030\,cm$^{-3}$ and {\te}({\oiii}) = 14100\,K. 
The extinction-free {\hb} line flux $I$({\hb}) is $(4.61 \pm 0.35){\times}10^{-12}$\,erg\,s$^{-1}$\,cm$^{-2}$. 
Table\,\ref{T-chb} summarises the resulting major {\hi} and {\heii} line fluxes. 
The measured ratios of these {\hi} and {\heii} lines to the respective {\hb} and {\heii}\,5411\,{\AA} in {\te} = 14100\,K and {\Ne} = 6030\,cm$^{-3}$ are in perfect accordance with the theoretically predicted values (seventh and eighth columns of the table). 
The four {\heii} line ratios around the 2200\,{\AA} bump are also in excellent agreement with the theoretical predictions, indicating that our assumed $R_{\rm V}$ provides an adequate fit to the data based on the fact that $f$($\lambda$) in UV wavelengths depends largely on the $R_{\rm V}$ value. 
The identified emission lines and their extinction fluxes relative to $I$({\hb}) are summarised in Table\,\ref{T:line}.

\subsubsection{Plasma diagnostics}
 \label{S-plasma}

\begin{table}
\centering
\renewcommand{\arraystretch}{0.75}
\caption{Summary of plasma diagnostics based the 1-D integrated spectra. 
\label{T-neTe}}
\begin{tabularx}{\columnwidth}{@{}@{\extracolsep{\fill}}
l@{\hspace{1pt}}D{p}{\pm}{-1}@{\hspace{-1pt}}D{p}{\pm}{-1}@{}}
    \midrule
CEL {\Ne}--diagnostic line & \multicolumn{1}{c}{Value} & \multicolumn{1}{c}{Result (cm$^{-3}$)}   \\ 
\midrule
{\sii}\,6716\,{\AA}/6731\,{\AA} & 0.596 ~p~ 0.026 & 3170 ~p~ 580 \\ 
{\cliii}\,5517\,{\AA}/5537\,{\AA} & 0.775 ~p~ 0.048 & 6030 ~p~ 1100 \\ 
{\ariv}\,4711\,{\AA}/4740\,{\AA} & 0.879 ~p~ 0.020 & 6240 ~p~ 450 \\  
\midrule
CEL {\te}--diagnostic line & \multicolumn{1}{c}{Value}
	 & \multicolumn{1}{c}{Result (K)}    \\ 
\midrule
{\nii}\,(6548\,{\AA} + 6583\,{\AA})/5755\,{\AA} & 61.340 ~p~ 6.121 & 11730 ~p~ 570 \\ 
{\siii}\,(9068\,{\AA} + 9530\,{\AA})/6312\,{\AA} & 32.593 ~p~ 3.780 & 12520 ~p~ 810 \\ 
{\ariii}\,(7135\,{\AA} + 7751\,{\AA})/5191\,{\AA} & 82.478 ~p~ 8.532 & 13650 ~p~ 700 \\ 
{\oiii}\,(4959\,{\AA} + 5007\,{\AA})/4363\,{\AA} & 77.844 ~p~ 0.815 & 14100 ~p~ 70 \\ 
{\cliv}\,8046\,{\AA}/5323\,{\AA} & 17.659 ~p~ 4.963 & 15410 ~p~ 2360 \\ 
{\ariv}\,4740\,{\AA}/7263\,{\AA} & 27.775 ~p~ 1.834 & 15640 ~p~ 830 \\ 
\midrule
RL {\te}--diagnostic line   & \multicolumn{1}{c}{Value}&\multicolumn{1}{c}{Result (K)}\\
	     \midrule 
{\hei}\,7281\,{\AA}/6678\,{\AA} & 0.224 ~p~ 0.025 &  8850 ~p~ 1200 \\ 
  Paschen jump        & 2.09{\times}10^{-2} ~p~ 3.46{\times}10^{-3} & 9990 ~p~ 2930 \\ 
\midrule
\end{tabularx}
\end{table}

The resultant {\Ne} and {\te} values are summarised in Table\,\ref{T-neTe}.

CEL {\Ne} and {\te} are determined at the intersection of the {\Ne} and {\te} diagnostic curves in each ionisation zone. 
In the neutral/low-ionisation zone, {\te}({\nii}) and {\Ne}({\sii}) are determined from their curves. 
In the moderate-ionisation zone, {\te}({\siii}), {\te}({\ariii}), {\te}({\oiii}), and {\Ne}({\cliii}) are derived from their curves. In the high-ionisation zone, {\te}({\ariv}), {\te}({\cliv}), and {\Ne}({\ariv}) are derived from their curves. {\te}({\nii}) and {\te}({\oiii}) values are calculated after subtracting the recombination contributions of N$^{2+}$ and O$^{3+}$ to the {\nii}\,5755\,{\AA} and {\oiii}\,4363\,{\AA} lines using Eqs.\,(1) and (3) of \citet{Liu:2000aa}, respectively. 
In Table\,\ref{T:line}, the recombination contributions of unsubtracted and subtracted {\oiii}\,4363\,{\AA} are denoted by $I$({\oiii}$_{\rm T}$\,4363\,{\AA}) and $I$({\oiii}$_{\rm P}$\,4363\,{\AA}), respectively. $I$({\oiii}$_{\rm R}$\,4363\,{\AA}) is 
the recombination contribution. It is a similar case for {\nii}\,5755\,{\AA}. 
The N$^{2+}$ contribution is extremely large, $\sim$18\,$\%$ of the observed $I$({\nii}\,5755\,{\AA}). Using the uncorrected $I$({\nii}\,5755\,{\AA}),  
a {\te}({\nii}) of $13010 ~\pm~ 330$\,K is obtained, which is higher by 1240\,K than that obtained using the uncorrected $I$({\nii}\,5755\,{\AA}).

\te}({\hei}) is calculated using the method explained in \S\,\ref{S-tene}. 
{\te}(PJ) is calculated from the Paschen continuum discontinuity using Eq.\,(7) of \citet{Fang:2011aa}.

\subsubsection{Ionic abundance derivations}

\begin{table}
\centering
\renewcommand{\arraystretch}{0.75}
\caption{The adopting {\Ne} and {\te} for each ionic abundance. \label{T-CELtn}}
\begin{tabularx}{\columnwidth}{@{}@{\extracolsep{\fill}}ccl@{}}
\midrule 
 {\te}&{\Ne}&Ion\\
\midrule 
{\te}({\hei}) &{\Ne}({\cliii})&He$^{+}$\\
{\te}({\nii}) &{\Ne}({\sii})  &C$^{+}$, N$^{0,+}$, O$^{0,+}$, S$^{+}$, Cl$^{+}$\\
{\te}({\ariii}) &{\Ne}({\cliii})&Ar$^{2+}$\\
{\te}({\siii})&{\Ne}({\cliii})&S$^{2+}$\\
{\te}({\oiii})&{\Ne}({\cliii})&He$^{2+}$, C$^{2+}$(CEL,RL), C$^{3+,4+}$(RL), N$^{2+}$, O$^{2+}$,\\ 
	      &               &Cl$^{2+}$, Fe$^{2+}$\\
{\te}({\ariv})&{\Ne}({\ariv}) &N$^{3+}$, O$^{3+}$, Ne$^{3+}$, Mg$^{4+}$, Cl$^{3+}$, Ar$^{3+,4+}$,\\
              &               &K$^{3+}$, Ca$^{4+}$, Fe$^{5+,6+}$, Kr$^{3+}$\\
\midrule 
\end{tabularx}
\end{table}

\begin{table*}
\renewcommand{\arraystretch}{0.75}
\caption{Ionic abundances based the 1-D integrated spectra. The values denoted by dagger (${\dagger}$) are excluded when calculating the average value. 
\label{T-ion}}
\begin{tabularx}{\textwidth}{@{}@{\extracolsep{\fill}}
lcD{p}{\pm}{-1}clcD{p}{\pm}{-1}c@{}}
 \midrule 
Ion(X$^{\rm m+}$)  &
$\lambda_{\rm lab.}$ ({\AA})&
\multicolumn{1}{c}{$n$(X$^{\rm m+}$)/$n$(H$^{+}$)}&
source&
Ion(X$^{\rm m+}$)  &
$\lambda_{\rm lab.}$ ({\AA})&
\multicolumn{1}{c}{$n$(X$^{\rm m+}$)/$n$(H$^{+}$)}&
source\\         
 \midrule 
He$^{+}$  & 4471.5 &  8.00{\times}10^{-2} ~p~ 3.04{\times}10^{-3}{^{\dagger}}  &  KOOLS-IFU  & S$^{2+}$  & 6312.1 &  1.63{\times}10^{-6} ~p~ 4.08{\times}10^{-7}  &  KOOLS-IFU  \\ 
  & 4921.9 &  6.96{\times}10^{-2} ~p~ 3.92{\times}10^{-3}  &  KOOLS-IFU  &    & 9068.6 &  1.60{\times}10^{-6} ~p~ 1.98{\times}10^{-7}  &  KOOLS-IFU  \\ 
  & 5875.6 &  6.67{\times}10^{-2} ~p~ 1.76{\times}10^{-3}  &  KOOLS-IFU  &    & 9538.0 &  1.61{\times}10^{-6} ~p~ 2.15{\times}10^{-7}  &  KOOLS-IFU  \\ 
  & 6678.2 &  6.90{\times}10^{-2} ~p~ 2.01{\times}10^{-3}  &  KOOLS-IFU  &   &    &  \bf{1.60{\times}10^{-6}} ~p~ \bf{2.15{\times}10^{-7}}  &    \\ 
  & 7281.4 &  6.54{\times}10^{-2} ~p~ 4.15{\times}10^{-3}  &  KOOLS-IFU  &  Cl$^{+}$  & 8578.7 &  \bf{3.83{\times}10^{-9}} ~p~ \bf{1.08{\times}10^{-9}}  &  KOOLS-IFU  \\ 
  &    &  \bf{6.77{\times}10^{-2}} ~p~ \bf{1.97{\times}10^{-3}}  &    &  Cl$^{2+}$  & 5517.7 &  2.58{\times}10^{-8} ~p~ 1.54{\times}10^{-9}  &  VIMOS  \\ 
He$^{2+}$  & 2512 &  4.64{\times}10^{-2} ~p~ 4.56{\times}10^{-3}  &  \emph{IUE}  &     & 5537.9 &  2.58{\times}10^{-8} ~p~ 1.38{\times}10^{-9}  &  VIMOS  \\ 
  & 2733.1 &  4.12{\times}10^{-2} ~p~ 2.46{\times}10^{-3}  &  \emph{IUE}  &    &    &  \bf{2.58{\times}10^{-8}} ~p~ \bf{1.03{\times}10^{-9}}  &    \\ 
  & 3202.3 &  4.17{\times}10^{-2} ~p~ 1.76{\times}10^{-3}  &  \emph{IUE}  &  Cl$^{3+}$  & 5323.3 &  2.74{\times}10^{-8} ~p~ 6.16{\times}10^{-9}  &  VIMOS  \\ 
  & 4541.6 &  4.35{\times}10^{-2} ~p~ 3.88{\times}10^{-3}  &  KOOLS-IFU  &     & 7530.8 &  2.89{\times}10^{-8} ~p~ 4.89{\times}10^{-9}  &  KOOLS-IFU  \\ 
  & 4685.7 &  4.31{\times}10^{-2} ~p~ 1.60{\times}10^{-3}  &  KOOLS-IFU  &     & 8046.3 &  3.02{\times}10^{-8} ~p~ 3.53{\times}10^{-9}  &  KOOLS-IFU  \\ 
  & 5411.5 &  4.01{\times}10^{-2} ~p~ 6.96{\times}10^{-4}  &  KOOLS-IFU  &    &    &  \bf{2.94{\times}10^{-8}} ~p~ \bf{2.60{\times}10^{-9}}  &    \\ 
  &    &  \bf{4.09{\times}10^{-2}} ~p~ \bf{2.22{\times}10^{-3}}  &    &  Ar$^{2+}$  & 5191.8 &  4.88{\times}10^{-7} ~p~ 1.08{\times}10^{-7}  &  VIMOS  \\ 
C$^{2+}$(RL)  & 4267.0 &  3.93{\times}10^{-4} ~p~ 3.42{\times}10^{-5}  &  VIMOS  &     & 7135.8 &  4.81{\times}10^{-7} ~p~ 4.85{\times}10^{-8}  &  KOOLS-IFU  \\ 
  & 6461.0 &  5.74{\times}10^{-4} ~p~ 1.13{\times}10^{-4}  &  KOOLS-IFU  &     & 7751.1 &  4.95{\times}10^{-7} ~p~ 5.29{\times}10^{-8}  &  KOOLS-IFU  \\ 
  &    &  \bf{4.08{\times}10^{-4}} ~p~ \bf{3.27{\times}10^{-5}}  &    &    &    &  \bf{4.88{\times}10^{-7}} ~p~ \bf{3.39{\times}10^{-8}}  &    \\ 
C$^{3+}$(RL)  & 4647.4 &  4.20{\times}10^{-4} ~p~ 3.30{\times}10^{-5}  &  VIMOS  &  Ar$^{3+}$  & 4711.4 &  4.26{\times}10^{-7} ~p~ 5.14{\times}10^{-8}  &  VIMOS  \\ 
   & 4650.3 &  4.50{\times}10^{-4} ~p~ 4.31{\times}10^{-5}  &  VIMOS  &     & 4740.2 &  4.26{\times}10^{-7} ~p~ 5.95{\times}10^{-8}  &  VIMOS  \\ 
  &    &  \bf{4.31{\times}10^{-4}} ~p~ \bf{2.62{\times}10^{-5}}  &    &     & 7170.5 &  5.03{\times}10^{-7} ~p~ 1.07{\times}10^{-7}{^{\dagger}}  &  KOOLS-IFU  \\ 
C$^{4+}$(RL)  & 4659.0 &  \bf{6.49{\times}10^{-5}} ~p~ \bf{7.69{\times}10^{-6}}  &  VIMOS  &     & 7262.7 &  4.30{\times}10^{-7} ~p~ 9.13{\times}10^{-8}  &  KOOLS-IFU  \\ 
C$^{+}$(CEL)  & 2325.0 &  \bf{4.46{\times}10^{-5}} ~p~ \bf{1.33{\times}10^{-5}}  &  \emph{IUE}  &    &    &  \bf{4.27{\times}10^{-7}} ~p~ \bf{3.58{\times}10^{-8}}  &    \\ 
C$^{2+}$(CEL)  & 1908.2 &  \bf{2.19{\times}10^{-4}} ~p~ \bf{2.16{\times}10^{-5}}  &  \emph{IUE}  &  Ar$^{4+}$  & 6435.1 &  1.04{\times}10^{-7} ~p~ 1.19{\times}10^{-8}  &  KOOLS-IFU  \\ 
N$^{0}$  & 5197.9 &  \bf{2.73{\times}10^{-7}} ~p~ \bf{3.93{\times}10^{-8}}  &  KOOLS-IFU  &     & 7005.4 &  9.23{\times}10^{-8} ~p~ 1.05{\times}10^{-8}  &  KOOLS-IFU  \\ 
N$^{+}$  & 5754.6 &  5.98{\times}10^{-6} ~p~ 1.47{\times}10^{-6}  &  KOOLS-IFU  &    &    &  \bf{9.74{\times}10^{-8}} ~p~ \bf{7.87{\times}10^{-9}}  &    \\ 
  & 6548.0 &  5.90{\times}10^{-6} ~p~ 6.87{\times}10^{-7}  &  KOOLS-IFU  &  K$^{3+}$  & 6101.8 &  8.85{\times}10^{-9} ~p~ 1.01{\times}10^{-9}  &  KOOLS-IFU  \\ 
  & 6583.5 &  5.93{\times}10^{-6} ~p~ 6.89{\times}10^{-7}  &  KOOLS-IFU  &     & 6795.1 &  1.08{\times}10^{-8} ~p~ 1.82{\times}10^{-9}  &  KOOLS-IFU  \\ 
  &    &  \bf{5.92{\times}10^{-6}} ~p~ \bf{4.62{\times}10^{-7}}  &    &    &    &  \bf{9.31{\times}10^{-9}} ~p~ \bf{8.83{\times}10^{-10}}  &    \\ 
N$^{2+}$  & 1751.6 &  \bf{3.08{\times}10^{-5}} ~p~ \bf{3.00{\times}10^{-6}}  &  \emph{IUE}  &  Ca$^{4+}$  & 5309.1 &  \bf{4.64{\times}10^{-9}} ~p~ \bf{8.65{\times}10^{-10}}  &  VIMOS  \\ 
N$^{3+}$  & 1484.2 &  \bf{2.28{\times}10^{-5}} ~p~ \bf{7.67{\times}10^{-6}}  &  \emph{IUE}  &  Fe$^{2+}$  & 5270.7 &  \bf{1.92{\times}10^{-8}} ~p~ \bf{3.69{\times}10^{-9}}  &  VIMOS  \\ 
O$^{0}$  & 6363.8 &  \bf{3.76{\times}10^{-6}} ~p~ \bf{6.39{\times}10^{-7}}  &  KOOLS-IFU  &  Fe$^{5+}$  & 5176.0 &  1.22{\times}10^{-8} ~p~ 2.77{\times}10^{-9}  &  VIMOS  \\ 
O$^{+}$  & 7325.0 &  \bf{2.32{\times}10^{-5}} ~p~ \bf{7.12{\times}10^{-6}}  &  KOOLS-IFU  &     & 5145.8 &  1.92{\times}10^{-8} ~p~ 5.48{\times}10^{-9}  &  VIMOS  \\ 
O$^{2+}$  & 1665.0 &  1.50{\times}10^{-4} ~p~ 1.45{\times}10^{-5}  &  \emph{IUE}  &     & 5278.0 &  2.74{\times}10^{-8} ~p~ 8.51{\times}10^{-9}  &  VIMOS  \\ 
    & 2321.0 &  1.57{\times}10^{-4} ~p~ 8.19{\times}10^{-6}  &  \emph{IUE}  &     & 5335.2 &  2.37{\times}10^{-8} ~p~ 5.33{\times}10^{-9}  &  VIMOS  \\ 
   & 4363.2 &  1.56{\times}10^{-4} ~p~ 8.19{\times}10^{-6}  &  KOOLS-IFU  &     & 5484.9 &  3.13{\times}10^{-8} ~p~ 7.07{\times}10^{-9}  &  VIMOS  \\ 
   & 4958.9 &  1.51{\times}10^{-4} ~p~ 2.80{\times}10^{-6}  &  KOOLS-IFU  &     & 5631.1 &  1.85{\times}10^{-8} ~p~ 3.82{\times}10^{-9}  &  VIMOS  \\ 
   & 5006.8 &  1.57{\times}10^{-4} ~p~ 2.85{\times}10^{-6}  &  KOOLS-IFU  &    &    &  \bf{1.77{\times}10^{-8}} ~p~ \bf{1.82{\times}10^{-9}}  &    \\ 
  &    &  \bf{1.55{\times}10^{-4}} ~p~ \bf{1.87{\times}10^{-6}}  &    &  Fe$^{6+}$  & 5720.0 &  1.31{\times}10^{-8} ~p~ 2.17{\times}10^{-9}  &  VIMOS  \\ 
O$^{3+}$  & 1403.3 &  \bf{1.39{\times}10^{-4}} ~p~ \bf{5.34{\times}10^{-5}}  &  \emph{IUE}  &     & 6086.4 &  1.24{\times}10^{-8} ~p~ 2.51{\times}10^{-9}  &  KOOLS-IFU  \\ 
Ne$^{3+}$  & 1602.4 &  2.21{\times}10^{-5} ~p~ 7.68{\times}10^{-6}  &  \emph{IUE}  &    &    &  \bf{1.28{\times}10^{-8}} ~p~ \bf{1.20{\times}10^{-9}}  &    \\ 
   & 2424.0 &  1.36{\times}10^{-5} ~p~ 3.17{\times}10^{-6}{^{\dagger}}  &  \emph{IUE}  &  Kr$^{3+}$  & 5346.0 &  2.37{\times}10^{-9} ~p~ 4.02{\times}10^{-10}  &  VIMOS  \\ 
   & 4724.2 &  2.61{\times}10^{-5} ~p~ 8.40{\times}10^{-6}  &  VIMOS  &     & 5867.8 &  1.61{\times}10^{-9} ~p~ 2.54{\times}10^{-10}  &  VIMOS  \\ 
  &    &  \bf{2.39{\times}10^{-5}} ~p~ \bf{2.77{\times}10^{-6}}  &    &    &    &  \bf{1.83{\times}10^{-9}} ~p~ \bf{2.15{\times}10^{-10}}  &    \\ 
Mg$^{4+}$  & 2783.2 &  \bf{2.49{\times}10^{-6}} ~p~ \bf{5.35{\times}10^{-7}}  &  \emph{IUE}  &            &  &  &  \\ 
S$^{+}$  & 6716.4 &  1.62{\times}10^{-7} ~p~ 1.07{\times}10^{-8}  &  KOOLS-IFU  &            &  &  &  \\ 
   & 6730.8 &  1.64{\times}10^{-7} ~p~ 1.05{\times}10^{-8}  &  KOOLS-IFU  &            &  &  &  \\ 
  &    &  \bf{1.63{\times}10^{-7}} ~p~ \bf{7.47{\times}10^{-9}}  &    &            &  &  &  \\ 
 \midrule 
\end{tabularx}
\end{table*}

\begin{table}
\caption{Ionisation correction factor ICF(X) for element X based the 1-D integrated spectra. \label{T-icf}}
\centering
\renewcommand{\arraystretch}{0.75}
\begin{tabularx}{\columnwidth}{@{}@{\extracolsep{\fill}}l@{\hspace{4pt}}lcD{p}{\pm}{-1}@{}}
 \midrule 
X      &Expression                      &ICF(X)               &\multicolumn{1}{c}{Value}\\ 
 \midrule 
He     &ICF(X)$\times$(He$^{+}$ + He$^{2+}$)&\multicolumn{1}{c}{1}                 &\multicolumn{1}{c}{1}\\ 
C(RL)  &ICF(X)$\times$(C$^{2+}$ + C$^{3+}$ + C$^{4+}$) &\multicolumn{1}{c}{1}                 &\multicolumn{1}{c}{1}\\ 
C(CEL) &ICF(X)$\times$(C$^{+}$ + C$^{2+}$)  &O/(O$^{+}$+O$^{2+}$)&1.78 ~p~ 0.31\\
N      &ICF(X)$\times$(N$^{+}$ + N$^{2+}$ + N$^{3+}$)  &\multicolumn{1}{c}{1}                 &\multicolumn{1}{c}{1}\\ 
O      &ICF(X)$\times$(O$^{+}$ + O$^{2+}$ + O$^{3+}$)  &\multicolumn{1}{c}{1}                 &\multicolumn{1}{c}{1}\\ 
Ne     &ICF(X)$\times$Ne$^{3+}$  &O/O$^{3+}$ &2.27 ~p~ 0.95\\
Mg     &ICF(X)$\times$Mg$^{4+}$  &Ar/Ar$^{4+}$ &10.38 ~p~ 0.98\\
S      &ICF(X)$\times$(S$^{+}$+S$^{2+}$)&Cl/(Cl$^{+}$+Cl$^{2+}$)&2.22 ~p~ 0.15\\
Cl     &ICF(X)$\times$Cl$^{3+}$+Cl$^{+}$+Cl$^{2+}$+Cl$^{3+}$&Ar$^{4+}$/Ar$^{3+}$&0.23 ~p~ 0.03\\
Ar     &ICF(X)$\times$(Ar$^{2+}$ + Ar$^{3+}$ + Ar$^{4+}$)&\multicolumn{1}{c}{1}                 &\multicolumn{1}{c}{1}\\ 
K      &ICF(X)$\times$K$^{3+}$&O/O$^{3+}$&2.27 ~p~ 0.95\\ 
Ca     &ICF(X)$\times$Ca$^{4+}$&Ar/Ar$^{4+}$&10.38 ~p~ 0.98\\  
Fe     &ICF(X)$\times$Fe$^{2+}$ + Fe$^{5+}$ + Fe$^{6+}$&3.22\,Ar/Ar$^{2+}$&6.67 ~p~ 0.57\\ 
Kr     &ICF(X)$\times$Kr$^{3+}$&Ar/Ar$^{3+}$&2.37 ~p~ 0.23\\ 
 \midrule 
\end{tabularx}
\end{table}

\begin{table*}
\renewcommand{\arraystretch}{0.75}
\caption{Elemental abundances based the 1-D integrated spectra. Selenium (Se) abundance is the result of \citet{Sterling:2015aa}. 
The value in the column (4) is 
 the derived abundances with respect to the solar abundances, defined as [X/H] =
  $\epsilon({\rm X})({\rm Ours}) - \epsilon({\rm X})_{\sun}$. 
 We take $\epsilon$(X)$_{\sun}$ of \citet{Asplund:2009aa}. In the columns (5-8), H94, P04, B13, and M19 mean 
\citet{Hyung:1994ab}, \citet{Pottasch:2004aa}, \citet{Bohigas:2013aa}, and \citet{Miller:2019aa}, respectively.
\label{T-elem}
}
\begin{tabularx}{\textwidth}{@{}@{\extracolsep{\fill}}lD{p}{\pm}{-1}D{p}{\pm}{-1}D{p}{\pm}{-1}
D{.}{.}{-1}
D{.}{.}{-1}
D{.}{.}{-1}
D{p}{\pm}{-1}
@{}}
\midrule
 \multicolumn{1}{c}{X}  &
  \multicolumn{1}{c}{X/H}   & 
  \multicolumn{1}{c}{$\epsilon({\rm X})$(Ours)} & 
  \multicolumn{1}{c}{[X/H]} &
  \multicolumn{1}{c}{$\epsilon({\rm X})$(H94)} &
  \multicolumn{1}{c}{$\epsilon({\rm X})$(P04)} &
  \multicolumn{1}{c}{$\epsilon({\rm X})$(B13)} &
  \multicolumn{1}{c}{$\epsilon({\rm X})$(M19)}\\
\midrule
He    & 1.09{\times}10^{-1}~p~2.97{\times}10^{-3} & 11.04~p~0.01 & +0.11~p~0.02 & 11.03 & 11.02 & 11.07 & 11.05 ~p~ 0.06\\ 
C(CEL)& 4.70{\times}10^{-4}~p~9.38{\times}10^{-5} & 8.67~p~0.09 & +0.24~p~0.10 & 8.63 & 8.68 &  & 8.62 ~p~ 0.09 \\ 
C(RL) & 9.04{\times}10^{-4}~p~4.26{\times}10^{-5} & 8.97~p~0.02 & +0.53~p~0.05 &   &   & 8.83 &  \\ 
N     & 5.95{\times}10^{-5}~p~8.24{\times}10^{-6} & 7.77~p~0.06 & -0.06~p~0.08 & 7.80 & 7.86 & 8.17 & 8.07 ~p~ 0.11 \\ 
O     & 3.17{\times}10^{-4}~p~5.39{\times}10^{-5} & 8.50~p~0.07 & -0.19~p~0.09 & 8.15 & 8.40 & 8.31 & 8.53 ~p~ 0.08\\ 
Ne    & 5.93{\times}10^{-5}~p~3.13{\times}10^{-5} & 7.77~p~0.23 & -0.16~p~0.25 & 7.81 & 7.76 & 7.70 & 7.73 ~p~ 0.10\\ 
Mg    & 2.58{\times}10^{-5}~p~6.07{\times}10^{-6} & 7.41~p~0.10 & -0.19~p~0.11 &  & 7.08 &  &  \\ 
S     & 3.92{\times}10^{-6}~p~5.48{\times}10^{-7} & 6.59~p~0.06 & -0.53~p~0.07 & 6.37 & 6.65 & 6.56 & 6.26 ~p~ 0.08\\ 
Cl    & 6.57{\times}10^{-8}~p~3.15{\times}10^{-9} & 4.82~p~0.02 & -0.68~p~0.30 & 4.94 & 5.04 & 4.81 &  \\ 
Ar    & 1.07{\times}10^{-6}~p~9.01{\times}10^{-8} & 6.03~p~0.04 & -0.37~p~0.13 & 6.09 & 6.08 & 6.06 & 6.00 ~p~ 0.15 \\ 
K     & 2.46{\times}10^{-8}~p~1.11{\times}10^{-8} & 4.39~p~0.20 & -0.64~p~0.22 & 4.85 & 4.79 & 4.52 &  \\ 
Ca    & 4.82{\times}10^{-8}~p~1.01{\times}10^{-8} & 4.68~p~0.09 & -1.66~p~0.10 & 5.06 &   & 4.58 &  \\ 
Fe    & 1.58{\times}10^{-7}~p~2.70{\times}10^{-8} & 5.20~p~0.07 & -2.30~p~0.08 &  &  & 5.31 &  \\ 
Se    & 1.70{\times}10^{-9}~p~8.86{\times}10^{-10} & 3.23~p~0.23 & -0.10~p~0.23 &  &  & &  \\ 
Kr    & 4.33{\times}10^{-9}~p~6.61{\times}10^{-10} & 3.64~p~0.07 & +0.39~p~0.09 &   &   & 3.82 &  \\ 
\midrule
\end{tabularx}
\end{table*}

Each CEL abundance is calculated, as in 2-D line map analysis, by solving atomic 
multiple energy-level population models given by the {\te} and {\Ne} values, which are determined by considering a targeting line's 2-D emission-line distribution and IP.
The adopting {\te}-{\Ne} pair for each CEL and RL ion is listed in Table\,\ref{T-CELtn}. 
The results are listed in Table\,\ref{T-ion}. 
When more than one line for a targeting ion is detected, the intensity-weighted average of the derived abundances is used. 
The adopted value is shown in boldface on the last line. 
The fourth and eighth columns indicate the data source that is used to calculate abundance.

Our abundance derivation results have advantages over others because 
(1) we extract the {\it IUE}'s PSF convoluted KOOLS-IFU and VIMOS spectra from the same region as observed by the \emph{IUE}, 
(2) we perform line-flux normalisation in UV-optical wavelengths with a single {\hb} line flux and can correct all line fluxes with a single $c$({\hb}) value, 
as demonstrated by the fact that the corrected {\hi} and {\heii} line fluxes coincide with the theoretical ones (Table\,\ref{T-chb}), and 
(3) Using the entire nebula's UV-optical spectrum, we obtain the representative physical and chemical 
parameters of IC2165. These cannot be performed without the IFU data.

Previous determinations were based on line-flux normalisation in every wavelength band 
and did not account for PSFs between instruments. 
A further major drawback was that these had sometimes dealt with spectra extracted from different regions. 
We completely overcome these problems causing incorrect elemental abundances of IC2165. 
This is supported by the fact that the He$^{2+}$ and O$^{2+}$ abundances derived from the {\it IUE} UV data 
are coincident with those from optical KOOLS-IFU data (Table\,\ref{T-ion}). 
Thus, the ionic/elemental abundances derived from our thorough analysis should be the most representative values of IC2165 determined so far. 
We provide brief comments on the CEL C$^{+}$, N$^{+}$, O$^{+,2+}$, and Kr$^{3+}$.

{\bf CEL C$^{+}$}  
We detect an emission line at 2326\,{\AA} identified as the complex of {\oiii}\,2321/30\,{\AA} and $[${\cii}$]$\,2322-28\,{\AA}. 
The {\oiii} $I$(2321/30\,{\AA})/$I$(4363\,{\AA}) (0.237) depends on their transition probabilities
because both lines radiatively transit from the same energy level. 
Thus, the $[${\cii}$]$\,2322-28\,{\AA} line flux corresponds to the residual of \mbox{$I$(2326\,{\AA}) - 0.237$I$({\oiii}\,4361\,{\AA})}. 

{\bf CEL N$^{+}$ and O$^{+,2+}$} 
For the N$^{+}$ and O$^{2+}$ estimates, the contaminations to the {\nii}\,5755\,{\AA} and the {\oiii}\,4363\,{\AA} are excluded. 
For the O$^{+}$ abundance, we eliminate the contribution from the {\ariv}\,7331\,{\AA} ($I$({\ariv}\,7331\,{\AA}) = $0.034 \pm 0.002$ in Table\,\ref{T:line}) 
using the {\ariv} $I$(7331\,{\AA})/$I$(7262\,{\AA}) (0.169, both lines have the same upper energy level) 
as well as the O$^{2+}$ recombination contribution using Eq.(2) of \citet{Liu:2000aa} ([O\,{\sc ii}]$_{\rm R}$\,7320/30\,{\AA} in Table\,\ref{T:line}). 
The contribution-free [O\,{\sc ii}]\,7320/30\,{\AA} line flux is indicated as [O\,{\sc ii}]$_{\rm P}$\,7320/30\,{\AA} in Table\,\ref{T:line}. 

{\bf CEL S$^{2+}$} 
The measurement of the {\siii}\,6312\,{\AA} line flux is performed through a slightly complicated technique due to the contribution of {\heii}\,6310\,{\AA} and {\oi}\,6300\,{\AA}. First, we subtract the {\oi}\,6300\,{\AA} line flux from the sum of the {\oi}\,6300\,{\AA} and {\siii}\,6312\,{\AA} line fluxes. 
The {\oi}\,6300\,{\AA} line flux is determined by subtracting the theoretical {\oi} $I$(6300\,{\AA})/$I$(6363\,{\AA}) (3.13). 
Then, we subtract {\heii}\,6310\,{\AA} from this residual: 
the {\heii}\,6310\,{\AA} can be estimated from the theoretical {\heii} $I$(6310\,{\AA})/$I$(5411\,{\AA}) (4.59$\times$10$^{-2}$) 
under the assumption of Case B in {\te}({\oiii}) and {\Ne}({\cliii}). 
In this manner, we obtain the {\siii}\,6312\,{\AA} line flux with fewer line contaminations. 

{\bf CEL Kr$^{3+}$} 
We subtract the contribution from the {\heii}\,5869\,{\AA} line by using the theoretical {\heii} $I$(5869\,{\AA})/$I$(5857\,{\AA}) (1.10) in {\te}({\oiii}) and {\Ne}({\cliii}).

\subsubsection{Elemental abundance derivations}

We calculate the elemental abundances using ICF(X) for element X.
To determine ICF(X), we refer to the ionisation fraction derived by our photoionisation model of IC2165 (\S\,\ref{S-PI}). The neutral species N$^{0}$ and O$^{0}$ 
are excluded for the N and O abundance calculations. 
We summarise the adopted ICF for each element and the derived elemental abundances in Tables\,\ref{T-icf} and \ref{T-elem}, respectively. 
In Table\,\ref{T-elem}, we list the measurements of $\epsilon$(X) by \citet{Hyung:1994ab}, 
\citet{Pottasch:2004aa}, \citet{Bohigas:2013aa}, and \citet{Miller:2019aa}. 
Excluding \citet{Bohigas:2013aa}, these studies did not investigate the average elemental abundances of the nebula as a whole, but rather those of the small regions that are limited by the slit dimension. Furthermore, each study focuses on different regions. Therefore, the abundances determined by each author are different 
from each other, implying that elemental abundances could be spatially varied. 
We discuss the spatial variation of elemental abundances in the next section.

\section{Discussion}
\label{S-discuss}

\subsection{Comparison with the AGB nucleosynthesis model}
\label{S-AGB}

\begin{figure}
\centering
\includegraphics[width=\columnwidth]{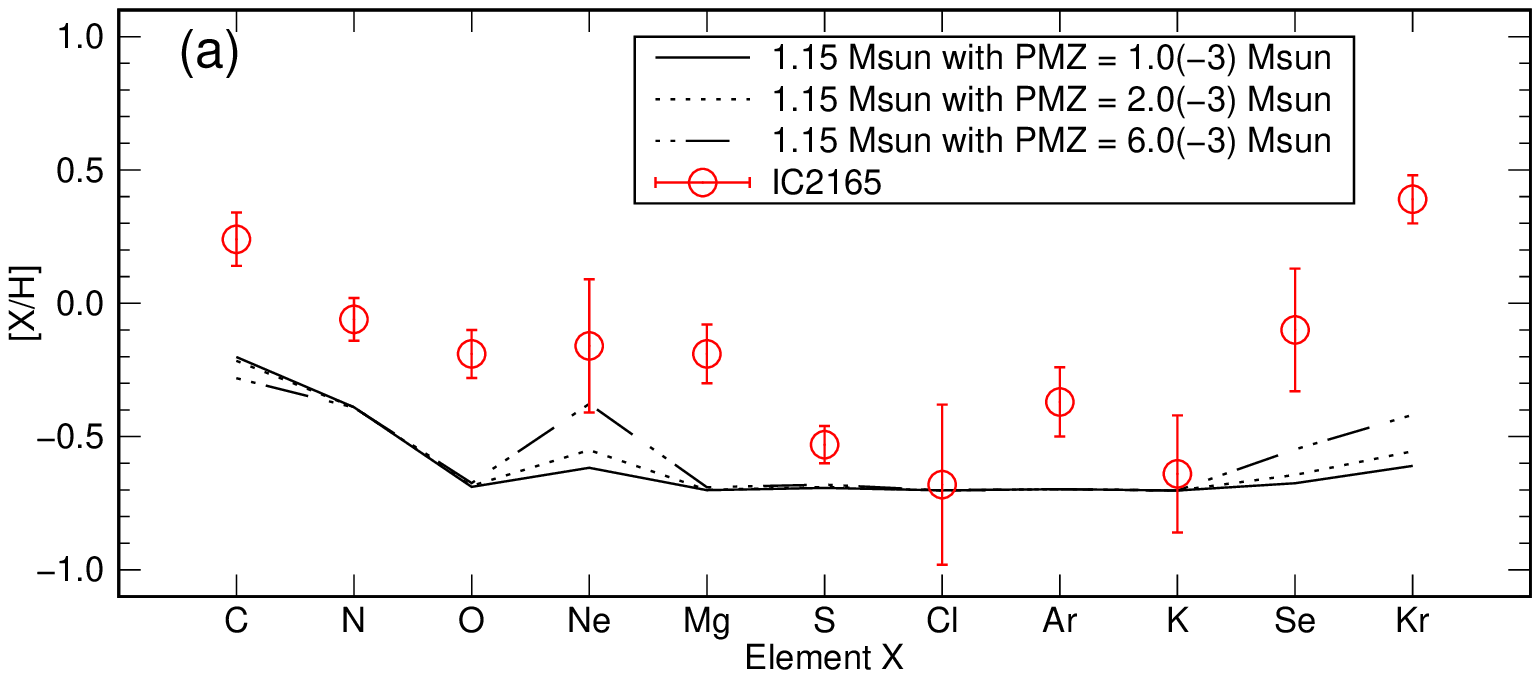}\\
\includegraphics[width=\columnwidth]{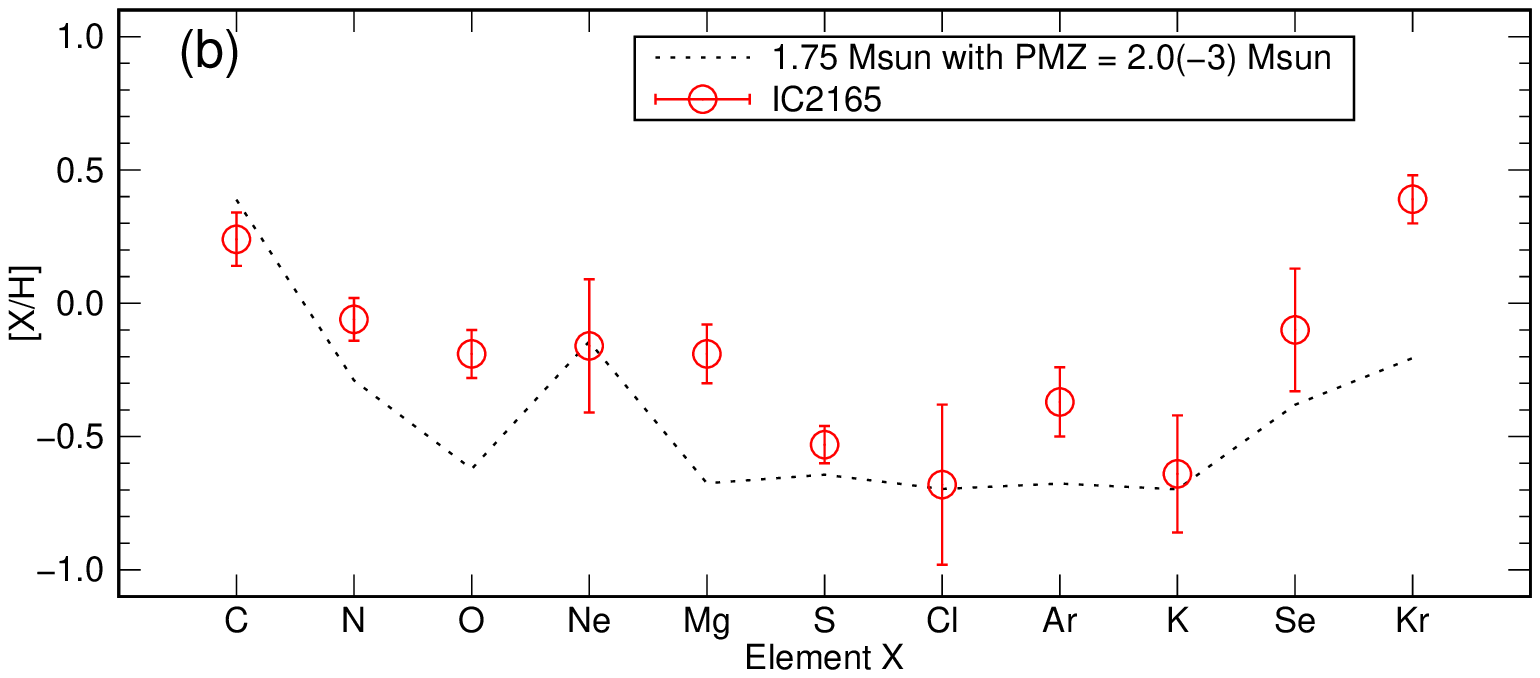}\\
\includegraphics[width=\columnwidth]{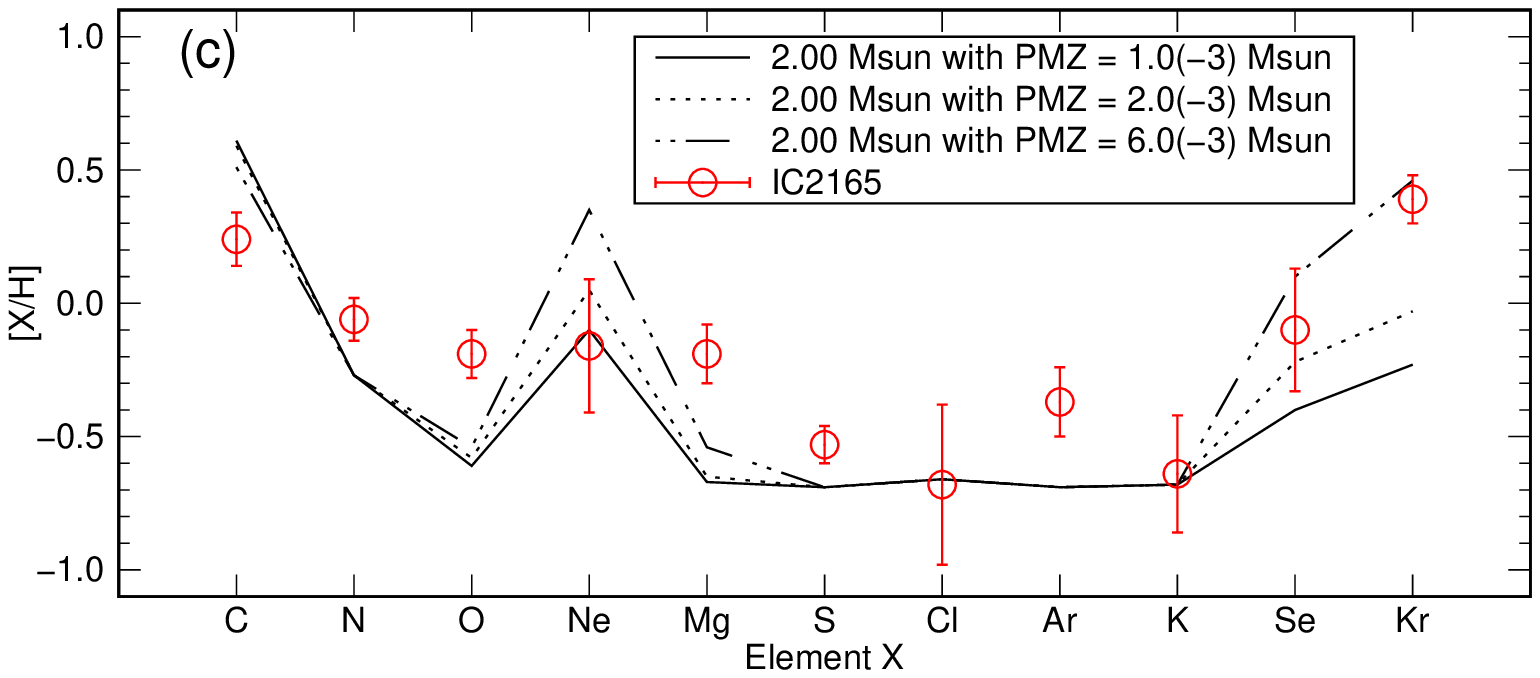}
\caption{Comparisons between the observed elemental abundances (red circles) 
and the AGB nucleosynthesis model predictions by \citet{Karakas:2018aa} 
for stars of initially $Z=0.003$ and 
1.15/1.75/2.00\,M$_{\sun}$ with different partial mixing zone (PMZ) masses. Here, we plot the CEL C value as the representative C abundance. 
}
\label{F-agb}
\end{figure}

\begin{figure*}
\centering
\includegraphics[width=\textwidth]{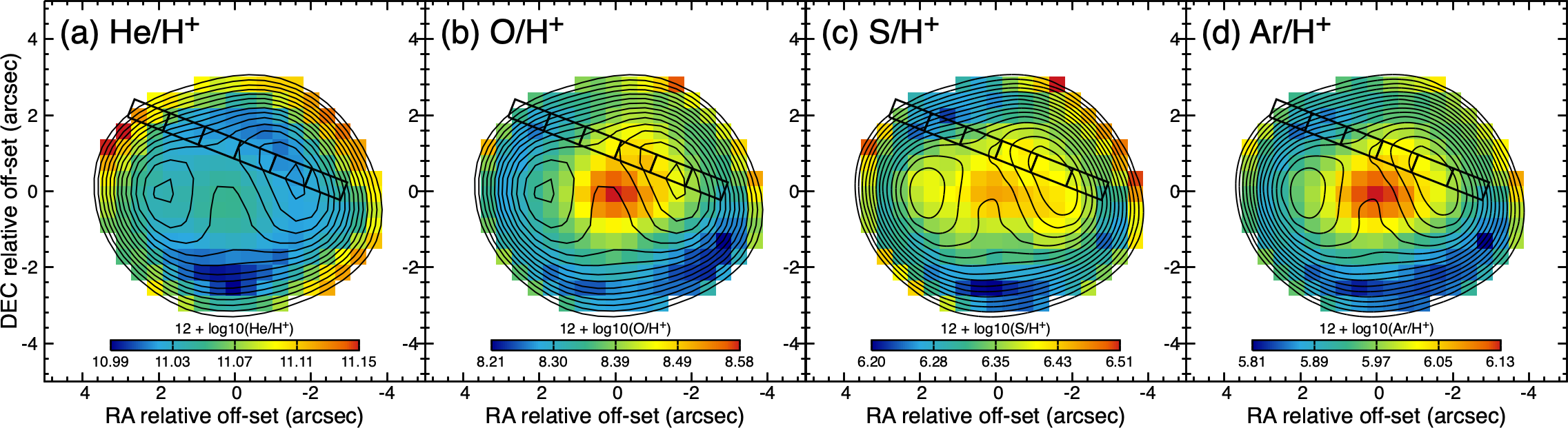}

\caption{The elemental abundance maps of He, O, S, and Ar and the location of 
the slit (black boxes) used in \citet{Miller:2019aa}. 
\label{F:ion-map4}}
\end{figure*}

We compare the derived elemental abundance pattern to the AGB nucleosynthesis models to determine how much of the progenitor has evolved into IC2165. 
To do so, we first calculate the metallicity $Z$. 
The O abundance is often used as a metallicity indicator. 
However, this element does not work, as explained below. 
Cl and Ar are the best metal indicators because they are hardly depleted by dust grains and molecules and are not increased during AGB nucleosynthesis. 
The average value between [Cl/H] and [Ar/H] indicates $Z$ of $0.003-0.005$. 

For comparisons, we select the results provided by \citet{Karakas:2018aa} for stars of initially 1.15, 1.75, and 2.00\,M$_{\sun}$ with $Z=0.003$. 
The model for 1.15\,M$_{\sun}$ stars are selected because many works concluded that the initial mass of IC2165 is $\sim$1.2\,M$_{\sun}$ 
\citep{Hyung:1994ab, Pottasch:2010aa, Bohigas:2013aa, Miller:2019aa}. 
These models artificially add a partial mixing zone (PMZ) at the bottom of the convective envelope during each third dredge-up (TDU). 
As a source of extra neutron ($n$), PMZ generates a $^{13}$C pocket.
In Fig.\,\ref{F-agb}, we plot the derived and the model-predicted abundances. 
These plots demonstrate that 
(1) C, Ne, and $n$-elements generally increase as the initial mass, and 
(2) Ne, Mg, and $n$-element production are extremely sensitive to the PMZ mass. 
The derived $n$-element abundances are inextricably linked to the PMZ mass.
The increased abundance of Ne, Se, and Kr suggests that the PMZ with $\ge 2.0{\times}10^{-3}$\, M$_{\sun}$ is required in all models.
Extra mixing from all convective borders, including the bottom of the convective shell, could explain the significant deviation of O from the model. 
Such extra mixing may result in enhancement of O, as \citet{2016MNRAS.458L.118G} provides observational and theoretical evidence for this enhancement of O in C-rich PNe. 
The derived Cl/H, O/Cl, and C/O values are in accordance with the models of \citet{2016MNRAS.458L.118G}. 
However, we cannot provide any explanations for the enhancement of Mg and Ar as per the models in the present.

Which mass model is more in accordance with the derived abundance pattern of IC2165? 
The 1.15\,M$_{\sun}$ model does not match with the derived abundance pattern at all, 
and the 2.00\,M$_{\sun}$ model predicts a large enhancement in C, implying that the initial mass is less than 2.00\,M$_{\sun}$. 
Thus, overall, the 1.75\, M$_{\sun}$ model provides a reasonable agreement up to light elements, except for O, Mg, and Ar, which should be excluded in the comparison. 
We could account for the derived $n$-element abundances with a larger PMZ mass 
(for stars with initially 1.75\,M$_{\sun}$, \citet{Karakas:2018aa} calculated for the models with PMZ = $2.0{\times}10^{-3}$\,M$_{\sun}$ only). 
Hence, we conclude that IC2165 is a descendant of a star with an initial mass of 1.75\, M$_{\sun}$.

\subsection{Comparison of the spatial distribution of physical and chemical parameters obtained from \emph{HST} and KOOLS-IFU}
\label{S-discuss1}

In principle, if the slit positions and its dimension are either 
exactly stated or can be inferred from the presented figures, 
one can recover the long-slit results from the IFU data to some degree. 
Here, we compare the results obtained from \emph{HST}/STIS by \citet{Miller:2019aa} 
and from KOOLS-IFU by us and discuss the spatial variation of elemental abundances.

\citet{Miller:2019aa} investigated the physical and chemical parameters at six locations 
by placing a single long-slit (slit width = 0.2{\arcsec} and 0.3{\arcsec}) 
through a northern part of the inner elliptical nebula's rim. 
Using Fig.\,1 in their report, we measure the coordinate of 
the center of the slit (RA(2000.0) = 06:21:42.79 and DEC(2000.0) = --12:59:12.97) 
and PA of +68.65$^{\circ}$. 
In each panel of Fig.\,\ref{F:ion-map4}, we illustrate the slit position on our abundance maps: each box has a dimension of 1.0{\arcsec} $\times$ 0.5{\arcsec}, and from west to east, each position is named as pos1, 2, 3, 4, 5 and 6 in \citet{Miller:2019aa}. 
Amongst them, the STIS spectrum extracted at pos6 seems to exhibit a low SNR; hence, 
the uncertainty of the obtained quantities is large.
Thus, we compare the quantities derived in pos1-5.
In comparison, we consider that the plate scale of STIS (0.03{\arcsec}\,pixel$^{-1}$) is significantly smaller 
than that of KOOLS-IFU (0.412{\arcsec}\,pixel$^{-1}$). 
Furthermore, the observation conditions and accuracy of telescope pointing are considerably different. 
Therefore, we measure the minimum/maximum and the average values within a pseudo slit corresponding to the STIS slit.

We summarise the comparison in Table\,\ref{T-miller}. 
In a similar manner to our report, \citet{Miller:2019aa} determined $c$({\hb}) using the H$\alpha$ and H$\beta$ lines 
in an iterative loop using {\te}({\oiii}) and {\Ne}(C\,{\sc iii}$]$). 
The {\heii} line contributions to {\ha} and {\hb} are subtracted. From 
the IP similarity and the same ionisation degree of C$^{2+}$ and Cl$^{2+}$ (24.4\,eV versus 23.8\,eV), 
{\Ne}(C\,{\sc iii}$]$) and {\Ne}({\cliii}) diagnose the same region. 
Thus, their derivation method of $c$({\hb}) is the same as ours, except for 
the number of the employed {\hi} lines. Taking into account the uncertainty of {\te}({\oiii}) 
reported in \S\,\ref{S-tene} (1000\,K), 
we can explain the offset of {\te}({\oiii}) between the report of \citet{Miller:2019aa} and ours. 
As expected, our maps reproduce the $c$({\hb}), {\te}, and {\Ne} variation found in \citet{Miller:2019aa}.

We specifically compare for He, O, S, and Ar, which are investigated by \citet{Miller:2019aa} and us. 
As shown in Table\,\ref{T-miller}, our maps reproduce their obtained spatial variations of these elements. 
However, it should be noted that \citet{Miller:2019aa} calculated all ionic abundances 
using {\te}({\oiii}) and {\Ne}(C\,{\sc iii}$]$), and in addition, 
could not measure {\te}({\nii}) and {\Ne}({\sii}) that are suitable for the O$^{+}$ derivation. 
Indeed, their O$^{+}$ abundance from the nebular {\oii}\,3726/29\,{\AA} lines are 
inconsistent with that from the auroral {\oii}\,7320/30\,{\AA} lines, 
strongly indicating incorrect adoption of {\te} for the O$^{+}$ abundance.
Most importantly, as we learn in Fig.\,\ref{F:2D-nete}, 
{\te}({\oiii}) and {\Ne}({\cliii}) distributions are 
quite different from {\te}({\nii}) and {\Ne}({\sii}). 
The uncorrected O$^{2+}$ contribution to the {\oii}\,7320/30\,{\AA} lines is also problematic. 
Their S and Ar abundances are calculated only on the basis of the S$^{2+}$ and Ar$^{2+}$ and large ICF values. However, we obtained the S and Ar maps using the S$^{+,2+}$ and Ar$^{2+,3+,4+}$ maps and determined the ICF maps for each element by considering the emission-line distribution and the ionisation fraction derived by our photoionisation model (\S,\ref{S-PI}). 
Thus, our O, S, and Ar abundances are not the same as \citet{Miller:2019aa}, even in the same region.

The uncertainties reported in \citet{Miller:2019aa} are 
$11-15$\,$\%$ in He, 
$18-34$\,$\%$ in O, 
$39-73$\,$\%$ in S, and 
$45-64$\,$\%$ in Ar. 
\citet{Miller:2019aa} argued that the reported sample PNe are chemically homogeneous 
and that observations taken anywhere across the PNe would accurately 
represent the nebula as a whole. Under such uncertainty owing to fewer diagnostics lines, inappropriate {\te}/{\Ne} selection, and high ICF values, this conclusion is expected, despite the fact that elemental abundances spatially vary. Another drawback in the study is that their observation did not cover the inner elliptical 
nebula where the elemental abundances are maximised. 

Our uncertainty in each map is reduced by a large number of diagnostic lines, appropriate {\te}/{\Ne} adoption, and low ICF values: 
$7-18$\,$\%$ in He, 
$25-49$\,$\%$ in O, 
$15-29$\,$\%$ in S, and 
$11-19$\,$\%$ in Ar. 
The spatial variation of elemental abundances except for He is larger than these uncertainties. 
Thus, based on our maps covering the entire nebula, we conclude that spatial variation of elemental abundances is not constant but varies, which is reflected by the central star's evolution. 
We will conduct detailed investigations on other PNe with IFU data to confirm the spatial variation of elemental abundances.

\begin{table}
\centering
\caption{Comparison of physical and chemical parameters obtained from \emph{HST}/STIS 
by \citet{Miller:2019aa} and KOOLS-IFU by us.
\label{T-miller}
}
\renewcommand{\arraystretch}{0.85}
\begin{tabularx}{\columnwidth}{@{}@{\extracolsep{\fill}}l@{\hspace{4pt}}rrrrrr@{}}
\midrule
           &\multicolumn{3}{c}{Ours}&\multicolumn{3}{c}{\citet{Miller:2019aa}}\\
Parameters &Ave. &Min. &Max. &Ave. &Min. &Max.\\
\midrule
$c$({\hb})                  &0.50   &0.41  &0.68   &0.46  &0.37 &0.65\\
{\te}({\oiii})  (K)         &14800  &12200 &16100  &13800 &13000 &14300\\
{\te}({\nii})   (K)         &12300  &7800  &16600  &\\
{\te}({\hei})   (K)         &10300  &6600  &16100  &\\
{\Ne}({\cliii}) (cm$^{-3}$) &7400   &5500   &8700  &\\
{\Ne}({\sii})   (cm$^{-3}$) &4300   &2600   &8700  &\\
{\Ne}(C\,{\sc iii}$]$) (cm$^{-3}$) &   &    &      &7500  &4500 &14200\\
He ($\times10^{-1}$) &1.06 &1.03 &1.16   &1.06  &1.03  & 1.14\\  
O  ($\times10^{-4}$) &2.53 &1.98 &3.14   &3.63  &3.41 &  3.96\\
S  ($\times10^{-6}$) &2.31 &1.67 &2.61   &1.77  &1.15 &  2.71\\ 
Ar ($\times10^{-7}$) &9.21 &7.09 &11.29  &9.91  &8.55 & 14.90\\
\midrule
\end{tabularx}
\end{table}

\subsection{Photoionisation modeling}
\label{S-PI}

The central star's luminosity ($L_{\ast}$) is one of the most important parameters for determining its current evolutionary status.
The masses of ejected gas and dust are also important parameters for studying stellar evolution.
These parameters cannot be determined without identifying the distance ($D$) towards the target object.
However, since the distance towards IC2165 is poorly measured, there remains no consistency between the observational and the theoretical results. 
Therefore, we first determine the exact distance that ensures consistency with both outcomes. 
Then, we construct the photoionisation model using {\sc Cloudy} to be consistent with all of the observational 
quantities using the derived distance. We verify the gas and dust masses, and GDR derived in the 2-D analysis.

We first establish the method to simultaneously determine $D$, $L_{\ast}$, $T_{\rm eff}$, and the total gas mass $m_{\rm g}$. 
We investigated the physical and chemical conditions within the mean radius of 4{\arcsec} in 2-D analysis. 
Thus, we maintain the outer radius of 4{\arcsec} through model iterations.

As the incident SED, we use the theoretical spectra calculated by non-LTE stellar atmosphere models for stars 
with [Z/H] = $-0.5$ and surface gravity $\log\,g$ = 6.3\,cm\,s$^{-2}$ provided by \citet{Rauch:2003aa}.
The flux density of their spectra is scaled with the extinction-free \emph{HST}/F547M band 
flux density, which is $6.76{\times}10^{-16}$\,erg\,s$^{-1}$\,cm$^{-2}$\,{\AA}$^{-1}$ at 5483.9\,{\AA}. 
For this correction, we adopt $c$({\hb}) of 0.44 around the position of the central star (\S\,\ref{S-chb}).
When adopting this type of incident SED, $D_{\rm kpc}$ as functions of $\log_{10}\,L_{\ast}$ and $T_{\rm eff}$ can be expressed in the following manner:
\begin{eqnarray}
D_{\rm kpc} &=& \left(\frac{L_{\ast}/L_{\sun}}{10^{0.8023T_{\rm eff}/10^{5} + 1.1338}}\right)^{0.5} ~~ \mbox{in kpc}.
\label{distance}
\end{eqnarray}

In terms of the elemental abundance pattern, we infer that the progenitor would be a single star of initially 1.75\,M$_{\sun}$ and $Z=0.003$.  
We assume that the progenitor is currently undergoing post-AGB H-burning evolution. However, no post-AGB evolution tracks are available for $Z=0.003$ stars.
We can write $\log_{10}\,L_{\ast}$ in $T_{\rm eff} = 110000 - 170000$\,K for 1.75\,M$_{\sun}$ 
by adopting linear interpolation for the H-burning post-AGB evolution tracks with $Z=0.004$ of \citet{Vassiliadis:1994ab} as below:
\begin{eqnarray}
\log_{10}\,(L_{\ast}/L_{\sun}) &=& -95.9808(\log_{10}\,T_{\rm eff})^{3} + \nonumber\\
                               && 1476.9947(\log_{10}\,T_{\rm eff})^{2} - \nonumber\\
                               && 7576.6797\log_{10}\,T_{\rm eff} + 12960.2188. 
\label{theolumi1}
\end{eqnarray}

Hence, given $T_{\rm eff}$, using Eqs.(\ref{distance}) and (\ref{theolumi1}), we can uniquely determine the distance and luminosity to be exactly 
along the theoretical post-AGB evolution track for such a mass star 
and in addition, $m_{\rm g}$ of $1.23{\times}10^{-2}$\,$D_{\rm kpc}$\,M$_{\sun}$ (\S\,\ref{S-gasmass}).

Although IC2165 is an elliptical nebula, we adopt a simple spherical shell in the model.
By employing Abel inverse transform as performed in \citet{Otsuka:2015aa}, we infer the respective hydrogen density radial profile $n_{\rm H}$($R$) ($R$ is the distance from the central star) along each minor and major axis from the 2-D hydrogen gas mass map (\S\,\ref{S-gasmass}). 
Then, we obtain the average $n_{\rm H}$($R$) after normalisation.

The abundances of C, O, Ne, S, Cl, and Ar are significant coolants in the ionised nebula. 
For C, we adopt the CEL $\epsilon$(C) value. 
The atomic data of elements up to Zinc (Zn) are prepared in {\sc Cloudy}. 
We replaced the originally used $A_{ji}$ and $\Omega$({\te}) with the values presented in Table\,\ref{T-atomf}. 
The uncalculated elements up to Zn are fixed to take the form of the predicted value by the AGB nucleosynthesis model of \citet{Karakas:2018aa} for stars 
of initially 1.75\,M$_{\sun}$. 
For the dust grains, we adopt the spherically shaped AC type amorphous carbon \citep{Rouleau:1991aa} with the radius of 0.1\,{\micron} (\S\,\ref{S-GDR}).

To determine $T_{\rm eff}$, $D$, and $L_{\ast}$, we run grid models with $T_{\rm eff}$ of $140000-160000$\,K at a constant interval of 2500\,K. 
While running these models, we provide allowance to vary the inner radius and AC grain mass to correspond with the observed/derived quantities, 
while we maintain $T_{\rm eff}$, $L_{\ast}$, gas filling factor ($f$, 0.45), and $\epsilon$(X). 
In Table\,\ref{T-modelin2}, we summarise the varied/fixed/derived parameters in this $D$-determination model. 
The quality of fit is computed from the reduced-$\chi^{2}$ value calculated from the 122 observational constraints; 
the {\hb} line flux $I$({\hb}) of $4.83{\times}10^{-11}$\,erg\,s$^{-1}$\,cm$^{-2}$, 
$m_{\rm g}$, and the average {\te}({\oiii}) of 14000\,K within 4{\arcsec} directly 
measured from their 2-D maps, outer radius (4{\arcsec}), 
110 atomic line fluxes relative to $I$({\hb}), 
five infrared band fluxes from 
\citet[\emph{WISE} 12{\micron} and 22{\micron} bands]{Cutri:2014aa}, \citet[\emph{AKARI}/IRC 18\,{\micron}]{Ishihara:2010aa}, and 
\citet[\emph{AKARI} FIS/65 and 90\,{\micron}]{Yamamura:2010aa}, and 
3 radio flux densities from \citet[31\,GHz]{Casassus:2007aa}, \citet[20\,GHz]{Chhetri:2015aa}, and \citet[1.4\,GHz]{Condon:1998aa}. 
Thus, we determine $T_{\rm eff} = 158640 \pm 4240$\,K, $D=4.68 \pm 0.31$\,kpc, $L_{\ast} = 5598 \pm 310$\,L$_{\sun}$, and $m_{\rm g} = 0.27 \pm 0.09$\,M$_{\sun}$ 
by researching $T_{\rm eff}$ to minimise the reduced $\chi^{2}$ value.

\begin{table}
\caption{
\label{T-modelin2}
Summary of the parameters in the $D$-determination model.
Note that $L_{\ast}$ and $D$ is the function of $T_{\rm eff}$. 
The optimised range of varied parameters and model constrains are explained in the text. 
}
\renewcommand{\arraystretch}{0.80}
\begin{tabularx}{\columnwidth}{@{}@{\extracolsep{\fill}}l@{\hspace{4pt}}l@{}}
\midrule
Varied parameters    &inner radius, AC grain mass ($m_{\rm d}$)\\
\midrule
Fixed parameters     &$T_{\rm eff}$, $L_{\ast}$, $D$\\
                     &nebular $\epsilon$(X) listed in Table\,\ref{T-elem}\\
                     &the unobserved nebular $\epsilon$(X): \citet{Karakas:2018aa}\\   
                     &$n_{\rm H}$($R$), $f$ (0.45), AC grain size (0.1${\micron}$)\\
\midrule
Derived parameters   &$\log_{10}\,I$({\hb}), inner radius, average {\te}({\oiii}), $m_{\rm g}$\\
                     &AC grain temperatures, $m_{\rm d}$, GDR\\
\midrule
\end{tabularx}
\end{table}

In the full modeling after determining $D$, $T_{\rm eff}$, $L_{\ast}$, and $\epsilon$(X), excluding all the unobserved elements, 
$n_{\rm H}$(H), $f$, and AC grain size are retained. The following 14 parameters are varied: 
the inner radius, 
12 observed elemental abundances, and AC grain mass. 
We did not set the optimal range of the inner radius and AC grain mass. 
We vary the abundances of the observed elements within two-$\sigma$ allowance, except for Mg, K, Ca, and Fe. 
Because the abundances of Mg, K, Ca, and Fe are determined using uncertain ICFs, their abundances are highly uncertain. 
Thus, we do not specify the optimal range of these four elements. 
In Table\,\ref{T-modelin3}, we summarise the varied/fixed/derived parameters in the full model. 
We terminate interactive calculations when either {\te} falls below 10000\,K or the gas mass exceeds 0.27\,M$_{\sun}$.

\begin{figure*}
\includegraphics[width=\textwidth]{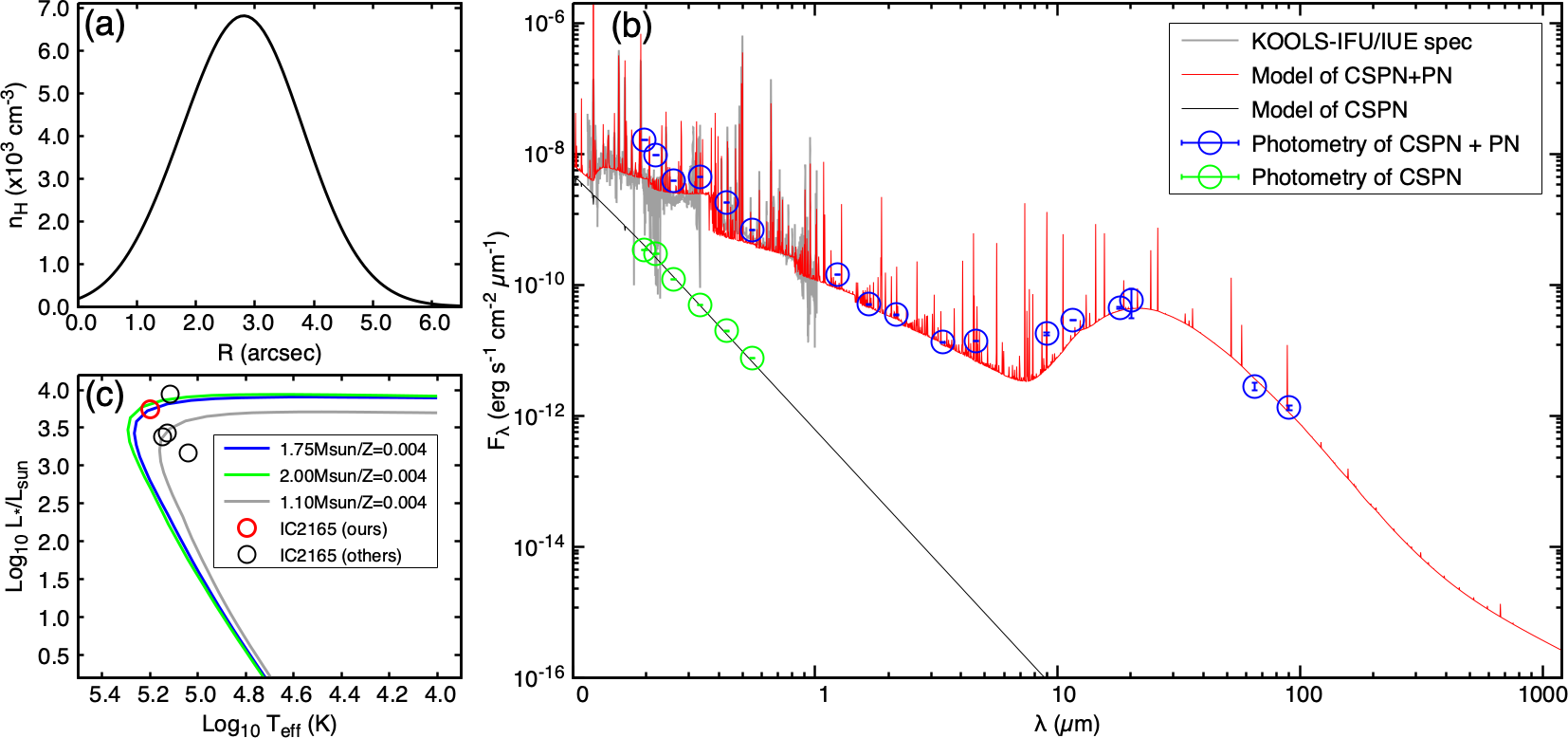}
\caption{
({\it panel (a)}) The adopted hydrogen density radial profile. $R$ is the distance from the central star.
({\it panel (b)}) The SED plots of the nebula (PN) and the central star (CSPN). The red and black lines are the model 
predicted SED of PN plus CSPN and CSPN, respectively. The UV-optical photometry are taken from Table\,\ref{T:HST}. The photometry points at $JHKs$ bands are 
the results of our own measurements for the relevant Two Micron All-Sky Survey \citep[2MASS,][]{Skrutskie:2006aa} images.
({\it panel (c)}) The location of IC2165 on the hydrogen burning post-AGB evolution tracks of stars of initially 1.10/1.75/2.00\,M$_{\sun}$ and $Z=0.004$.}
\label{F-sed}
\end{figure*}

\begin{table}
\caption{
\label{T-modelin3}
Summary of the parameters in the full modelling. 
The optimised range of varied parameters and model constrains are explained in the text. 
}
\renewcommand{\arraystretch}{0.80}
\begin{tabularx}{\columnwidth}{@{}@{\extracolsep{\fill}}l@{\hspace{4pt}}l@{}}
\midrule
Varied parameters    &nebular $\epsilon$(X) listed in Table\,\ref{T-elem}\\
                     &nebula inner radius, AC grain mass ($m_{\rm d}$)\\
\midrule
Fixed parameters     &$T_{\rm eff}$ (158640\,K), $L_{\ast}$ (5598\,L$_{\sun}$), $D$ (4.68\,kpc)\\    
                     &the unobserved nebular $\epsilon$(X): \citet{Karakas:2018aa}\\                        
                     &$n_{\rm H}$($R$), $f$ (0.45), AC grain size (0.1${\micron}$)\\
\midrule
Derived parameters   &$\log_{10}\,I$({\hb}), inner radius, average {\te}({\oiii}), $m_{\rm g}$\\
                     &AC grain temperatures, $m_{\rm d}$, GDR\\
\midrule
\end{tabularx}
\end{table}

\begin{table}
\renewcommand{\arraystretch}{0.80}
\caption{
\label{T-modelin}
The adopted and derived parameters in the best model of IC2165.}
\begin{tabularx}{\columnwidth}{@{}@{\extracolsep{\fill}}l@{\hspace{4pt}}l@{}}
\midrule
Central star&Values\\ 
\midrule
$T_{\rm eff}$ / $\log_{10}\,g$ / $L_{\ast}$ &158640\,K / 6.3\,cm\,s$^{-2}$ / 5598\,L$_{\sun}$\\
$[$Z/H$]$ / $M_{\rm V}$ / $D$     &--0.5 / 3.5367 / 4.68\,kpc    \\
\midrule
Nebula&Values\\
\midrule 
$\epsilon$(X)    &He: 11.04, C: 8.58,  N: 7.89, O: 8.42, Ne: 7.77, Mg: 7.51\\
                 & S: 6.47, Cl: 4.82, Ar: 6.07, K: 4.70, Ca: 4.58, Fe: 5.27\\
                 &Others: \citet{Karakas:2018aa}\\
Geometry         &spherical shell nebula\\
                 &inner radius = 4231\,AU (0.91{\arcsec})\\
                 &outer radius = 18752\,AU (4.00{\arcsec})\\
$n_{\rm H}$($R$) &   See Fig.\,\ref{F-sed}(a)\\
Filling factor ($f$)                  &0.45\\
$\log_{10}\,I$({\hb})&--10.316\,erg\,s$^{-1}$\,cm$^{-2}$ (within 4.00{\arcsec})\\
Average {\te}({\oiii}) &14054\,K\\
Gas mass ($m_{\rm g}$)               &0.270\,M$_{\sun}$\\
\midrule
AC grain &Values\\
\midrule 
Particle size                &0.1\,{\micron} (radius)\\
Temperature                  &$84 - 132$\,K\\ 
Dust mass ($m_{\rm d}$)                        &$3.77{\times}10^{-4}$\,M$_{\sun}$\\
GDR             &709\\
\midrule
\end{tabularx}
\end{table}

\begin{table*}
\caption{Comparison between the {\sc Cloudy} model prediction and the observation. $I$({\hb}) in both the model and observation is $4.83\times10^{-11}$\,erg\,s$^{-1}$\,cm$^{-2}$.
\label{T-modelC}
}
\centering
\renewcommand{\arraystretch}{0.75}
\begin{tabularx}{\textwidth}{@{}@{\extracolsep{\fill}}D{.}{.}{-1}lD{.}{.}{-1}D{.}{.}{-1}D{.}{.}{-1}lD{.}{.}{-1}D{.}{.}{-1}
@{}}
\midrule
\multicolumn{1}{c}{$\lambda$} & Ion & \multicolumn{1}{c}{Model} & \multicolumn{1}{c}{Obs} & \multicolumn{1}{c}{$\lambda$} & Ion & \multicolumn{1}{c}{Model} & \multicolumn{1}{c}{Obs} \\ 
\multicolumn{1}{c}{({\AA})} &  & \multicolumn{1}{c}{[$I$({\hb})=100]} & \multicolumn{1}{c}{[$I$({\hb})=100]} & \multicolumn{1}{c}{({\AA})} &  & \multicolumn{1}{c}{[$I$({\hb})=100]} 
& \multicolumn{1}{c}{[$I$({\hb})=100]} \\ 
\midrule
1402.00 & [O\,{\sc iv}] & 24.987 & 53.868 & 6073.99 &  {\heii} & 0.092 & 0.099 \\ 
1486.00 & [N\,{\sc iv}] & 42.442 & 45.711 & 6086.97 & {\fevii} & 0.078 & 0.062 \\ 
1602.00 & [Ne\,{\sc iv}] & 3.496 & 5.113 & 6101.83 & [K\,{\sc iv}] & 0.295 & 0.285 \\ 
1666.00 & O\,{\sc iii}] & 46.152 & 42.094 & 6118.06 &  {\heii} & 0.107 & 0.089 \\ 
1750.00 & [N\,{\sc iii}] & 39.054 & 29.244 & 6170.48 &  {\heii} & 0.126 & 0.152 \\ 
1909.00 & [C\,{\sc iii}]+C\,{\sc iii}] & 978.009 & 879.855 & 6233.61 &  {\heii} & 0.150 & 0.152 \\ 
2320.95 & {\oiii} & 5.781 & 5.174 & 6310.64 &  {\heii} & 0.180 & 0.167 \\ 
2326.00 & [C\,{\sc ii}] & 48.585 & 58.312 & 6312.06 & {\siii} & 2.168 & 1.948 \\ 
2385.35 & {\heii} & 4.459 & 6.117 & 6363.78 & {\oi} & 0.130 & 1.095 \\ 
2424.00 & [Ne\,{\sc iv}] & 62.522 & 68.473 & 6406.17 &  {\heii} & 0.219 & 0.223 \\ 
2511.15 & {\heii} & 6.803 & 7.227 & 6435.12 & {\arv} & 0.883 & 0.663 \\ 
2733.24 &  {\heii} & 11.120 & 10.721 & 6548.05 & {\nii} & 14.532 & 14.567 \\ 
2782.76 & [Mg\,{\sc v}] & 9.636 & 5.883 & 6559.91 & {\heii} & 5.797 & 6.213 \\ 
3203.04 &  {\heii} & 20.194 & 20.637 & 6562.81 &  {\hi} & 278.729 & 273.150 \\ 
4199.71 &  {\heii} & 0.911 & 0.769 & 6583.45 & {\nii} & 42.838 & 43.340 \\ 
4267.00 & C\,{\sc ii} & 0.240 & 0.380 & 6678.15 &  {\hei} & 2.654 & 2.815 \\ 
4338.55 &  {\heii} & 1.227 & 1.129 & 6682.98 &  {\heii} & 0.341 & 0.343 \\ 
4340.46 & {\hi} & 47.430 & 46.680 & 6716.44 & {\sii} & 2.131 & 2.343 \\ 
4363.00 & {\oiii} & 23.132 & 21.820 & 6730.82 & {\sii} & 3.972 & 3.928 \\ 
4387.93 &  {\hei} & 0.438 & 0.598 & 6795.16 & [K\,{\sc iv}] & 0.063 & 0.087 \\ 
4471.49 &  {\hei} & 3.588 & 4.164 & 6890.67 &  {\heii} & 0.437 & 0.392 \\ 
4541.46 &  {\heii} & 1.715 & 1.716 & 7005.83 & {\arv} & 1.877 & 1.248 \\ 
4685.64 & {\heii} & 48.089 & 49.681 & 7135.79 & [Ar\,{\sc iii}] & 12.450 & 9.978 \\ 
4701.62 &  {\feiii} & 0.027 & 0.043 & 7263.33 & {\ariv} & 0.170 & 0.179 \\ 
4711.26 & {\ariv} & 4.091 & 4.836 & 7281.35 &  {\hei} & 0.767 & 0.630 \\ 
4725.00 & {\neiv} & 0.538 & 0.937 & 7323.00 & {\oii} & 5.183 & 3.342 \\ 
4740.12 & {\ariv} & 4.955 & 5.502 & 7332.00 & {\oii} & 4.160 & 2.632 \\ 
4859.18 &  {\heii} & 2.346 & 2.315 & 7332.15 & {\ariv} & 0.029 & 0.034 \\ 
4921.93 &  {\hei} & 0.928 & 0.951 & 7530.54 & {\cliv} & 0.332 & 0.376 \\ 
4958.91 &  {\oiii} & 447.330 & 420.501 & 7592.50 &  {\heii} & 0.780 & 0.627 \\ 
5006.84 &  {\oiii} & 1334.643 & 1260.607 & 7751.11 & [Ar\,{\sc iii}] & 2.986 & 2.463 \\ 
5015.68 &  {\hei} & 2.137 & 2.131 & 8045.62 & {\cliv} & 0.767 & 0.907 \\ 
5047.64 &  {\hei} & 0.178 & 0.210 & 8236.51 & {\heii} & 0.963 & 1.238 \\ 
5145.76 & {\fevi} & 0.131 & 0.076 & 8333.74 &  {\hi} & 0.138 & 0.063 \\ 
5158.89 & {\fevii} & 0.025 & 0.047 & 8345.51 &  {\hi} & 0.154 & 0.103 \\ 
5176.05 & {\fevi} & 0.122 & 0.055 & 8358.96 &  {\hi} & 0.174 & 0.171 \\ 
5191.82 & [Ar\,{\sc iii}] & 0.193 & 0.151 & 8392.35 &  {\hi} & 0.225 & 0.177 \\ 
5199.00 & [N\,{\sc I}] & 0.043 & 0.385 & 8413.28 &  {\hi} & 0.260 & 0.221 \\ 
5270.40 &  {\feiii} & 0.054 & 0.056 & 8437.91 &  {\hi} & 0.303 & 0.311 \\ 
5276.38 & {\fevii} & 0.006 & 0.023 & 8467.21 &  {\hi} & 0.357 & 0.262 \\ 
5309.11 & [Ca\,{\sc v}] & 0.070 & 0.076 & 8502.44 &  {\hi} & 0.425 & 0.419 \\ 
5323.28 & {\cliv} & 0.045 & 0.051 & 8545.34 &  {\hi} & 0.514 & 0.496 \\ 
5335.19 & {\fevi} & 0.067 & 0.057 & 8578.70 & {\clii} & 0.080 & 0.113 \\ 
5411.37 &  {\heii} & 3.567 & 3.633 & 8598.35 &  {\hi} & 0.630 & 0.675 \\ 
5484.85 & {\fevi} & 0.037 & 0.042 & 8664.98 &  {\hi} & 0.786 & 0.825 \\ 
5517.71 & {\cliii} & 0.410 & 0.402 & 8746.57 &  {\heii} & 0.035 & 0.033 \\ 
5537.87 & {\cliii} & 0.622 & 0.519 & 8750.43 &  {\hi} & 1.000 & 1.024 \\ 
5631.08 & {\fevi} & 0.050 & 0.035 & 8858.83 &  {\heii} & 0.046 & 0.043 \\ 
5720.71 & {\fevii} & 0.052 & 0.062 & 8862.74 &  {\hi} & 1.300 & 1.320 \\ 
5755.00 & {\nii} & 1.397 & 1.155 & 9068.62 & {\siii} & 18.024 & 16.035 \\ 
5875.64 &  {\hei} & 10.965 & 10.095 & 9224.90 &  {\heii} & 0.084 & 0.078 \\ 
5952.73 &  {\heii} & 0.053 & 0.050 & 9228.97 &  {\hi} & 2.392 & 2.294 \\ 
5976.82 &  {\heii} & 0.060 & 0.048 & 9344.62 &  {\heii} & 1.380 & 1.317 \\ 
6004.52 &  {\heii} & 0.069 & 0.060 & 9530.62 & {\siii} & 45.274 & 39.778 \\ 
6036.58 &  {\heii} & 0.079 & 0.053 & 10049.30 &  {\hi} & 5.130 & 5.337 \\ 
\midrule
\multicolumn{1}{c}{$\lambda$} & Band & \multicolumn{1}{c}{Model} & \multicolumn{1}{c}{Obs} &   & Band & \multicolumn{1}{c}{Model} & \multicolumn{1}{c}{Obs} \\ 
\multicolumn{1}{c}{({\micron})} &  & \multicolumn{1}{c}{[$I$({\hb}=100]} & \multicolumn{1}{c}{[$I$({\hb}=100]} &  &  & \multicolumn{1}{c}{$F_{\nu}$ (Jy)} & \multicolumn{1}{c}{$F_{\nu}$ (Jy)} \\ 
\midrule
11.56 & WISE band3 & 324.394 & 332.157 &  & 31\,GHz & 0.165 & 0.147 \\ 
18.00 & AKARI & 966.551 & 916.045 &  & 20\,GHz & 0.174 & 0.148 \\ 
22.09 & WISE band4 & 385.482 & 415.514 &  & 1.4\,GHz & 0.218 & 0.180 \\ 
66.70 & AKARI & 160.593 & 118.662 &  &  &  &  \\ 
89.20 & AKARI & 120.438 & 110.366 &  &  &  &  \\ 
\midrule
\end{tabularx}
\end{table*}

The input parameters and the derived quantities for the full model are summarised in Tables\,\ref{T-modelin}. 
The adopted $n_{\rm H}$($R$) is displayed in Fig.\,\ref{F-sed}(a). 
The quality of fit is computed from the reduced-$\chi^{2}$ value calculated from the same constraints used in the $D$-determination model. 
The best model has a reduced-$\chi^{2}$value of 21.2. 

Our model fairly reproduces the derived quantities of the gas and dust in our 2-D analysis (\S\,\ref{S-GDR}). 
Although it is expected that the gas mass and the average {\te}({\oiii}) as model constrains are 
perfectly reproduced, it is remarkable that the GDR within 4{\arcsec} is perfectly reproduced: 
678 in our 2-D map analysis in contrast to 709 in the model prediction. 
This implies that the model effectively reproduces the dust mass within the same region: $3.92{\times}10^{-4}$\, M$_{\sun}$ versus $3.77{\times}10^{-4}$\, M$_{\sun}$. 

The observed/model-predicted line fluxes and band fluxes are compared in Table\,\ref{T-modelC}. 
Our model underestimates the observed {\oi} and {\NI} line fluxes because it is weighted on the reproduction of the dust and ionised gas components.
This is due to the total gas mass exceeding 0.27\, M$ {\sun}$ before entering the PDRs. 
Basically, the larger $n_{\rm H}$($R$) value beyond 4{\arcsec} yields higher {\oi} and {\NI} fluxes, 
while it results in the increase of [{\cii}], {\nii}, and {\oii} fluxes.

We display the modelled SED of each central star and central star plus nebula in Fig.\,\ref{F-sed}(b). 
The observed \emph{WISE} band 1 (3.4\,{\micron}) and 2 (4.6\,{\micron}) flux densities slightly exceed the modelled one. 
The near-IR band excess could be explained by including dust grains of smaller size, and also, polycyclic aromatic hydrocarbon (PAH) molecules. 
Indeed, \citet{Ohsawa:2016aa} reports the detection of a neutral PAH band at 3.3\,{\micron}. 
However, since the PAH mass fraction concerning the dust mass would be very small, as confirmed in other C-rich PNe \citep[e.g.,][]{Otsuka:2017aa, Otsuka:2020aa}, 
the derived GDR and dust mass are not largely modified even while doing so.

\subsection{Post-AGB evolution}

In Fig.\,\ref{F-sed}(c), we plot the derived luminosities ($L_{\ast}$) and $T_{\rm eff}$ of the central star 
on the post-AGB evolutionary tracks for stars of initial values of 1.10, 1.75, and 2.00\,M$_{\sun}$ and $Z=0.004$. 
For comparison, we plot the results of \citet{Hyung:1994ab}, \citet[based on \citet{Pottasch:2004aa}]{Pottasch:2010aa}, \citet{Bohigas:2013aa}, and \citet{Miller:2019aa}. 

\citet{Bohigas:2013aa} derived $T_{\rm eff}$ of 130\,700\,K and $L_{\ast}$ of 8770\,L$_{\sun}$ without adopting any distances.
They concluded that the initial mass of the progenitor is $1.2-2.0$\,M$_{\sun}$ by plotting the derived $T_{\rm eff}$ and $L_{\ast}$ on theoretical post-AGB evolutionary tracks. 
Conversely, \citet{Miller:2019aa} derived $T_{\rm eff}$ of 110\,000\,K and $L_{\ast}$ of 2400\,L$_{\sun}$ at a distance of 2.4\,kpc in the same manner and estimated it to be 0.99\,M$_{\sun}$ in the same way. 
The former estimate is based on very uncertain derivations. And the latter is incompatible with the origin of IC2165 
because TDU is required for the evolution of C-rich objects, and it occurs in stars of initial 
$\gtrsim 1.2-1.5$\, M$_{\sun}$ predicted by theoretical models \citep[e.g.,][]{1987ApJ...313L..15L,1988ApJ...328..671B}. 
As we explained in \S\,\ref{S-AGB}, any AGB nucleosynthesis models for initial values of 1.15\,M$_{\sun}$ cannot indeed account for 
the observed abundance pattern.

Three plot points below the evolution tracks for the 1.75 and 2.00\,M$_{\sun}$ stars 
are taken from \citet[$D=2.4$\,kpc]{Hyung:1994ab}, 
\citet[$D=1.5$\,kpc]{Pottasch:2010aa}, and 
\citet[$D=2.4$\,kpc]{Miller:2019aa}. Of course, no other models except ours are in accordance with the H-burning 
post-AGB evolution track for 1.75\,M$_{\sun}$ stars. These results might suggest that the central star is in the course of the 
He-shell burning track. However, the model with $D=2.4$\,kpc, with much less hydrogen density and smaller radius are needed 
to match the observed {\hb} line flux. The average {\te}({\oiii}) within the radius of 4{\arcsec} 
would become higher than that in the model with $D=4.68$\,kpc, resulting in the overprediction of high-excitation line fluxes such as {\oiii}, {\ariv}, and {\arv}. 
Because less dust is required to match the infrared fluxes, the predicted GDR would be greater than the observed one. 
Thus, the model with $D=2.4$\, kpc cannot account for all the observed/derived quantities. 

Even if the most recent Gaia distance \citep[2.89\,kpc;][]{Chornay:2021aa} is adopted, the initial mass of IC2165 is estimated to be 
$\sim$1.2\,M$_{\sun}$. 
While adopting Gaia and pre-Gaia distances, there is still a discrepancy between the estimated initial masses via observation and theoretical prediction. 
Thus, our model with $D=4.68$\,kpc offers a better fit to all the observational data and is consistent with the predictions of the AGB theoretical model.
That is, the central star is in the course of the H-burning track, and it is about to reach its maximum effective temperature.

\subsection{Carbon production of IC2165}

\begin{table}
\caption{Comparison of gas and dust masses between 2D emission analysis and {\sc CLOUDY} modelling. 
See text for details.}
\centering
\renewcommand{\arraystretch}{0.85}
\begin{tabularx}{\columnwidth}{@{}@{\extracolsep{\fill}}lcc} 
\midrule  
Element & 2D emission analysis (M$_{\sun}$) & {\sc CLOUDY} modelling (M$_{\sun}$)\\
\midrule
H       &$1.93{\times}10^{-1} \pm 8.41{\times}10^{-2}$ &$1.85{\times}10^{-1}$\\
He      &$7.29{\times}10^{-2} \pm 2.28{\times}10^{-2}$ &$8.27{\times}10^{-2}$\\
C       &$2.35{\times}10^{-3} \pm 7.48{\times}10^{-4}$&$8.45{\times}10^{-4}$\\
O       &$6.59{\times}10^{-4} \pm 2.63{\times}10^{-4}$&$7.71{\times}10^{-4}$\\
\midrule 
AC dust &$3.92{\times}10^{-4} \pm 1.96{\times}10^{-4}$&$3.77{\times}10^{-4}$\\
\midrule 
\end{tabularx}
\label{T:Cmass}
\end{table}

\citet{Karakas:2018aa} predict that stars of initially 1.75\,M$_{\sun}$ and $Z=0.003$ eject $\sim$1.1\,M$_{\sun}$ during their life. 
The ejected mass in the final thermal pulse is $\sim$0.54\, M$_{\sun}$, indicating that we trace approximately half of the gas mass ejected during this event. 
However, it is not well known when and how much mass the PN progenitors eject. 
Despite the metallicity being only slightly different, stars of initially 1.75\,M$_{\sun}$ and $Z=0.004$ eject 
$\sim$0.24\,M$_{\sun}$ in the last thermal pulse, according to \citet{Karakas:2007aa}. 

The total ejected gas during the AGB phase is a reliable prediction compared with the mass-loss rate.
Direct calculations of gas and dust masses in IC2165 can answer critical questions regarding the amount of carbon atoms and carbonaceous dust synthesised by the low-mass PN progenitors during their lives. 
In Table\,\ref{T:Cmass}, we list gas and dust masses in 2-D emission analysis and {\sc Cloudy} modeling. 
Note that the atomic C mass measured from the 2D emission analysis is based on the C RLs, while the mass measured from {\sc CLOUDY} modeling is based on the C CELs. 
The total gas mass in both results is consistent because the majority is the H and He gas masses, although the C mass in each result is calculated from different types of emissions. When adopting the {\sc Cloudy} result (which adopts the CEL C abundance), the sum of the atomic and dust C masses is $1.23{\times}10^{-3}$\,M$_{\sun}$ within 4{\arcsec}. 
That is, about 31\,$\%$ of the total C might exist as dust grains. 
Adopting the RL C mass derived from our 2-D analysis, this fraction is down to $\sim$14\,$\%$. 
In either case, the mass fraction of carbon existing as dust grains is extraordinarily higher than we expected. 
C mass and its abundance to H could be underestimated if the C mass locked in dust grains is not counted.

According to \citet{Karakas:2018aa}, stars of initially 1.75\,M$_{\sun}$ and $Z=0.003$ ejected the C mass of $6.02{\times}10^{-3}$\,M$_{\sun}$. 
Assuming that 30\,$\%$ of the synthesised C exists as dust grains, the progenitor might have produced the carbonaceous dust of $1.81{\times}10^{-3}$\,M$_{\sun}$ in total. 
If that is the case, we detect $\sim$20\,$\%$ of the dust mass ejected from the progenitor. 
The remaining dust might be detected beyond 4{\arcsec}.

\section{Summary}
\label{S-sum}

We used multiwavelength spectra from the Seimei 3.8-m telescope/KOOLS-IFU and archived spectra to conduct a comprehensive study of the gas and dust components in IC2165. 
This is the first time in the spatial study of the entire area of this PN. 
Using the KOOLS-IFU and ESO/VIMOS data, we created emission-line maps with their spatial resolution of $\sim$1.3{\arcsec}.

We derived the extinction map via a self-consistent way. 
The extinction map clearly shows a wide variation in spatial. 
We solely derived the circumstellar GDR map. 
It is critical to note that GDR varies radially from 1210 in the central nebula filled with hot gas plasma to 120 near the ionisation front. The obtained GDR is comparable to the values generally adopted for C-rich AGB stars and ISM. 
Except for the inner regions, GDR in IC2165 is nearly the same as that 
in such AGB stars, indicating that most dust grains endure the harsh radiation field without being destroyed.

We derived the spatial distribution of the 16 ionic abundances and their elemental maps. 
These maps indicate that elements regularly vary in space and are reflected by the central star’s evolution.

We succeeded in deriving the spatial distribution of the gas and dust masses. 
Their masses have peaks at the edges of the minor axis of the inner elliptical nebula. 
Such gas and dust mass distributions may be related to the nonisotropic mass loss during the AGB phase and nebula shaping.

We determined the representative abundance value of 13 elements using PSF-matched spatially integrated spectra extracted from the same aperture. 
Their values are consistent with the theoretical model for stars of initially 1.75\,M$_{\sun}$ and $Z=0.003$ with PMZ mass of 2.0$\times$10$^{-3}$\,M$_{\sun}$. 
The O enhancement confirmed in IC2165 would be due to extra mixing.

Finally, we constructed the photoionisation model to be consistent with all quantities derived from the 1-D and 2-D analyses 
and the post-AGB evolution track for 1.75\,M$_{\sun}$ stars. 
We found that $\sim$$30$\,$\%$ of the C mass synthesised in the progenitor is locked as carbonaceous grains. 
The C mass and its abundance are probably underestimated without counting the C atom contained in dust grains. 

It has been believed that the circumstellar extinction should be isotropic and equal to the ISM value. 
Furthermore, it is sometimes problematic that the $c$({\hb}) value varies according to the authors, although their $c$({\hb}) measurements are performed using the spectra obtained at different positions. Moreover, we tend to regard the error of measurement as the origin of the $c$({\hb}) differences. 
Our $c$({\hb}) map demonstrates that such an interpretation is possibly not valid. 
It is natural that $c$({\hb}) spatially varies because it reflects circumstellar dust and gas distribution.  
Applying the $c$({\hb}) value measured in a part of the nebula to the other parts is not preferred.
We cannot accurately understand morphology-kinematics using velocity channel maps or plasma diagnostics without correcting internal extinction variation because these investigations must be performed using extinction-free line-flux distributions.

Through the present work, we demonstrated the capability of Seimei/KOOLS-IFU and how the spatial variation of the gas and dust components in PNe 
derived from IFU observations with extra efforts can shed light on the evolution of the circumstellar/interstellar medium. 
We will continue to study the sample PNe taken with Seimei/KOOLS-IFU.

\section*{Acknowledgements}

I sincerely thank all the people involved in the Seimei project. 
In particular, I greatly thank 
K.~Isogai, F.~Iwamuro, H.~Izumiura, M.~Kino, M.~Kurita, D.~Kuroda, H.~Maehara, K.~Matsubayashi, and Y.~Nakatani. 
I would like to express my sincere thanks to Beth Sargent for her kind help to improvement of the paper. 
This study was supported by JSPS Grants-in-Aid for Scientific Research(C) (JP19K03914). 
This research is in part based on observations obtained using \emph{AKARI}, a JAXA project with the participation of ESA. 
I dedicate this paper to my parents for their deep love.

\section*{DATA AVAILABILITY}
The Seimei KOOLS-IFU, VLT VIMOS, and \emph{HST} data used in this article are available in their respective public archives.

\bibliographystyle{mnras}

\appendix

\section{Flux calibration of KOOLS-IFU data}
\label{A:flux}

\begin{figure}
\centering
\includegraphics[height=\columnwidth,angle=-90]{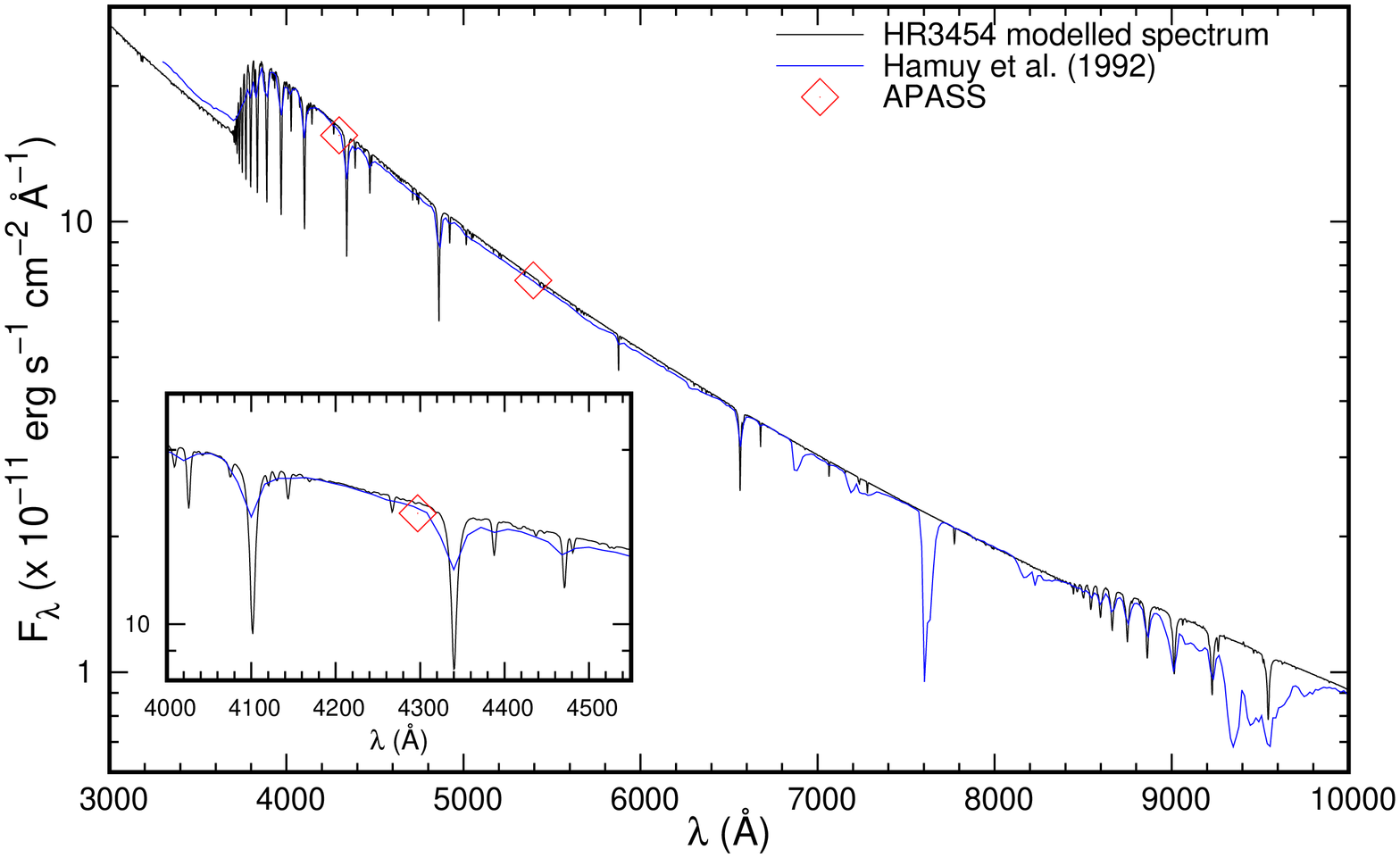}
\caption{
The $3000-10000$\,{\AA} spectrum of HR3454 synthesised by the non-LTE theoretical stellar atmospheres modelling code {\sc TLUSTY}. 
The flux density is reddened and scaled using the interstellar extinction function computed by the reddening law of
\citet{Cardelli:1989aa} with $E(B-V) = 0.01$ at $R_{V} = 3.1$ and 
the APASS $V$-band photometry \citep[$m_{B} = 4.052$, $m_{V} = 4.256$;][]{Zacharias:2012aa}. 
The spectral resolution ($R = \lambda/{\rm FWHM}$) is set to be a constant 2100 corresponding to the VPH-683 resolution. 
The spectrum obtained by \citet{Hamuy:1992aa} is plotted as a comparison. 
The plots in $4000-4500$\,{\AA} are closed-up in the inset. See text for details.
\label{F:h3454}}
\end{figure}

In the flux calibration, one might sometimes use the database of flux standard star spectra prepared in {\sc IRAF} \citep[e.g.,][]{Hamuy:1992aa}. 
As we know, however, the recorded telluric absorption (whose depth is naturally different in each observatory site) and the sparse spectral bin (e.g., 16\,{\AA} in HR3454) makes it harder to calibrate wavelengths in Balmer/Paschen jump and metal absorption forest around $4000-5000$\,{\AA} (Fig.\,\ref{F:h3454}). Worse, the data of several standards show strange SED and artefact discontinuity between blue and red spectra. 
These issues would finally propagate incorrect derivations of important physical and chemical parameters. 
Hence, to improve such difficulties in flux calibration and telluric removal, we produced a 
modelled spectrum of HR3454 by fitting the archived $3690-10480$\,{\AA} high-dispersion spectrum ($R$$\sim$68000) 
taken by Canada-France-Hawaii telescope/ESPaDOnS \citep{Donati:2003aa} with the B-type star grid model BStar2006 \citep{Lanz:2003aa} using 
the non-LTE model stellar atmospheres modeling code {\sc TLUSTY} \citep{Hubeny:1988aa}. 
We determined 15 photospheric abundances: the respective 
$\epsilon$(He/C/N/O/Ne/Mg/Al/Si/P/S/Cl/Ar/Cr/Fe/Zn) are 10.80, 8.25, 8.20, 8.70, 7.86, 6.93, 6.00, 7.00. 5.45, 6.90, 5.40, 6.55, 5.53, 7.35, and 6.10, where $\epsilon$(X) is the number density ratio with respect to H in the form of 12 + $\log_{10}$\,$n$(X)/$n$(H). 
$T_{\rm eff}$, surface gravity ($\log\,g$), microturbulence, and rotational velocity ($v\,{\rm sin}(i)$) are 17\,500\,K, 3.68\,cm\,s$^{-2}$, 2\,{\kms}, and 95\,{\kms}, respectively. 
Since this modelled spectrum is not dereddened by the interstellar medium, we need to estimate the value of $E(B-V)$ towards HR3454.  
We determined $E(B-V) = 0.01$ at the selective extinction $R_{V} = 3.1$ (the average value in the Milky Way) by applying 
minimum $\chi^{2}$ method in comparison with the reddened spectrum of HR3454 registered 
in the Indo-U.S. Library of Coud\'{e} Feed Stellar Spectra \citep{2004ApJS..152..251V}.
Here, we adopted interstellar extinction law by \citet{Cardelli:1989aa}. 
We adopted the AAVSAO $V$-band magnitude \citep[4.256;][]{Zacharias:2012aa}. 
The resultant modelled spectrum is presented in Fig.\,\ref{F:h3454}. 
Thus, we simultaneously performed flux calibration and telluric removal by using this synthesised {\it telluric-free} and {\it reddened} 
HR3454 spectrum with the same spectral resolution and plate scale as each VPH-blue/red and 683. 
Finally, we obtained a single $4200-10100$\,{\AA} datacube by sensitivity-function weighted combining of the VPH-blue and red spectra after 
PSF matching as explained in \S\,\ref{S-reduc}.

\section{Supporting Results}

\begin{table}
\centering
\renewcommand{\arraystretch}{0.85}
\renewcommand{\thetable}{B\arabic{table}}
\caption{The adopted atomic data of CELs. \label{T-atomf}}
\begin{threeparttable}
\begin{tabularx}{\columnwidth}{@{}@{\extracolsep{\fill}}l@{\hspace{0pt}}ll@{}}
\midrule
Line &transition probability $A_{ji}$ &effective collision strength $\Omega$({\te})\\
\midrule
$[${\cii}$]$ &\citet{Nussbaumer:1981aa};   &\citet{Blum:1992aa}\\
             &\citet{Froese:1994aa}\\
{\ciii}$]$   &\citet{Wiese:1996aa} &\citet{Berrington:1985aa}\\
{\NI}        &\citet{Wiese:1996aa} &\citet{Pequignot:1976aa}; \\
             &                     &\citet{Dopita:1976aa}\\
{\nii}       &\citet{Wiese:1996aa} &\citet{Lennon:1994aa}\\
{\niii}      &\citet{Wiese:1996aa} &\citet{Blum:1992aa}\\
N\,{\sc iv}$]$       &\citet{Wiese:1996aa}                     &\citet{Ramsbottom:1994aa}\\
{\oi}        &\citet{Wiese:1996aa} &\citet{Bhatia:1995aa}\\
{\oii}       &\citet{Wiese:1996aa} &\citet{McLaughlin:1993aa}; \\
             &                     &\citet{Pradhan:1976aa}\\
{\oiii}      &\citet{Wiese:1996aa} &\citet{Lennon:1994aa}\\
{\oiv}       &\citet{Wiese:1996aa} &\citet{Blum:1992aa}\\
{\neiv}      &\citet{Becker:1989aa}; &\citet{Ramsbottom:1998aa}\\
             &\citet{Bhatia:1988aa} \\
{\mgv}       &\citet{Mendoza:1983aa}; &\citet{Butler:1994aa}\\
             &\citet{Kaufman:1986aa}\\
{\sii}       &\citet{2009JPCRD..38..171P} &\citet{Ramsbottom:1996aa}\\
{\siii}      &\citet{2009JPCRD..38..171P} &\citet{Galavis:1995aa}\\
{\clii}      &CHIANTI Data Base$^{\rm a}$ &\citet{Tayal:2004ab}\\
{\cliii} &\citet{Mendoza:1982aa};  &\citet{Ramsbottom:2001aa}\\
         &\citet{Kaufman:1986aa}\\
{\cliv}  &\citet{Mendoza:1982ab};&\citet{Galavis:1995aa}\\
         &\citet{Ellis:1984aa};\\
         &\citet{Kaufman:1986aa}\\
{\ariii} &\citet{Mendoza:1983aa}; &\citet{Galavis:1995aa} \\
         &\citet{Kaufman:1986aa}\\
{\ariv}  &\citet{Mendoza:1982aa};  &\citet{Zeippen:1987aa}\\
         &\citet{Kaufman:1986aa}\\
{\arv}   &CHIANTI Data Base$^{\rm a}$&\citet{Galavis:1995aa}\\
{\kiv}   &\citet{Mendoza:1983aa};&\citet{Galavis:1995aa}\\
         & \citet{Kaufman:1986aa}\\
$[$Ca\,{\sc v}$]$       &\citet{Mendoza:1983aa};                      &\citet{Galavis:1995aa}\\
        &\citet{Kaufman:1986aa}\\
{\feiii} &\citet{Garstang:1957aa}; & \citet{Zhang:1996aa}\\
         &\citet{Nahar:1996aa}\\
$[$Fe\,{\sc vi}$]$       &\citet{2000AaAS..147..111C}                     &\citet{1999AaAS..136..395C}\\
$[$Fe\,{\sc vii}$]$       &\citet{2008AaA...481..543W}                     &\citet{2008AaA...481..543W}\\
{\kriv}  &\citet{Biemont:1986aa} & \citet{Schoning:1997aa}              \\
\midrule
\end{tabularx}
 \begin{tablenotes}[flushleft] 
\item[a] \url{http://www.chiantidatabase.org}
\end{tablenotes}
\end{threeparttable}
\end{table}

\bsp 
\label{lastpage}
\end{document}